\documentclass[epj]{svjour}
\usepackage{graphicx}
\usepackage{ulem}
\begin{document}
\title{A New Radioactive Decay Mode,
   True Ternary Fission, the Decay of Heavy
  Nuclei Into Three Comparable Fragments.}
\author{W. von Oertzen\inst{1}
 A. K. Nasirov\inst{2}
}
\authorrunning{W. von Oertzen and A. K. Nasirov}
\titlerunning{A New Radioactive Decay Mode}
\institute{Helmholtz-Zentrum Berlin, Hahn-Meitner Platz 1,
14109 Berlin, Germany,
and Fachbereich Physik Freie Universitaet Berlin,
\and {Bogoliubov Laboratory of
Theoretical Physics, JINR, Dubna 141980 Russia,
Institute of Nuclear Physics, Ulugbek, Tashkent,100214, Uzbekistan}}

\date{Received: date / Revised version: date}

\abstract{
The  ternary  cluster decay of heavy nuclei (e.g. in spontaneous fission of $^{252}$Cf(sf,fff)),
has been observed in several experiments with binary coincidences between two fragments using detector telescopes (the FOBOS-detectors) with very large solid angles  and placed on the opposite sides from the source of fissioning nuclei. The binary coincidences at a relative angle of 180$^0$ deg. correspond to binary fission or  to the decay into three cluster fragments by registration of two of them in coincidences with missing
 nuclei of different masses (e.g.$^{132}$Sn,$^{52-48}$Ca,$^{68-72}$Ni). This marks a new step in the physics of fission-phenomena of heavy nuclei. The new decay mode has been observed more  then 10 years ago by
 the FOBOS group in JINR (Dubna) Russia. These experimental results for the collinear cluster tripartition
 (CCT), refer to the decay into three clusters of comparable  masses. In the present work  we  discuss  the various  aspects of this ternary fission (FFF)   mode, with  different mass partitions.The question of collinearity is analysed on the basis of recent publications. Further  insight into the possible decay modes is obtained by the discussion of the path towards larger deformation, towards \textit{hyper-deformation} and
 by inspecting details of the potential energy surfaces (PES). The PES is determined as the total sum of the masses,  including the shell effects, which enter with  $Q_{ggg}$, the three-body Q-value  for the separation into three fragments, and the total interaction (nuclear and Coulomb) between all three nuclei. In the path towards  the extremely deformed states leading to ternary fission, the concept of deformed shells is most important. At the scission configuration the phase space determined by the PES's leads to the final mass distributions. The possibility of formation of fragments of almost equal size ($Z_i$ = 32, 34, 32, for  $Z$=98) and the observation of several other  fission modes  in the same system  can be  predicted by the PES. The PES's  show  pronounced minima and valleys, where the phase space for the decay reaches maximum values, namely for  several mass/charge combinations of ternary fragments, which correspond to  a variety of collinear ternary fission  (multi-modal) decays. The case of the decay of $^{252}$Cf(sf,fff) turns out to be unique due to the presence of deformed shells in the total system and of closed shells in all three nuclei in the decay.
}

%
\authorrunning{W. von Oertzen}
\titlerunning{Tue ternary Fission}
\maketitle

\section{Introduction, Binary Fission}
Binary fission of nuclei after irradiation of Uranium  with slow neutrons has been observed
by Otto Hahn and Fritz Strassman in 1938, \cite{Hahn}. The fragments of
Barium (charge $Z$ = 56)  have been identified  by methods of radio-chemistry.
This group from Berlin,
together with Lise Meitner, who had to escape from Nazi-Germany in 1938,  had previously studied
a variety of nuclear processes induced by neutrons.
 Actually L. Meitner urged Hahn
repeatedly to study in more detail the reaction products observed from the
 irradiation of
 Uranium-foils with neutrons. Different groups, in particular
Fermi (see Ref.~\cite{Fermi}) had published results, where new radio-activities have
 been observed and attributed to new
 elements heavier then Uranium. Very  controversial results of other groups followed
in the 1936-38's. Fermi actually received the Nobel-price for this work of finding new elements
(incorrect!) in 1938.

Lise Meitner had worked together with
O. Hahn  in Berlin  for many years. L. Meitner as an  Austrian citizen   of Jewish origin,
 had to escape from
 Nazi-Germany, she went  to Stockholm, Sweden  in the fall of 1938,
when Hitler invaded Austria (``Anschluss'').
Due to these circumstances she could not participate in the final phases of the
work of her  colleagues
in Berlin by O. Hahn and F. Strassman. They  published their results without
Lise Meitner,
 in  \textit{Naturwissentschaften}, {\bf 27} p11, (1939).
 However, O. Hahn was eager to tell L.Meitner
about his final result, already earlier and
with an exchange of letters in the fall of 1938 and in  Jan. 1939 between Stockholm,
Copenhagen and Berlin. Lise Meitner
 was able to produce a paper with her
nephew Otto Frisch (at that time in Copenhagen) which gave the first interpretation of
 the results of fission: L. Meitner und O. Frisch,  Ref.~\cite{Meitner}:
``\textit{Disintegration of Uranium by Neutrons,
a new type of Nuclear Reaction}''.
 Frisch actually introduced here the concept/expression of ``fission'' for
the splitting of Uranium-nuclei. Meitner and Frisch were the first to realize the very large
energy-release involved in fission:
$>$ 200 MeV, a very large (enormous!) value for the energy-release in a nuclear decay, at that time
such a high value has  never been observed  before.
 With this result an avalanche of publications started in 1939. For  their first
publication in the Nature-journal,  the draft-paper had been
presented by Frisch to Niels Bohr (one of the giants of atomic/nuclear physics at
that time) in  January 1939 in Copenhagen, before his departure to the USA.
Reading their draft, Bohr exclaimed ``what fools we have been''! In fact the result
could have been obtained many years  before, using the by that time well known
formula for the mass(energy) of nuclei with the liquid-drop model.

 The liquid drop model for the calculation of the binding energy of nuclei has been
 developed in the years $1933 -- 35$.
The total binding energy of nuclei is
 illustrated in Fig.~\ref{fig1},
as function of the binding energy  per nucleon.
From this figure we deduce that the values
reach their maximum for nuclear masses
 of $A = (60 - 80)$. This fact was of course also well known by most nuclear physicists
 in 1938. The heavy fragments
(Kr, Ba) with their high electric charges and large values of the kinetic energy
released in the fission process give rise to very large signals
  in the ionisation chambers used at that time (often misinterpreted as ``failures''
of the system).
 Soon after the news spread in the  beginning of the year 1939, several groups
were able to register directly the fission fragments with ionisation chambers,
 because they found  large signals due
to the large charges of the fragments (Ba, Z = 56 and Kr, Z= 36).
Arriving in the USA in the beginning of 1939 in Princeton
Niels Bohr worked with J. Wheeler and they produced within six month a monumental
 publication, title: ``The Mechanism of Nuclear Fission''
 in: Physical Review Vol. 56 (1939) 426 \cite{Bohr39}.
Still today a bible for nuclear physicists studying fission.
Nuclear fission is a fascinating object of study, because it
involves the interplay of macroscopic and microscopic  quantum
 properties of nuclear matter.
For the following  decades and  years many groups have studied binary fission in detail,
 for an intermediate stage of our knowledge
we can look into the proceedings of conferences for the 50'ties anniversary
for the discovery of fission:

\begin{figure*}[t]  
\vspace{+4 mm}
\begin{center}
\includegraphics[width=0.48\textwidth]{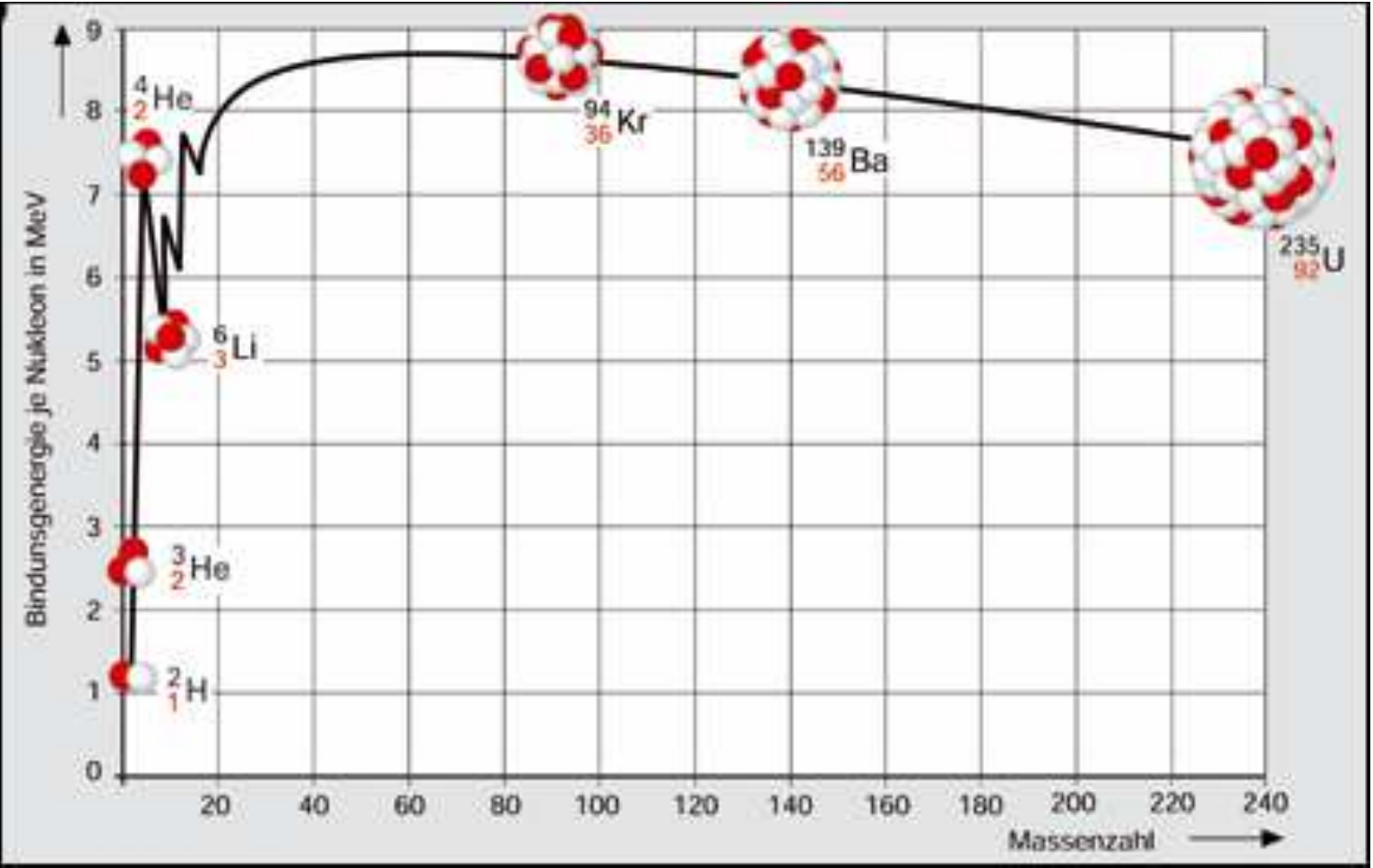}
\vspace{+4 mm}
\caption{The binding  energies per nucleon ($E_B/{nucleon}$) in  nuclei as function
of mass A.
The maximum is reached for masses with A = 60-110. Ternary fission of heavy nuclei
into three fragments
with these masses is possible. The liquid drop model
describes the variation of the total binding energy as shown in the figure. }
\label{fig1}
\end{center}
\end{figure*}

Four conferences  on fission have been devoted in 1989 to this anniversary:
 50 Years of Nuclear Fission~\cite{ff_1989},
 in the Journal:  Nuclear Physics  A 502 (1989), 1-639,
eds. D. Hilscher, H. Krappe and  W. von Oertzen.
Further: \textit{50 years with nuclear fission}, National Academy of Science,
Washington D.C., USA, and National Institute of Standards and Technology,
Gaithersburg, Maryland USA Ref.~\cite{ff_1989y}.
Another conference has been held in Leningrad (Soviet Union) October 16-20, 1989:
Proceedings ed. L.W. Drabchinski,
\textit{Fission of Nuclei, 50 Years}, Chlopin Radium Institute, 1992,
Vol. 1 and 2~\cite{ff_1989x}, (dominantly in Russian).

The study of binary fission over the last decades  has produced a large body of experimental
results with emphasis on a detailed description of several  parameters such as the mass and charge
distribution of fission fragments and their kinetic energies. For
the heavier nuclei the fission decay may  become the dominant decay channel.
 An important step for the understanding of
the fission process is the  appearance of shell structures
in heavy deformed nuclei, as introduced by
 Strutinsky~\cite{Strut89,Strut89x}.
The shell effects play decisive role in the yield of fragments, which is varying as function
 of charge and neutron number.
 The variation of the mass distribution of fission fragments with the given proton and neutron
 numbers has been found to depend mainly on the spherical and deformed shells.
These phenomena are described with  a
large variety of theoretical approaches, many of
 these  have been
put forward during the last decades. In the recent book by H. Krappe and K. Pomorski~\cite{Krappe2012}
many of these established approaches are described.

\begin{figure*}[t]  
\vspace{+6 mm}
\begin{center}
\includegraphics[width=0.58\textwidth]{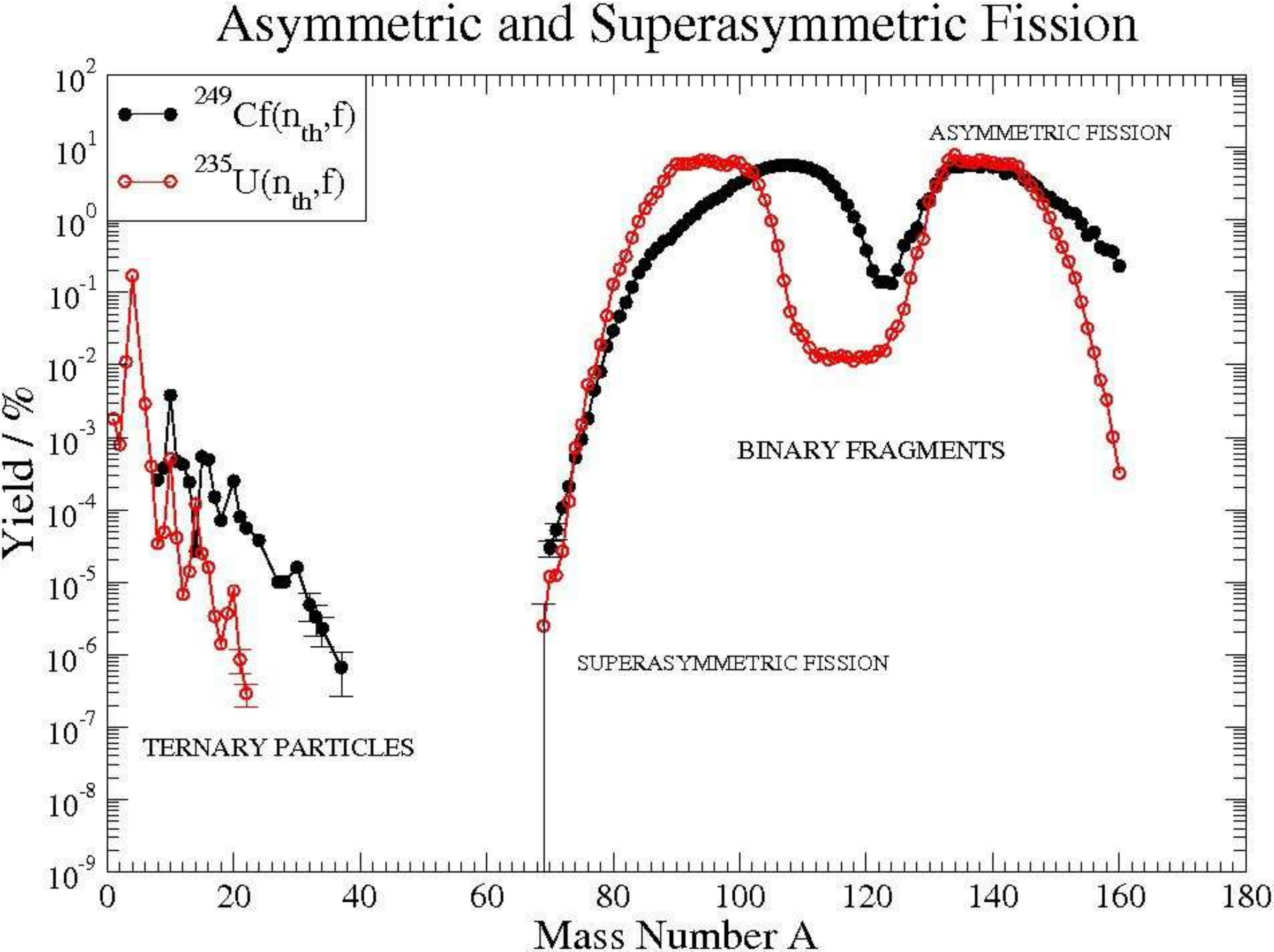}
\vspace{+1 mm}
\caption{ Mass distribution of binary fission fragments
 from the spontaneous decay of  $^{252}$Cf(sf),
as function of fragment mass. The shell model
explains the two hump distribution with enhanced yields for masses
around $A$ = 80 -- 110 and $A$ = 120--140.
 The left side of the figure
 shows the yields of lighter third fragments in coincidence with two binary fission fragments,
 for two systems as indicated, from Ref.~\cite{Gon}. }
\label{fig2}
\end{center}
\end{figure*}

 For an earlier overview on scientific aspects of (binary) fission we  suggest
 the book edited by C. Wagemans 1991: ``The Nuclear Fission Process''~\cite{wag91}, covering all
important aspects of this process. The fact which dominates the
mass-distributions
 of the fission fragments
are the shell effects for the charges Z = 20, 28, 50, and for the neutrons with
 N = 20, 28, 50 and 82.
 Thus the mass distribution
of binary fission
 is asymmetric as shown in Fig.\ref{fig2}.

 In addition this figure (from Ref.~\cite{Gon}) shows the
yield for  light fragments
 with masses from $A = 4$ up to $A = 30$, observed in  coincidence with binary fission channels.
 These are typically emitted perpendicular to the
fission axis, determined by the vectors of the two heavy fragments.
 These events usually have been defined as ``ternary fission''~\cite{Gon}.
 The  ``true ternary fission'' of heavy nuclei discussed in the present survey,
 has been predicted many times
 in theoretical works since the early 1950-60'ties, a decay with an increasing probability
for increasing total charge of the nuclei.

\section{Binary and Ternary Fission}
\subsection{General considerations, shell effects, hyper-deformation, sequential decay}

\begin{figure}[t]  
\begin{center}
\vspace{+3 mm}
\includegraphics[width=0.48\textwidth]{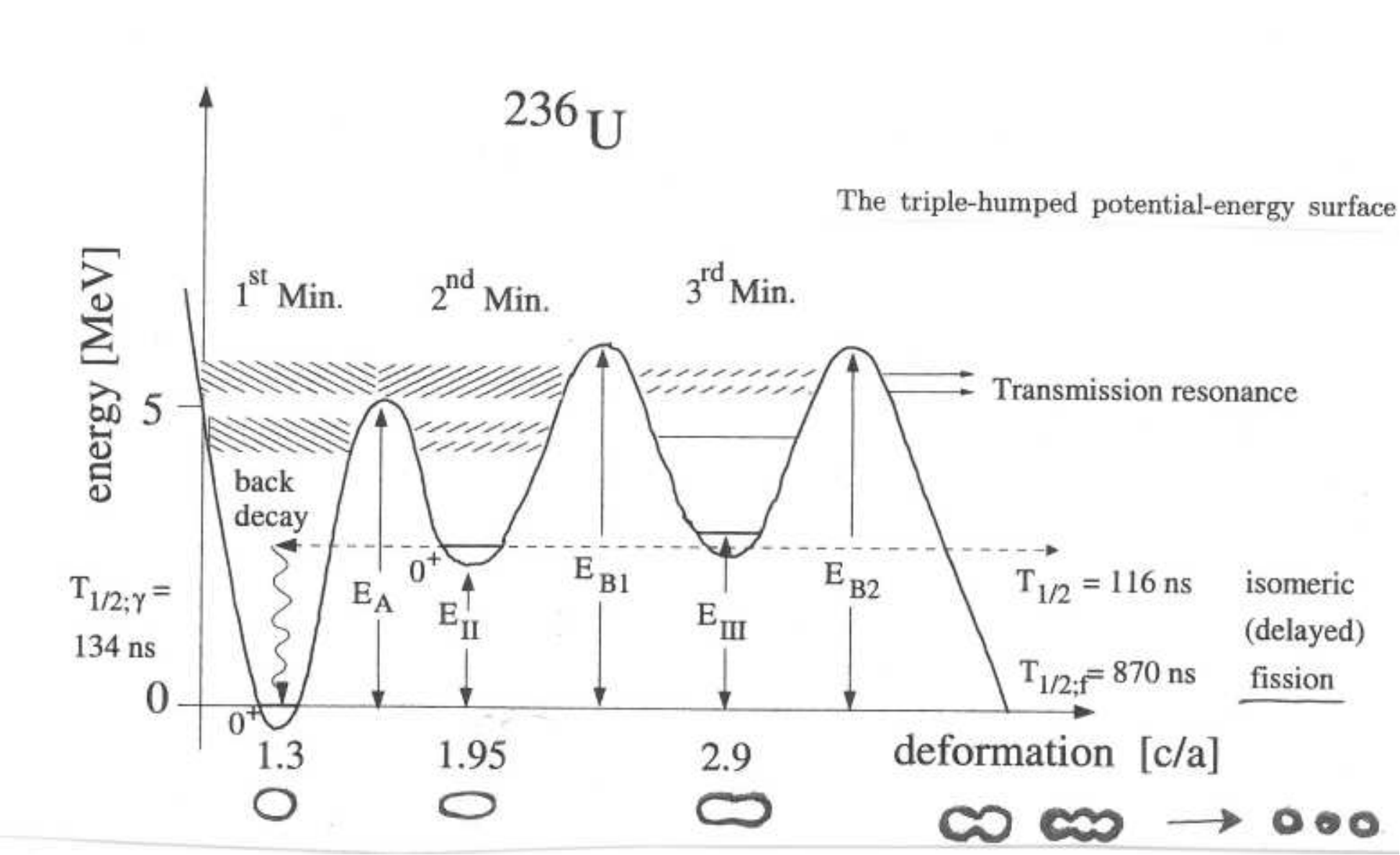}
\vspace{+4 mm}
\caption{The potential energy as function of deformation for a  hyper-deformed nuclear
 shape, which shows the double humped fission barrier
 suggesting a path to the collinear ternary
fission decay in   $^{236}$U, adapted from Ref.\cite{U236Hyperd}.}
\label{fig:Fig3}
\end{center}
\end{figure}

Fission involves a rearrangement of nucleons in a collective macroscopic motion
(evolution in shapes)  towards an elongated, deformed
structure with energies dominated by the liquid drop aspects and the quantal properties of
nuclei, the shell effects.
These aspects have been described earlier, as an example we cite the work of Swiatecki~\cite{Swiat58}
and of Diehl and Greiner~\cite{DG}. In the first it has been indicated
that apart of binary decays, ternary (and multiple) decays are energetically possible in heavy nuclei.
The probability for
 the decay depends on the barriers
for the individual combinations of fragments and on the phase space in these decays.
Quite important
is the variation of fission properties in
dependence on the total mass and charge of the fissioning nucleus and the appearance of
shells, particularly for protons. For the heavier nuclei the fission
decay may  become the dominant decay channel. As already mentioned an important step for the understanding of
the fission process is the appearance of shell structures in heavy
\textit{deformed nuclei},
 as introduced by Strutinsky~\cite{Strut89,Strut89x}.
This very unique approach has been essential  for
nuclear structure studies by describing the \textit{super - and hyper-deformed} shapes.
 Further super- and hyper-deformation played an important role in
the more recent studies in $\gamma$-spectroscopy of high spin states \cite{Riley2016}.
 The hyper-deformed configurations
at low spin show important
 manifestations with the occurrence of fission isomers \cite{fissionIs1,fissionIs2,fissionIs3},
 which have been studied extensively 40 years ago \cite{fissionIs0,fissionIs,U236Hyperd}.
 Thus it appears that the ternary decay passes through hyper-deformed
shapes of the fissioning nucleus (see Fig.~\ref{fig:Fig3}).

  In the following we list the factors governing the  probabilities (the \textit{phase space})
 for these decays:\\
 i)  the details of the  potential energy surface, the PES,\\
 ii) its valleys and hills,\\
 iii) the internal barriers at the two necks,\\
 iv) the $Q_{ggg}$-values, the latter determining  the kinetic energies and the
number of possible fragment (isotope) combinations,\\
 v) the excitation energy range in the individual fragments,\\
 vi) the momentum range for these,\\
 vii) the number of excited states (or the density of states) in each of the fragments, the
 combinations consisting of 3 isotopes,  and by \\
 viii) the spin ($J$)-multiplicity (phase space factor $(2J+1)$) in
these excited states,  with  spins expected up to values of (6-8)$^+$ .

\begin{figure}  
\begin{center}
\vspace{+2 mm}
\includegraphics[width=0.47\textwidth]{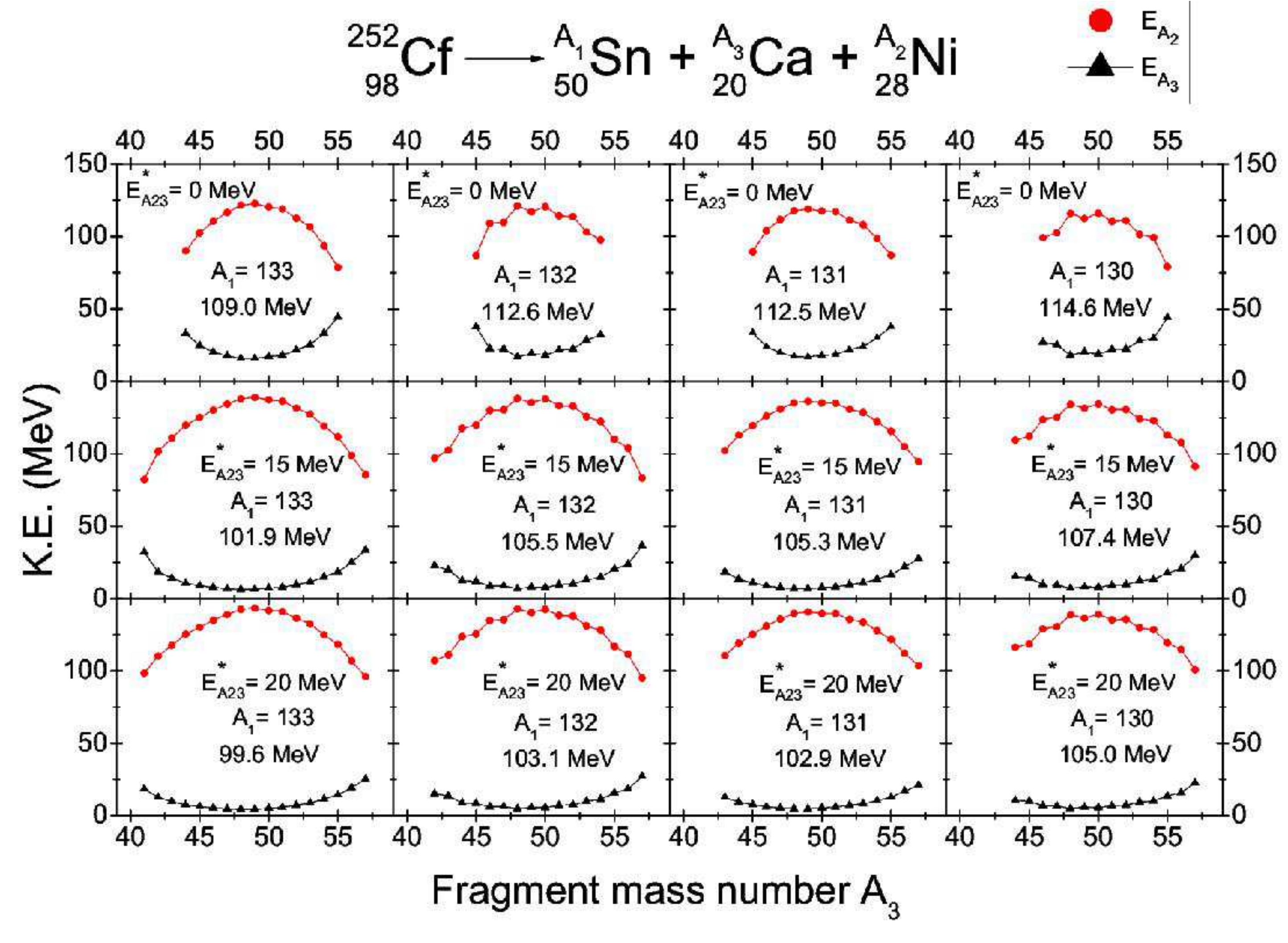}
\vspace{-3 mm}
\caption{Maximum values of the kinetic energies of fragments $A_3$ and $A_2$ in
 the  model with sequential decay (as shown in Fig.~\ref{fig:Fig19})
leading to ternary decays,
 (from Ref.~\cite{vijay14}). A similar  result has been obtained in Ref.~\cite{Holm17}}.
\label{fig:Fig4}
\end{center}
\end{figure}

 More general considerations for a three-body decay have been
analysed in Ref.~\cite{Garr2006}.  From this work we can state that the
 radioactive ternary fission  decay,
 being allowed from $Q$-value considerations,  will proceed sequentially,
because the two barriers and the phase space
favour the individual steps, the first steps are determined by the lower barriers and their
phase space.
The first step usually will start with the higher probability and
 proceed with the lighter mass combination of fragments with a
 lower barrier and the higher energy balance,
which has the larger preferred phase space. For this case of sequential decay
we can calculate the
kinetic energies of the ternary fission decay
of  $^{252}$Cf(sf,fff) with the decay mode  into the three fragments:
 $^{70}$Ni + $^{48}$Ca + $^{132}$Sn, which are
most likely. The result  is shown
in Fig.\ref{fig:Fig4}, and discussed further in sect.{\ref{Sequential}, and
illustrated in Fig.~\ref{fig:Fig19}. Actually in this simple  model the kinetic energies
of the outer fragments  are high, the figure shows obtained
 maximum values, they are too high (as indicated
in Ref.\cite{Pyat17}) because no excitation energy of the fragments is considered.

 An important approach is to consider the connection between ternary fission
and extreme deformations, namely hyper-deformation, illustrated in Fig.~\ref{fig:Fig3}.
In this context the work of Brosa et al.\cite{Brosa} is of importance,
 in this work different fission paths
 are obtained, there extremely long (``super long'') shapes with very
large deformations have been predicted.
 Fission must be considered as a
 macroscopic rearrangement of all nucleons into
a hyper-deformed shape with two necks.
 There will be two neck ruptures in a very short time sequence.

\begin{figure} 
\begin{center}
\vspace{4 mm}
\includegraphics[width=0.38\textwidth]{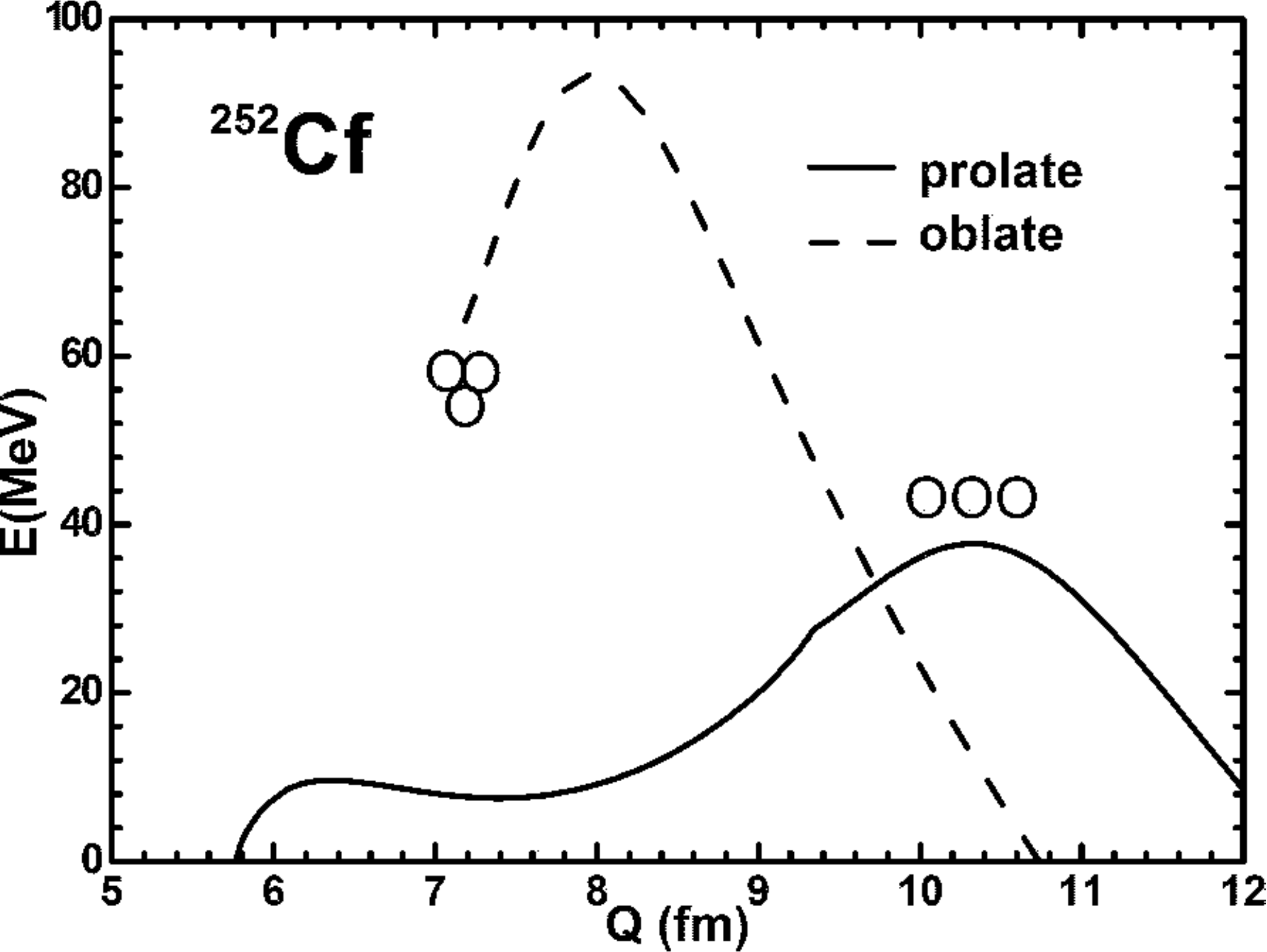}
\vspace{1 mm}
\caption{Comparison of the total potential energies, dominated by the Coulomb
 interaction, of three fragments
in an oblate configuration compared to a prolate (collinear) shape for three cluster decay,
courtesy of Royer~\cite{Royer}.}
\label{fig:Fig5}
\end{center}
\end{figure}
The term
``ternary fission'' has been used for binary fission  accompanied by light particle
(like $\alpha$-particles) emission.
For such  decays with  a third light particle
 emitted perpendicular to the binary fission axis,
compilations are available in Ref.~\cite{Gon}.
These must be considered  to be  emitted
from an prolate-binary configuration with a neck, the corresponding  yields decrease
 strongly (see Fig.~\ref{fig2}) as
 function of increasing  mass(charge) of the third particle, from Ref.~\cite{Gon}.
For the general survey of the possible ternary decays (FFF), and
 the discussion of the relative probabilities of oblate to prolate ternary fission,
we must look into the Coulomb energy of the total system.
 In Fig.~\ref{fig:Fig5}  we show the comparison, from Ref.~\cite{Royer},
of the PES of an oblate versus a prolate arrangement of three equal (comparable)
 sized fragments,
which can  be related to   the ternary decay of $^{252}$Cf(fff).
\begin{figure}  
\begin{center}
\vspace{-2 mm}
\includegraphics[width=0.23\textwidth]{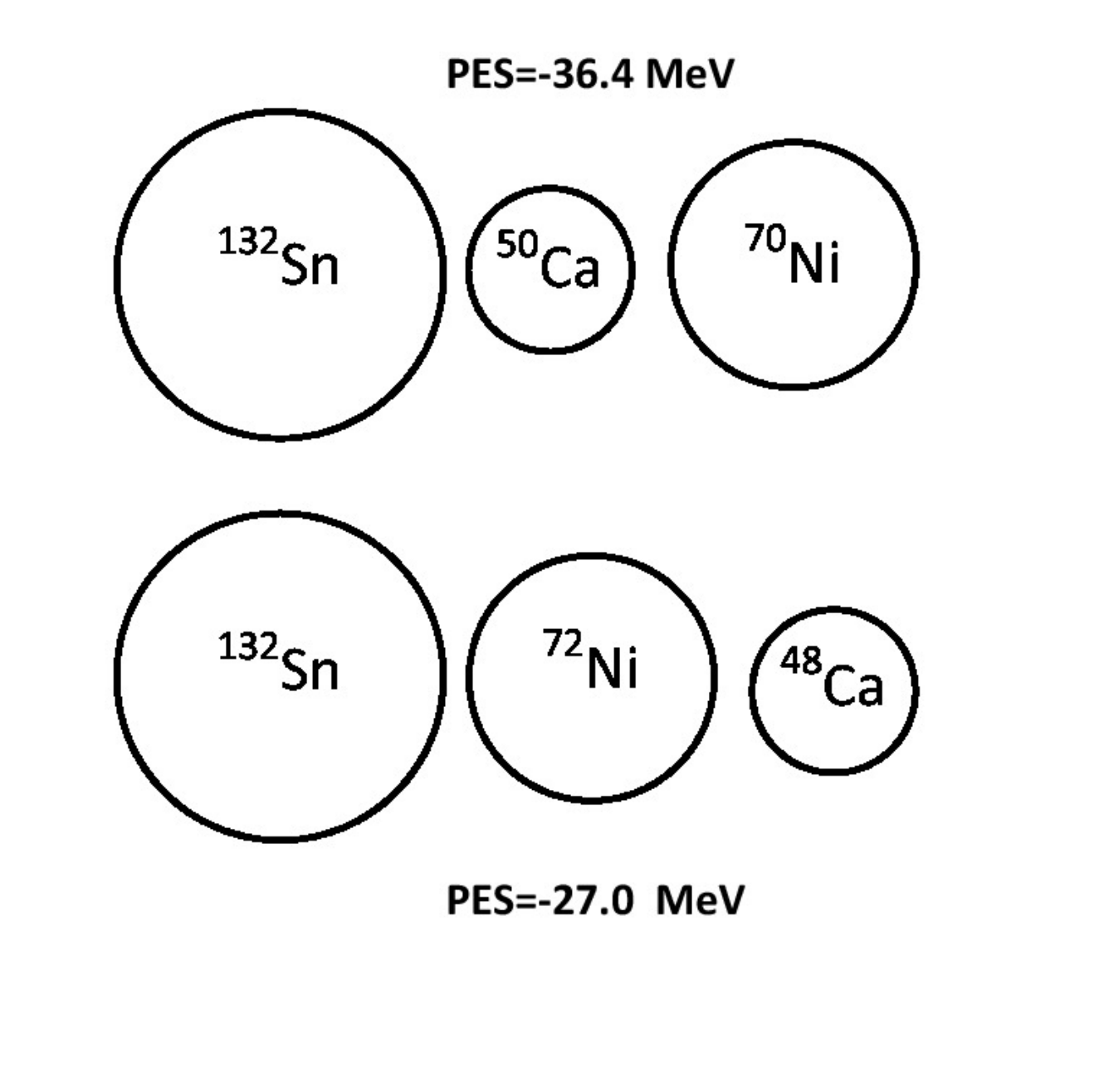}
\vspace{-4 mm}
\caption{The total potential energy of three fragments
in a collinear configuration, it is smallest with  the smaller fragment at the center.
The collinear configuration
dominates the  three cluster decay.}
\label{fig:Fig6}
\end{center}
\end{figure}

Clearly the linear arrangement will be preferred. Further, the potential energies, actually the PES as a function of mass asymmetry,  contain the most relevant   quantities
determining the phase space of the decays (discussed in detail later). The dominance
 of the Coulomb interaction points to distinct geometrical  arrangements of
the three clusters as shown in Fig.~\ref{fig:Fig5}
and Fig.~\ref{fig:Fig6}. The latter implies that the collinear arrangements with the
smallest fragment in the center
will have the largest phase space and therefore the highest probability.

The almost symmetric combination of three fragments gives maximal $Q_{ggg}$
values, as indicated in Fig.~\ref{fig:Fig7},
however, as shown later, this mode has a very low yield,
because the corresponding potential energy barriers are high, and the probability of its population
is small, because the  highest internal barriers occur,  making  the decay difficult
from these configurations. Still the symmetric decays
into three comparable fragments with the largest $Q$-value, can be extracted from
the data (see sect.~\ref{sect4}).
\begin{figure} 
\vspace{-1 mm}
\begin {center}
\includegraphics[width=0.42\textwidth]{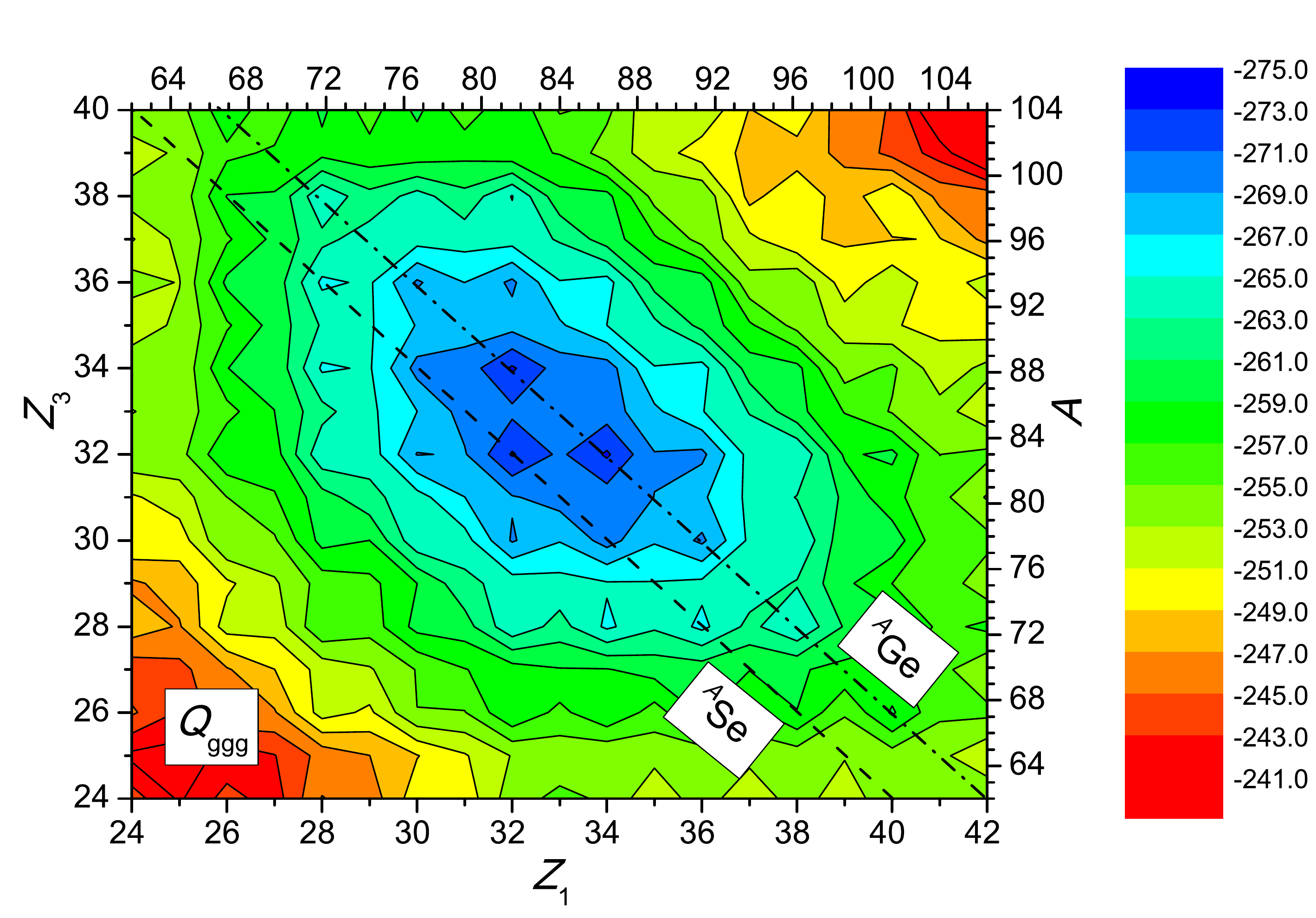}
\caption{The $Q_{ggg}$-values (-2n) for  $^{252}$Cf, the ternary
fragments with  $Z_i = 32-34$,
 correspond to the highest Q-values. the most symmetric decays, however,
  with the highest internal barriers.  }
\label{fig:Fig7}
\end{center}
\end{figure}

We have already introduced the concept of hyper - deformation, which must  be
considered in
the dynamical path towards a three fragment channel. In this the combination of macro- and micros-copic
aspects of the potential energies as function of smoothly varying nuclear shapes become essential for the
 understanding of super- and hyper-deformation. Related to this
quite important and convincing results concerning the physical circumstances for ternary fission
have been  obtained in the  work of Karpov~\cite{Karpov16}. In this work
the asymmetric three-center shell model is used, which is related to the two-center shell model,
 originally developed in 1972  by Greiner et.al. by  the Frankfurt group~\cite{Maruhn72}.
 We will see that the structure of this PES is dominated by the shell
 effects in all three fragments
 like in   $^{132}$Sn,  $^{48}$Ca and  $^{72}$Ni
   see Figs.~\ref{fig:Fig9Karpov}, \ref{fig:Fig10Karpov} and Fig.~\ref{fig:Fig11Karpov}.
In this work  appropriate shapes and coordinates are introduced  in order to describe the ternary decay,
they are shown in Fig.\ref{fig:Fig8Karpov}.
\begin{figure}  
\vspace{-1 mm}
\begin {center}
\includegraphics[width=0.45\textwidth]{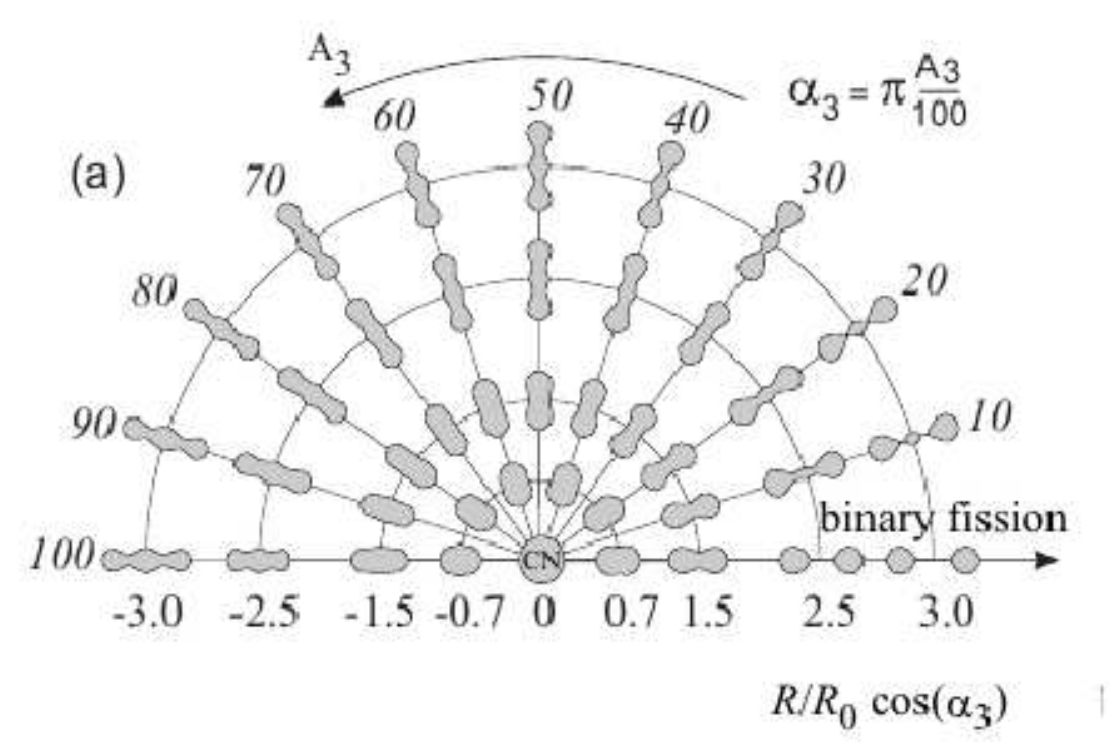}
\caption{The macroscopic geometry and coordinates chosen for the ternary fission
 study of Karpov in the three center-shell model.
The coordinates are defined appropriate to the ternary decay, from Ref.~\cite{Karpov16}.
The formation of shapes reminiscent of super- and
hyper-deformation in nuclei is observed. The radius of the spherical nucleus (CN) is denoted as R$_0$.}
\label{fig:Fig8Karpov}
\end{center}
\end{figure}

\begin{figure}[h]  
\hspace{-5 mm}
\vspace{+5 mm}
\includegraphics[width=0.5\textwidth]{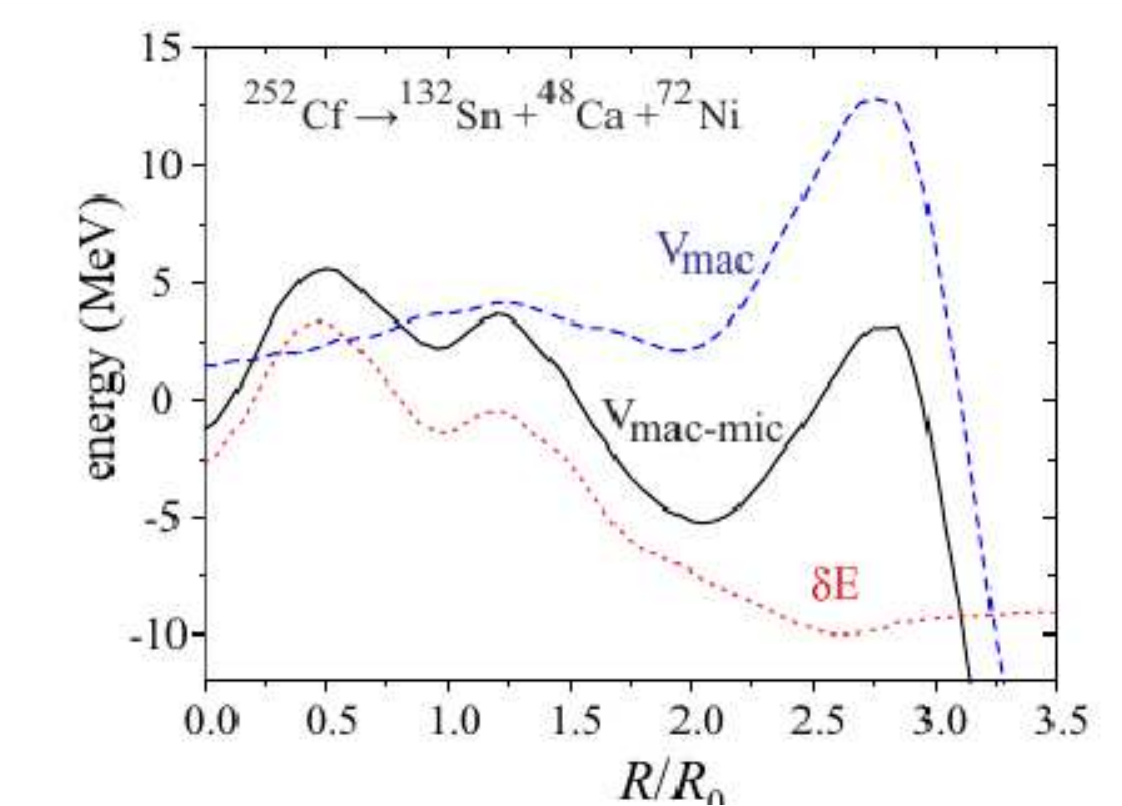}
\vspace{-4mm}
\caption{
 The macroscopic (dashed line) potential energy of the three-center system as function
 of the reduced elongation. The shell correction ($\delta E$, red dotted line),
and the full macro-micro-scopic potential energy of $^{252}$Cf with a
shape corresponding to the   splitting into
 ($^{132}$Sn,  $^{48}$Ca and  $^{72}$Ni), from Ref.~\cite{Karpov16}.
The third barrier (in the mac-mic approach) is strongly reduced.}
\label{fig:Fig9Karpov}
\end{figure}

Macroscopic potential energy (dashed line), shell correction
(dotted line), and total macro-microscopical potential energy
(solid line) of the $^{252}$Cf nucleus corresponding to the
$^{132}$Sn+$^{48}$Ca+$^{72}$Ni ternary splitting.

  In these calculations the favored fission path
(dashed line in Fig.~\ref{fig:Fig11Karpov}
 passes through two barriers. The first
 barrier is higher
the second barrier appears at rather large elongations(deformations). In the discussion of
 the results, the author
analyses the macroscopic and microscopic features of the PES. The barriers are shown
in Fig.~\ref{fig:Fig9Karpov}.   The second neck
appears in the macroscopic
calculation, with a very high second barrier (see Fig.~\ref{fig:Fig9Karpov}),
 ternary fission is not possible with
the liquid drop shapes.
 The inclusion of the shell effects in all three fragments produces the important effects:
the fission barrier becomes double-humped. The hyper-deformation
appears (see Fig.~\ref{fig:Fig3}), a feature giving rise to the
fission-isomers in the actinides.
 The ternary fission barrier is reduced dramatically,
the barriers exist in the elongation parameter (deformation) and in the
mass  asymmetry. Actually the binary fission is suppressed and the probability
 for the ternary decay increases.
Apart from the case discussed, other channels will be  favored like
 (``tin-sulfur-germanium'') and
the combination (``nickel-oxygen-samarium''). The ternary decay of $^{252}$Cf
observed in the experiments is due to the unique configurations with deformed shells and
the shells in the fragments.
\begin{figure}[t] 
\begin{center}
\vspace{+2 mm}
\includegraphics[width=0.52\textwidth]{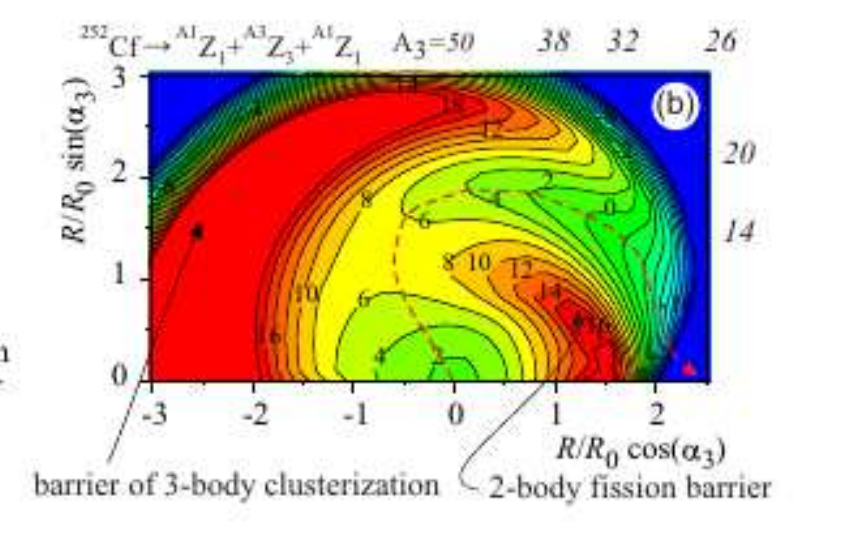}
\vspace{-5 mm}
\caption{ The potential energy surface for  $^{252}$Cf as
 function of the   elongation of  the system, it
 shows the binary fission path, this figure is without the shell effects,
see the next figure.
 Results obtained with the three-center shell
model. The dominant decay is always with  $^{132}$Sn.
  The  approach also secribes (dashed line) the binary fission path, from Ref.\cite{Karpov16}.}
\label{fig:Fig10Karpov}
\end{center}
\end{figure}
\begin{figure}[t]  
\hspace{+4 mm}
\includegraphics[width=0.45\textwidth]{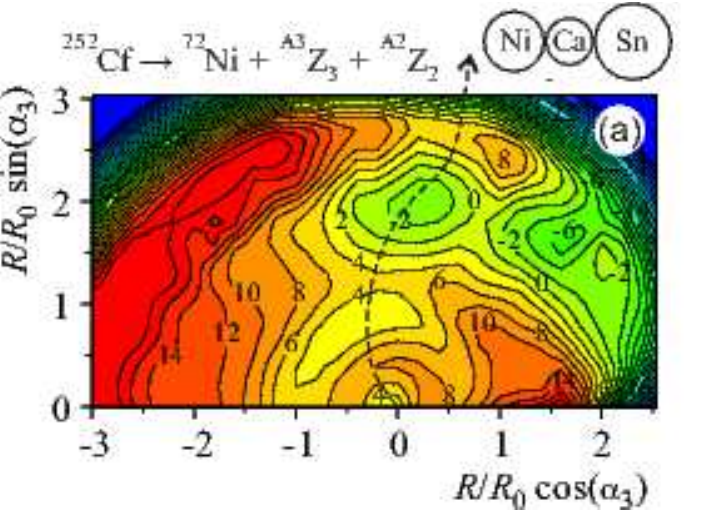}
\vspace{2 mm}
\caption{As in figure \ref{fig:Fig10Karpov} the potential energy
surface for ternary mass split of  $^{252}$Cf as
 function of the elongation of  the system and the
mass of the third (central)  fragment, from Ref.\cite{Karpov16}. This
figure compared to the previous figure shows the influence of the shell effects,
which distinctly change the landscape.
 Results obtained with the three-center shell
model, with the shell effects included. The dominant decay is with  $^{132}$Sn as
 the  heaviest fragment with the smallest ``Ca''-fragment at the center.
 The case with a closed-shell Ni-isotope
is shown. The dashed line shows the ternary fission path with two barriers.}
\label{fig:Fig11Karpov}
\end{figure}

\subsection{Potential Energy Surfaces,  Different  Ternary Decays}
\label{sect:3}
With the evolution of the nuclear shape towards the ternary mass split through
 a hyper-deformed shape with the formation of fragments, the further decay
(from the scission point) is governed
by the phase space.
The fission process can proceed towards  several mass partitions, their individual
 phase space will be different  according to the peculiarities.

With the PES's we obtain an overview   of the different decay channels.
The PES's, which are discussed in Ref.~\cite{Tashk2011}, are obtained by
 calculating the interaction between all
fragments:
\begin{eqnarray}
 \label{PES3}
&&U(R_{13},R_{23},Z_1,Z_3,A_1,A_3)=Q_{ggg} +\nonumber\\
&&+ V_{12}^{(Coul)} (Z_1,Z_2,R_{13}+R_{23})\nonumber\\
&& +V_{13}(R_{13},Z_1,Z_3,A_1,A_3)\nonumber\\
&&+V_{23} (R_{23},Z_3,Z_2,A_3,A_2 ),
\end{eqnarray}
where   $Q_{\rm ggg}=B_1+B_2+B_3-B_{\rm CN}$ is the balance of the fragments
 binding energies in  the ternary fission.
  The values of the binding energies are obtained
  from the mass tables in Ref.~\cite{Wapstra2003}; $V_{13}$ and $V_{23}$ are
  the nucleus-nucleus interaction of the middle cluster ``3'' ($A_3$ and $Z_3$)
  with the other two, their mass and charge numbers, with the left ``1''
 ($A_1$ and $Z_1$)  and right ``2'' ($A_2$ and $Z_2$) fragments of the ternary system; $V_{12}^{(\rm Coul)}$
 is the Coulomb interaction between the two border fragments  ``1'' and ``2'',
  which are separated by the distance $R_{13}+R_{23}$, where $R_{13}$ and $R_{23}$ are the distances
  between the middle cluster and two outer clusters placed on the left and right sides, respectively.
  The interaction potentials $V_{13}$ and $V_{23}$ consist of the Coulomb and nuclear parts:
 \begin{eqnarray}
   &&V_{3i}(R_{3i},Z_i,Z_3,A_i,A_3)=V_{3i}^{(\rm Coul)}(Z_i,Z_3,R_{i3})\nonumber\\
   &&+V_{3i}^{(\rm Nucl)}(Z_i,A_i,Z_3,A_3,R_{3i}),
\hspace*{0.25cm} {\rm where}
\hspace*{0.25cm} i=1,2.
 \end{eqnarray}
  The nuclear interaction is calculated by the double folding procedure with the effective
  nucleon-nucleon forces depending on the nucleon density distribution  (see Ref.~\cite{Tashk2011}).
  The Coulomb interaction is determined by the Wong formula ~\cite{Wong1973},
 which also allows us to take into account the deformed shape of fragments and the
 possibility to consider interactions
 under different angles of their axial symmetry axis.
\begin{figure}[t] 
\hspace{-0 mm}
\vspace{-3 mm}
\includegraphics[width=0.60\textwidth]{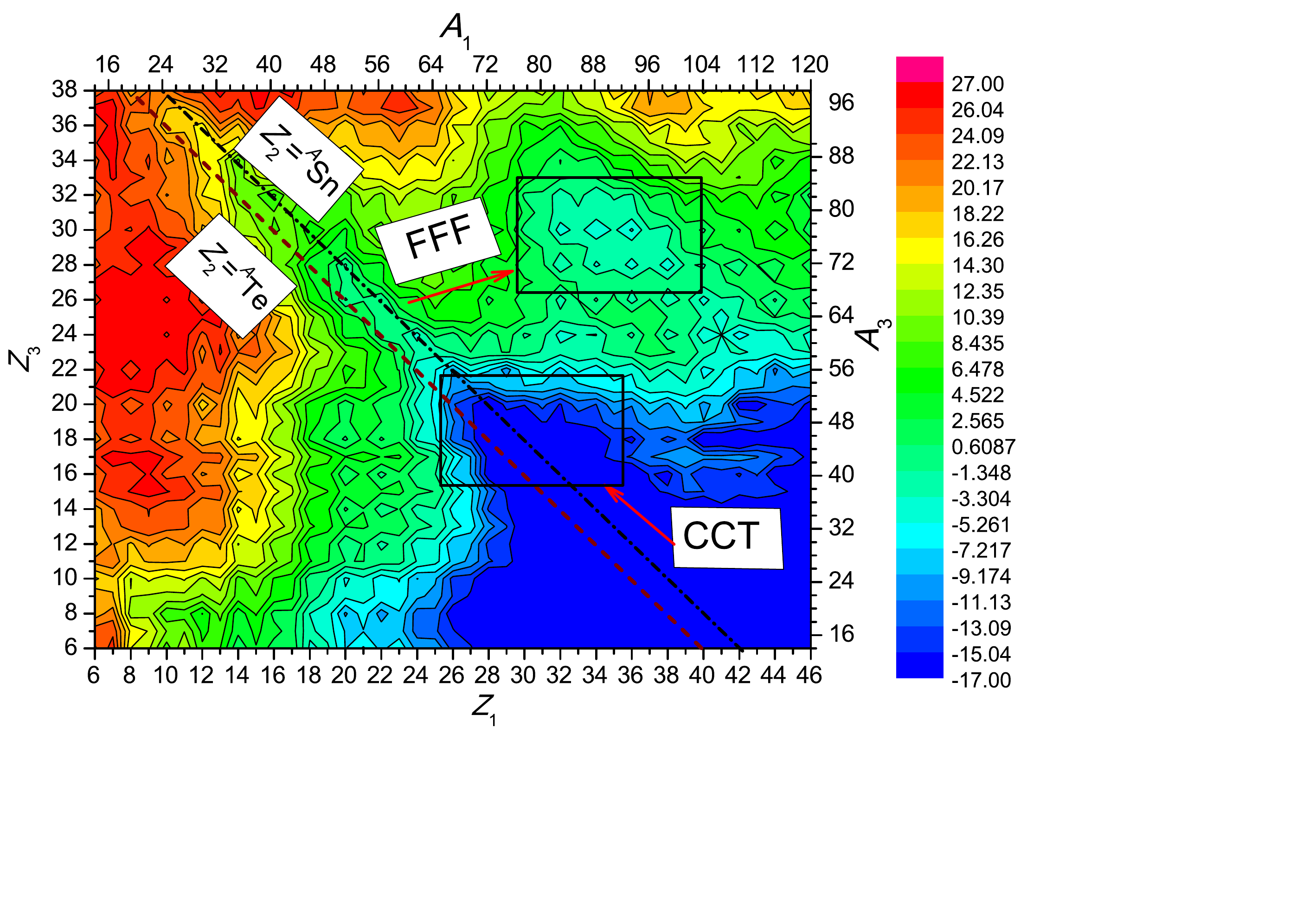}
\vspace{-17 mm}
\caption{
The contour plot of the potential energy surface (PES) it gives an overview of the different modes
 for the collinear ternary decays characterized by two values of  $Z_3$ and $Z_1$
 of the ternary fragments for
 $^{252}$Cf. The pre-scission configurations of the ternary system appear with the isotopes of tin,
 $^{132–-128}$Sn, $Z_{1}$=50, and Tellurium $^{134–-138}$Te, $Z_1$=52. The upper rectangle shows
 the area of the expected mass (charge) distribution of the symmetric ternary FFF-decay
 products with almost equal fragment masses
 $A_1\approx A_2$ and $A_3\approx A_2$. The lower rectangle $Z_3$=16--20 corresponds to the expected region
 for ternary decays of the two products with the charge numbers  $Z_1$=28--34 and
 mass numbers $A_1=$72--88 of the ternary configurations leading to the collinear cluster
 decays (CCT) observed in Refs.\cite{Pyat10,Pyat12}.
}
\label{fig:Fig12_252Cf0n}
\end{figure}\\

 Potential energy surface\\

We inspect the PES's for the two cases of  $^{252}$Cf(sf) and $^{236}$U(n,f) in
two figures, Fig.~\ref{fig:Fig12_252Cf0n} and Fig.~\ref{fig:Fig13_PESU236}, respectively.
 We note the ``lower''(blue)  regions connected to the combinations with the Sn-fragments.
Apart from the already known CCT and FFF - decays we observe a  pronounced region with a
 minimum for ${A}_3$, with  ${Z}_3$ = 18, this must be
combined as shown in the PES, with two fragments with ${Z}_1$ = 40 (Zr),
 or the adjacent value of  ${Z}_1$ = 42,
combined with ${Z}_1$ = 38(Sr). The particular structure of the isotopes with ${Z}$ = 18,
have been observed in Refs.~\cite{Forn2000,Sorlin2001}, they correspond
 to the new ``shells'' observed in neutronrich isotopes, as in $^{44-46}$Ar (but also in $^{26}$Ne).
The same depletion in the PES's is observed in the PES for
the case of Uranium fission as  shown in Fig.~\ref{fig:Fig13_PESU236}. These effects on the masses,
which are seen here, are already contained in the recent mass table in Ref.~\cite{Wapstra2003}.
The decays corresponding to these structures are actually contained in the data
and can be extracted as discussed in
the chapter on multimodal fission (in sect.~\ref{multi}).

With these figures
  we illustrate the role of the PES
in the formation of the ternary  fragments, they point to the differences of the two cases.
Clearly the charges have the strongest influence
on the phase space  in the decays.
The effect of the closed shells for protons (${Z}_{1,2,3}$ = 20, 28 and 50) is
clearly visible with the low (deep blue) valleys in the PES's.
 The case of the almost symmetric decays
with charges of ${Z}_{1,2,3}$ = 32, 34, 32 is seen as an depletion and points to
the possible symmetric decays, these have been
 discussed in Refs.~\cite{voe14,voe15}, and are later illustrated
in the chapter on multimodal fission in the present survey.

\begin{figure}[t]  
\begin{center}
\hspace{6 mm}
\vspace{+22 mm}
\includegraphics[width=0.58\textwidth]{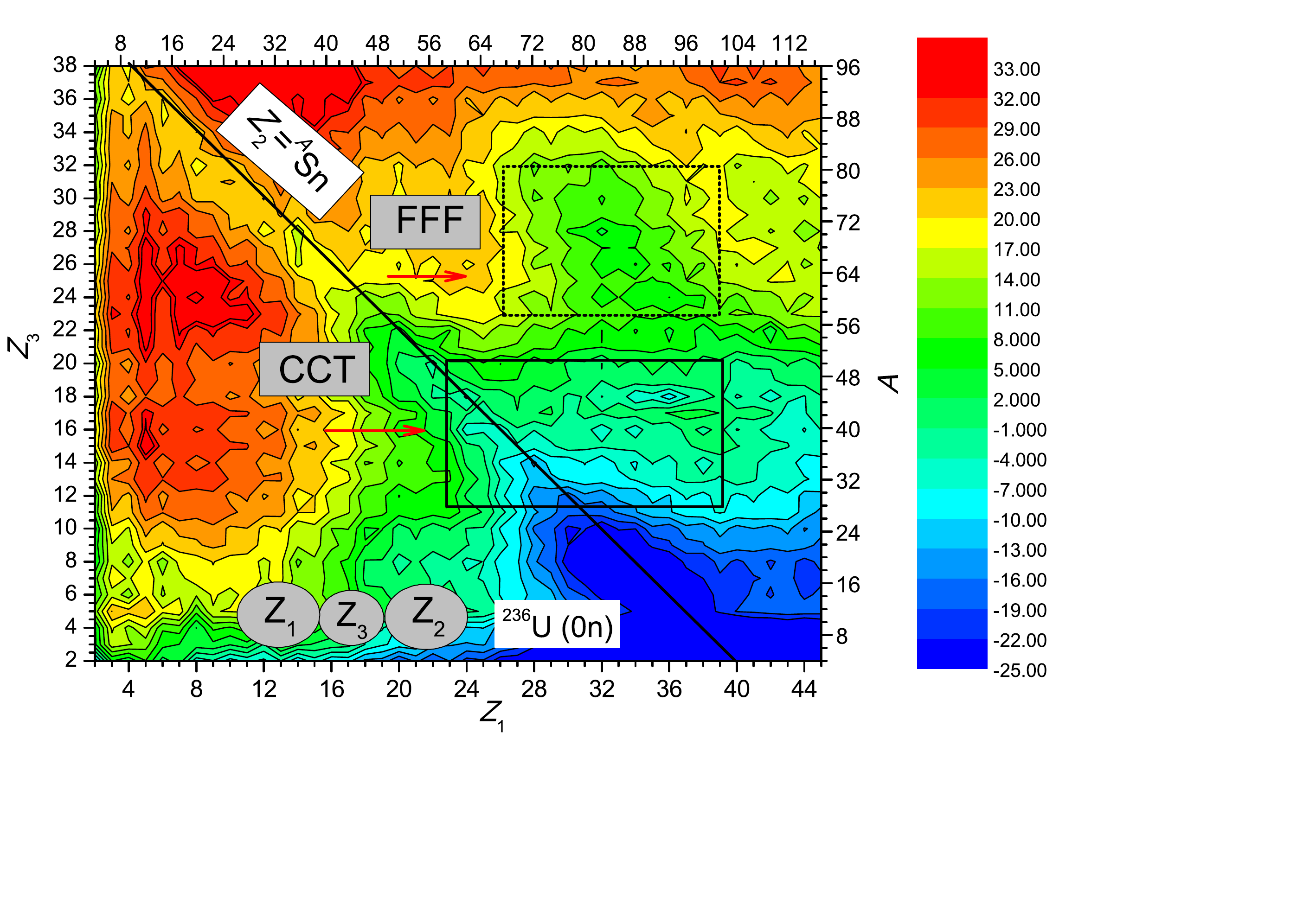}
\vspace{-40mm}
\caption{ The potential energy surface for the ternary decays of $^{235}$U(n,f).
 The general features are similar
  to the  $Z_3 - Z_1$ pattern for $^{252}$Cf,
with a valley dominated by Sn-fragments, $Z_2$ = 50. However,
 we observe the absence of pronounced effects due $Z_3$ = 20, 28
(as in the case of $^{252}$Cf), because for the case of U these fragments are absent.
Similarly a region with a lower (green) valley for symmetric decays with comparable
three fragments with $Z$ = 32+28+32
can be favoured (total charge of U = 92). Again the ternary decays for $Z$ = 18
(as for the case of $^{252}$Cf) can be favoured,
 where a proton shell in the fragment makes favorable $Q$-values.
These decays are discussed as multi-modal fissions later.}
\label{fig:Fig13_PESU236}
\end{center}
\end{figure}

Quite remarkable is
the structure of the PES shown in Fig.~\ref{fig:Fig14} for the ternary decay
of the  super-heavy nucleus $^{298}$Fl,
a neutron rich isotope of the recently (Ref.~\cite{Oganess}) observed element with $Z$=114.
 However, with a much
larger neutron-number, then observed.
\begin{figure}[t]   
\begin{center}
\vspace{+3 mm}
\includegraphics[width=0.41\textwidth]{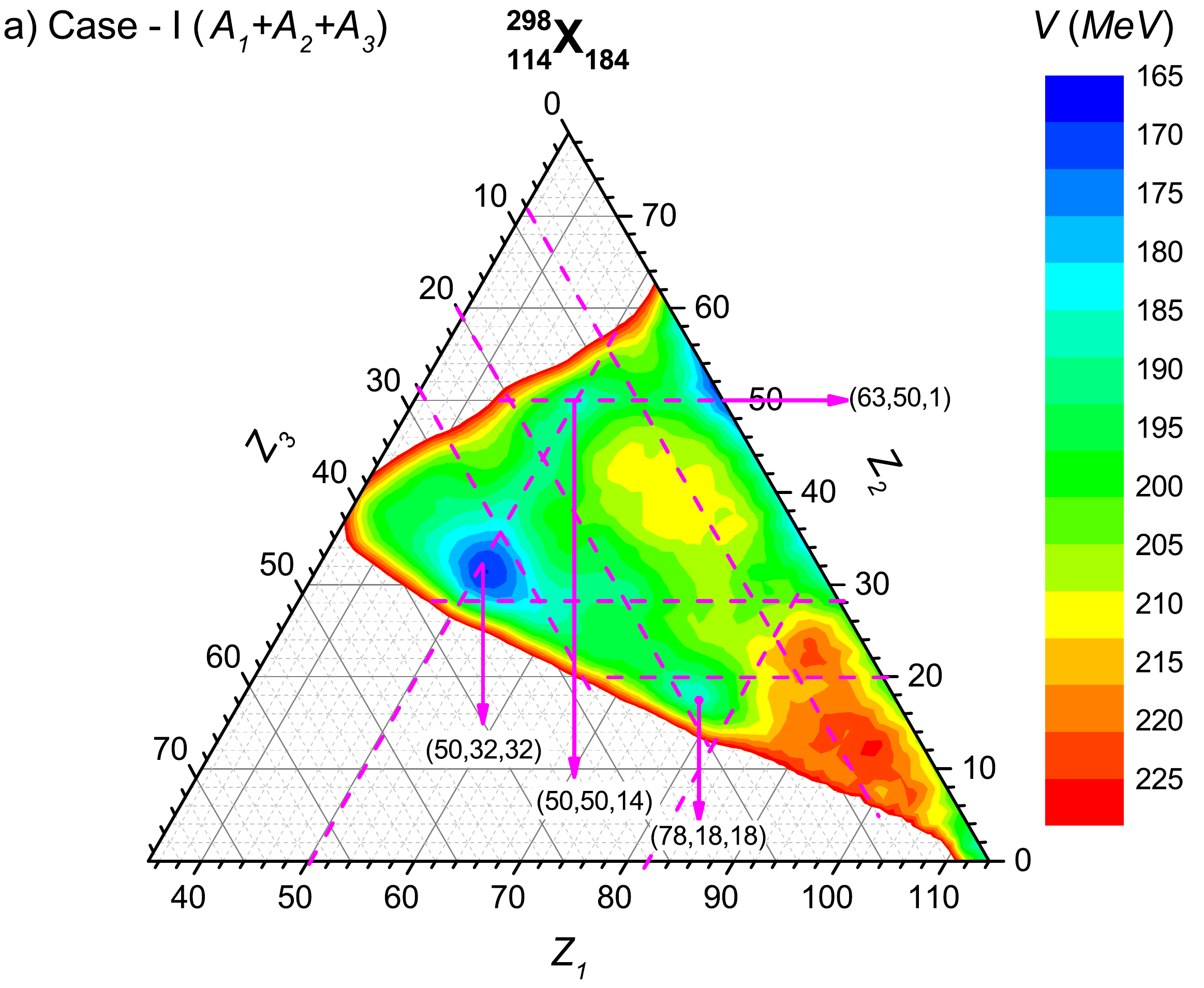}
\caption{ The potential energy surface for the spontaneous fission decay of a superheavy
        nucleus $^{298}$Fl,
        which has been calculated by Balasubramanian et al.~Ref.\cite{Bala2016}.
        With the increase of
        the total charge one region dominates, there are no closed shells (except for Z = 50)
         in the fragments.}
\label{fig:Fig14}
\end{center}
\end{figure}
We observe the change of the favoured decays with  the change
 of the total charge, and connected to this the influence of the variation of the
closed shells in the three fragments.

\begin{figure}[t]   
\begin{center}
\vspace{+1 mm}
\includegraphics[width=0.48\textwidth]{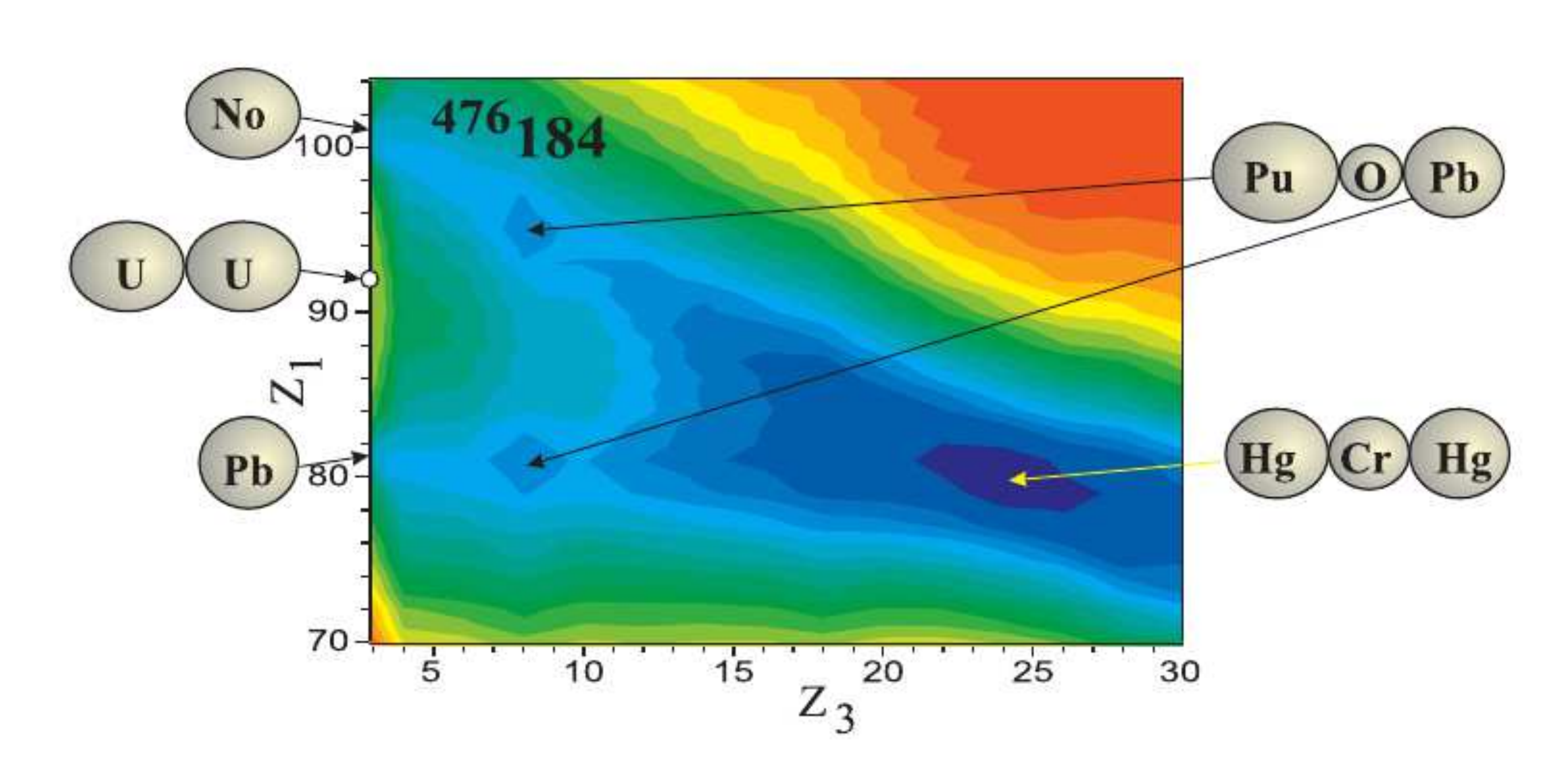}
\vspace{-3 mm}
\caption{The potential energy surface for the three-body configurations
       formed in a collision of two U-nuclei:
      $^{238}$U + $^{238}$U, which has been calculated by Zagrebaev et al.~Ref.\cite{Zagreb10a}.
        With the increase of  the tree-body clusterizations become more probable, the shell
         effect of $Z$=82  is still visible.}
\label{fig:Fig15}
\end{center}
\end{figure}

It is interesting to analyse the situation for  the ternary fission for even heavier
 systems. In heavy systems  shell effects in the fragments
are not dominant any more, the decay is determined by
the liquid drop energies. However, in the fragments
the shell with the $Z$ = 50 remain to be present.
Further consequences concerning ternary decays can be deduced from the study of
the PES's of  very heavy systems like in Refs.\cite{Bala2016,Zagreb10a}. In the first case
the maximal heavy system
is studied, which will have no barrier for binary  fission. In the reaction path
of very heavy systems, like for $^{238}$U + $^{238}$U in Ref.~\cite{Zagreb10a},
only a few exit channels appear. A binary channel with $^{276}$No$_{102}$ and a Pb-isotope
could be observed. Further decay channels
are shown in Fig.~\ref{fig:Fig15}, where the tree-body clusterizations
with the formation of neutron rich O-isotopes and some heavy
clusters ($^{240-246}$Pu + $^{212}$Pb + $^{18-22}$O) are shown.
Looking into these cases it becomes evident, that in the studies of ternary fission
the case of Cf-isotopes  (and possibly other heavy nuclei/isotopes in the vicinity),
 which we will discuss,
is unique. Many aspects of nuclear structure
connected to shells in the fragments and to deformed shells in the total system
are present in this case. More cases are discussed by Balasubramanian et al. in Ref.~\cite{Bala2016},
where the ternary decay of heavy nuclei and for a variety of super-heavy elements
have been studied. In this case
shells in the neutron number of the super-heavy elements can
 be important for the ternary decay.

In very heavy  systems, i.e. in  reactions of very heavy ions leading to
binary and ternary reaction channels,  colli-near decays have been reported.
 The dominance of the Coulomb interaction leads to aligned (collinear)
ternary fragmentations in higher energy collisions of heavy ions, in particular
  in deep inelastic collisions in the selection
of central impact parameters as described for $^{197}$Au + $^{197}$Au collisions
in Ref.~\cite{Wilc2010}.

\subsection{Sequential decay, shell effects, hyper-deformation, Collinearity}
\label{Sequential}
As already mentioned a three-body decay can be considered to proceed sequentially, each step
being determined by the
phase space and $Q$-values of each step. From  considerations of the PES's the
collinear decays are predicted, actually the experimental results have been interpreted as
collinear cluster tripartion CCT). This claim from the experimental
 observations has been contested repeatedly.
Most theoretical arguments are based on the strong deep valleys
    in the potential energy surfaces, which give a favoured phase space
for the collinear decays.
The collinear decays  are actually suggested by the hyper-defor-med shapes observed in heavy
 nuclei. This has been studied experimentally in  Ref.~\cite{U236Hyperd}, for the case of
  $^{236}$U and shown
in Fig.~\ref{fig:Fig3}. These observations favor
the collinear geometry relative to ``oblate'' shapes.  In the collinear
tripartition of the hyper-deformed
nuclear state
one of the deformed fragments in the first binary fission starts the sequential decay
defining the axial symmetry axis, the following fission process preserves this axis,
thus both fission axes are parallel.

These strongly deformed, hyper-deformed, configurations are favored because of the
binding energy including shell effects already mentioned, which  depend non-linearly
on the degree of elongation (shape deformation).  This phenomenon has been
 introduced by Strutinsky~\cite{Strut89} and studied experimentally extensively 40 years ago
and described in Refs.~\cite{fissionIs3,fissionIs0,fissionIs}.
 These deformed shells  are quite pronounced at larger values of the quadrupole deformation.
The hyper-deformed shapes (axis ratios of 3:1),
  appear in most of the heavier nuclei due to the deformed
shell effects in the total system, and give rise
 to the observation of fission isomers~\cite{fissionIs1,fissionIs2,fissionIs3}.
 For spontaneous fission and fission  induced  by gamma-rays, neutrons or charged
 particles, the  nucleus on its way to fission undergoes extreme deformations with a
 stretching due the Coulomb forces and
 inevitably passes through super-deformed and hyper-deformed shapes before splitting.
 The shell effects in nuclei will delay fission due
  to the structure  with one and two  barriers in the collective potential.

In the recent theoretical analysis of ternary fission by Karpov~\cite{Karpov16} the shell effect
discussed previously by Strutinsky~\cite{Strut89}, suggests that deformed shells are most important
for the ternary decays, as well for the binary fissions.
In this work \cite{Karpov16}, based on the three-center shell model, the shell structures in {\it all}
 fragments give rise
to a complex deformation of the total fissioning system, which is equivalent to hyper-deformation.
 The potential energies for
 the fission path have been shown
in Figs.~\ref{fig:Fig10Karpov} and ~\ref{fig:Fig11Karpov}.
With this work the most complete description for the dynamics of the decay,
particularly related to the
barriers has been obtained. There still remains the final dynamics of the
 ternary decay,
which will be governed by the phase space, the latter being determined by the potential
energy surfaces  PES'S (generally obtained with an approach with three deformed
 clusters).

 The ground state shape of the actinide nuclei is prolate ~\cite{Moeller},
 the shape is the result of the balance between the repulsive Coulomb and
 attractive surface tension forces. The prolate shape in the ground state
is favorable  energetically for the three massive fragments of the  collinear ternary fission.
  Authors of Ref.~\cite{Manim1} found that the potential
 energy of the collinear shape is significantly lower than
 the one for the oblate shape and the relative yield of the ternary fission products
 is much larger for the prolate case than for the latter shape.

Actually with the formation of the necks due to deformed shells,  quantum mechanical
fluctuations are expected, an
almost simultaneous (a sequential within a very small time delay) break up of the two
necks is expected for the ternary scission as described in   Refs.~\cite{Tashk15,Tashk16}).
 In these calculations
also the more asymmetric binary fission is predicted and finally a three cluster
configuration due to the  shell structures
for protons and neutrons appears in the exit channel. Actually the shell effects give
for both,  binary and ternary decays a dominance for the decay with fission
fragments with mass $A$ = 132 ($Z$ = 50, $N$ = 82).

  \begin{figure}   
\resizebox{0.58\textwidth}{!}{\includegraphics{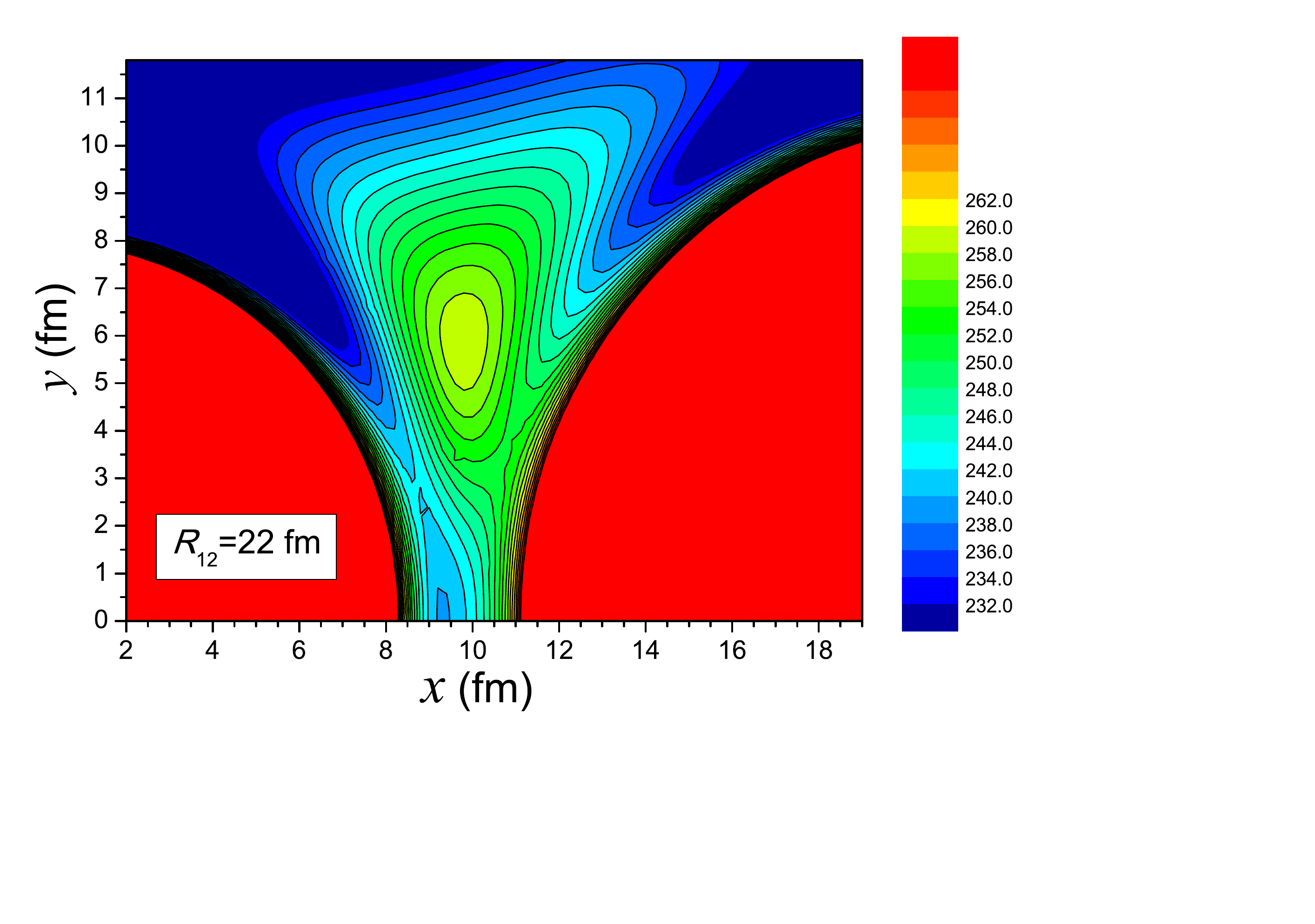}}
\vspace*{-1.9 cm}
\caption{
The  potential energy surface $V(R_{12},x_3,y_3)$ as a function of the position $x_3$
of the middle fragment ``3'' (Ca) along axis $R_{12}$ connecting the centres
of the outer fragments ``1'' and ``2'' of the tri-nuclear system and $y_3$, which is
 perpendicular to $R_{12}$ at the value of $R_{12}=22$ fm.
The minimum value of the potential well for Ca is
$V_{\rm min}(R_{12}$=22 fm, $x_3$=9.4 fm, $y_3$=0 fm)=239.75 MeV.}
\label{fig:Fig16}
\end{figure}

In the work of Tashkhodjaev presented in Refs.~\cite{Tashk15,Tashk16} an explicit quantum mechanical
 calculation of the motion of
the fragments towards scission and separation has been done.
According to results of this calculations in  Ref.~\cite{Tashk2016}, the nuclear interaction
 between fragments plays a decisive role in the pre-scission geometry of the tri-nuclear
 system. The collinear configuration of the tri-nuclear system (TNS) is preferable for
 the values of
 the distance between outer nuclei ``1'' and ``2'' being in region of
$R_{12} = 21 - 22$ fm, but the
fluctuation of the position of the middle Ca  cluster (nucleus ``3'') from
the collinearity axis can be observed due to the extension of the potential well in
the interaction potential between the fragments of the TNS up to 2 fm
around the axis connecting the centres of the outer nuclei (see Fig.~\ref{fig:Fig16}).
At the  distance $R_{12} > 22$ fm the overlaps of the nucleon densities of the
TNS nuclei decrease and, as a result of the  decrease in the nuclear attraction and the
Coulomb repulsion the depth of the potential well decreases. In that way the condition
for the separation of the middle fragment Ca from Ni can arise.
  As a result the
   potential well for Ca-fragments  moves away (perpendicular)
 from the $x$-axis  at values of $R_{12}$=23 fm
   and its minimum value decreases:
$V_{\rm min}(R_{12}=23$ fm, $x_3$=9.0 fm, $y_3$=2.2 fm)=236.83 MeV (see Fig.~\ref{fig:Fig17}).

 \begin{figure}  
\resizebox{0.58\textwidth}{!}{\includegraphics{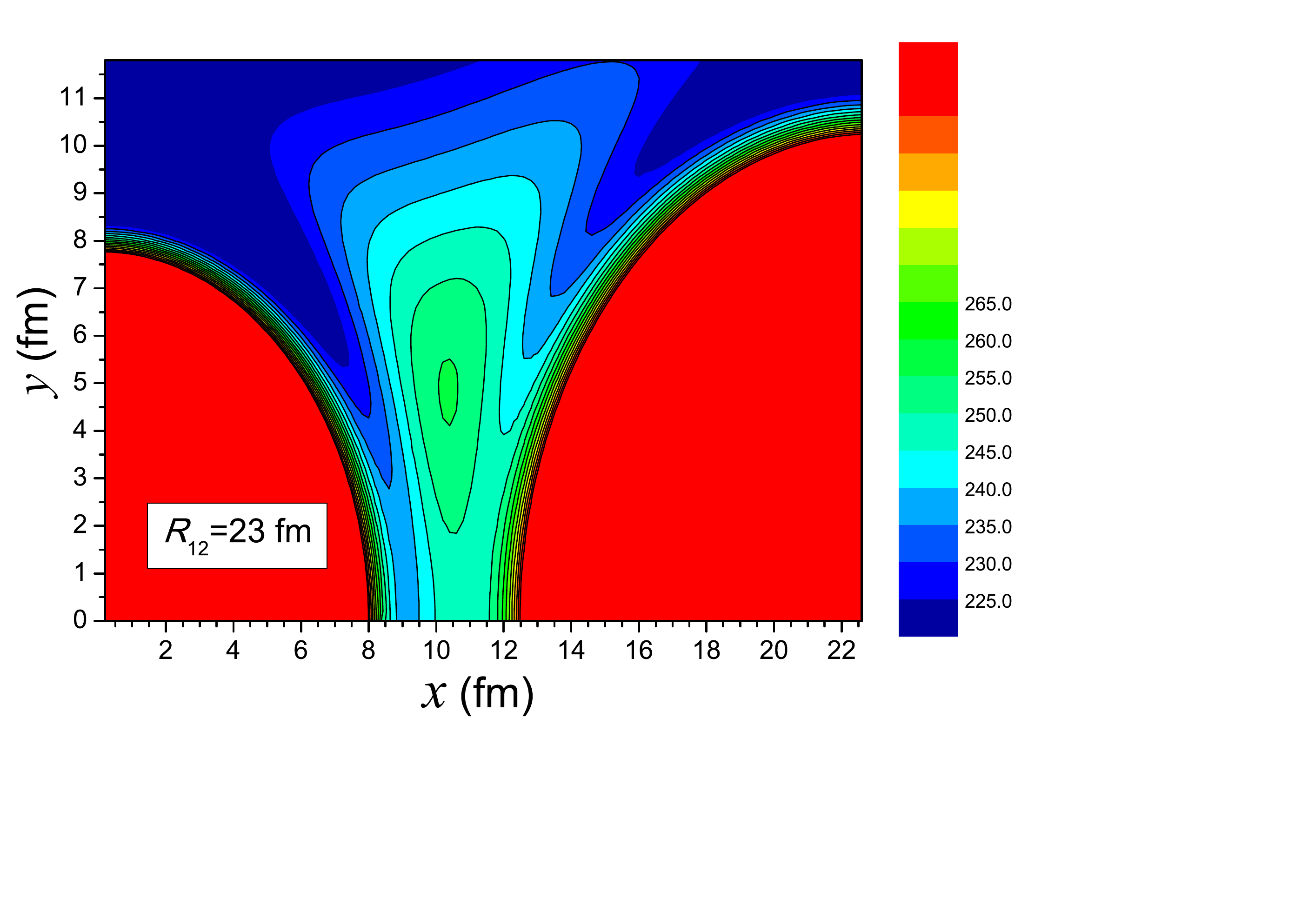}}
\vspace*{-1.9 cm}
\caption{
The same as in Fig.~\ref{fig:Fig18} but for the larger distance $R_{12}= 23$ fm.
The minimum value of the potential well for Ca is now
$V_{\rm min}(R_{12}=23$ fm, $x_3$=9.0 fm, $y_3$=2.2 fm)= 236.83 MeV.}
\label{fig:Fig17}
\end{figure}

\begin{figure}[t]  
\begin{center}
\hspace{-5 mm}
\includegraphics[width=0.53\textwidth]{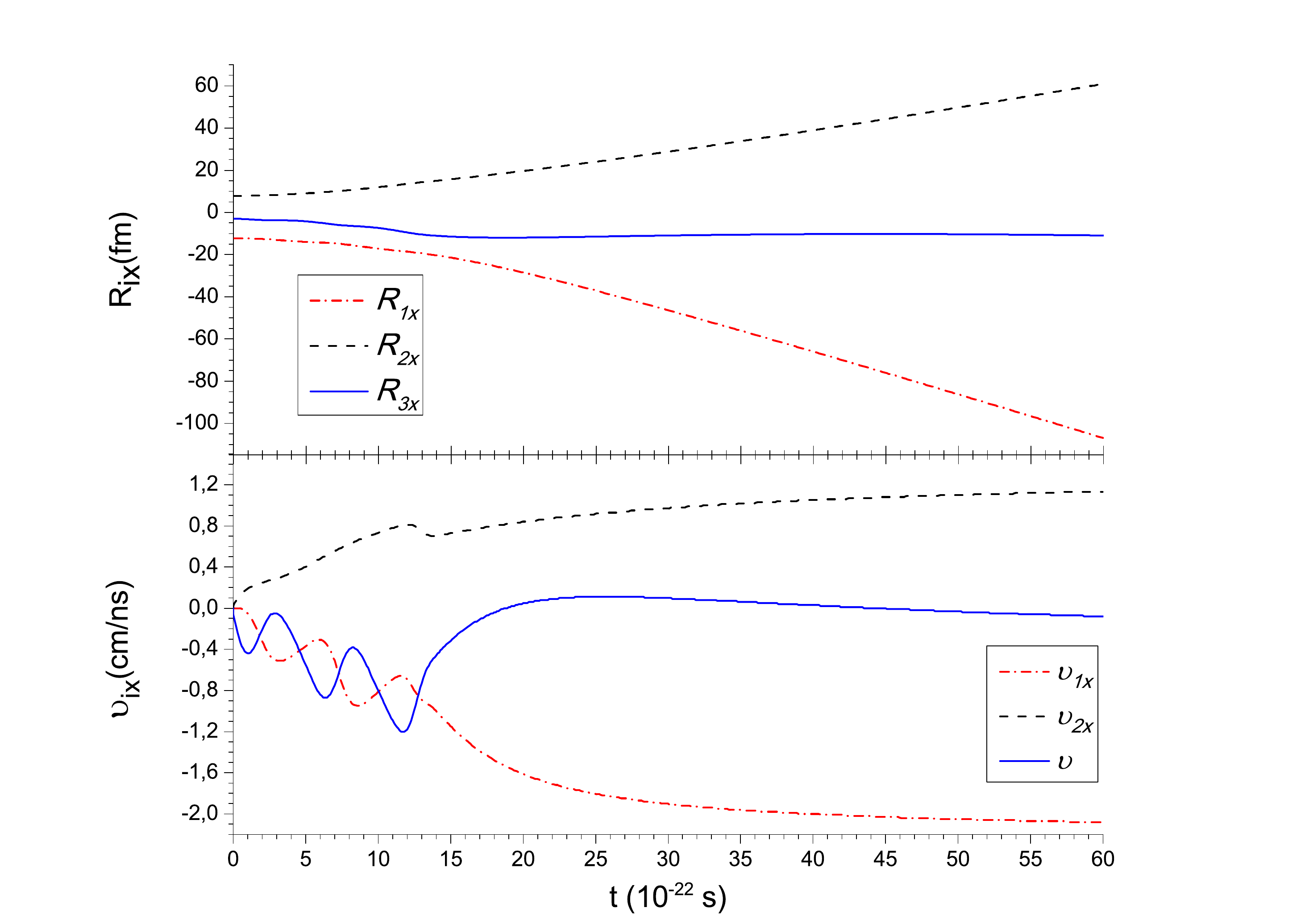}
\caption{
The $x$ component of the coordinates (upper part)
 and the velocities (lower part) in the decay into
 three fragments as functions of time, from Ref.~\cite{Tashk2016}.}
\label{fig:Fig18}
\end{center}
\end{figure}

The local minimum at point $R_{x_3}=9.4$ fm and $R_{3y}=0$ fm (see Fig.~\ref{fig:Fig17})
moves to the left (to the side of the  Ni nucleus), and starting from $R_{12}=23$ fm this minimum point
  is transferred to a saddle point.
 In Ref.~\cite{Tashk16} the authors have considered the case when
 initially all nuclei are placed in one line, which means that
 $R_{y_1}(t=0)=R_{y_2}(t=0)=R_{y_3}(t=0)=0$, since
 the energy of the collinear configuration in the pre-scission state is the smallest, and the
values of the $x$-coordinates of these nuclei (with the smaller relative distance between nuclei)
 correspond to the local minimum in the Fig.~\ref{fig:Fig17}. Both components ($x$ and $y$) of the
 initial velocities of the three nuclei are zero. In other words the formation of fragments of the TNS is
 so slow that the fragments have zero (or very  small) velocities. Nevertheless, the assumption that all
 initial velocities are zero, means that there is no net force,  which acts on the  nuclei in the
equilibrium state. Results of the  calculations of the equations of motion  with the above  mentioned
 initial conditions are shown in the Fig.~\ref{fig:Fig18}.  There it is shown that from the beginning,
 the Sn nucleus is going to breakup into the Ni+Ca system, and then at a time of
 $t\approx13.5\times10^{-22}~s$ the Ni+Ca system has decayed. Moreover, an important
result has been obtained, namely
that the central third nucleus (Ca)  has almost not changed it's coordinate, because it's velocity is
 about zero. This means, that the detection of the middle nucleus (Ca)
is almost impossible in an experiment.
 All together these conditions lead to collinear fission of the tri-nuclear system.
\begin{figure}[t]  
\begin{center}
\vspace{+1 mm}
\includegraphics[width=0.32\textwidth]{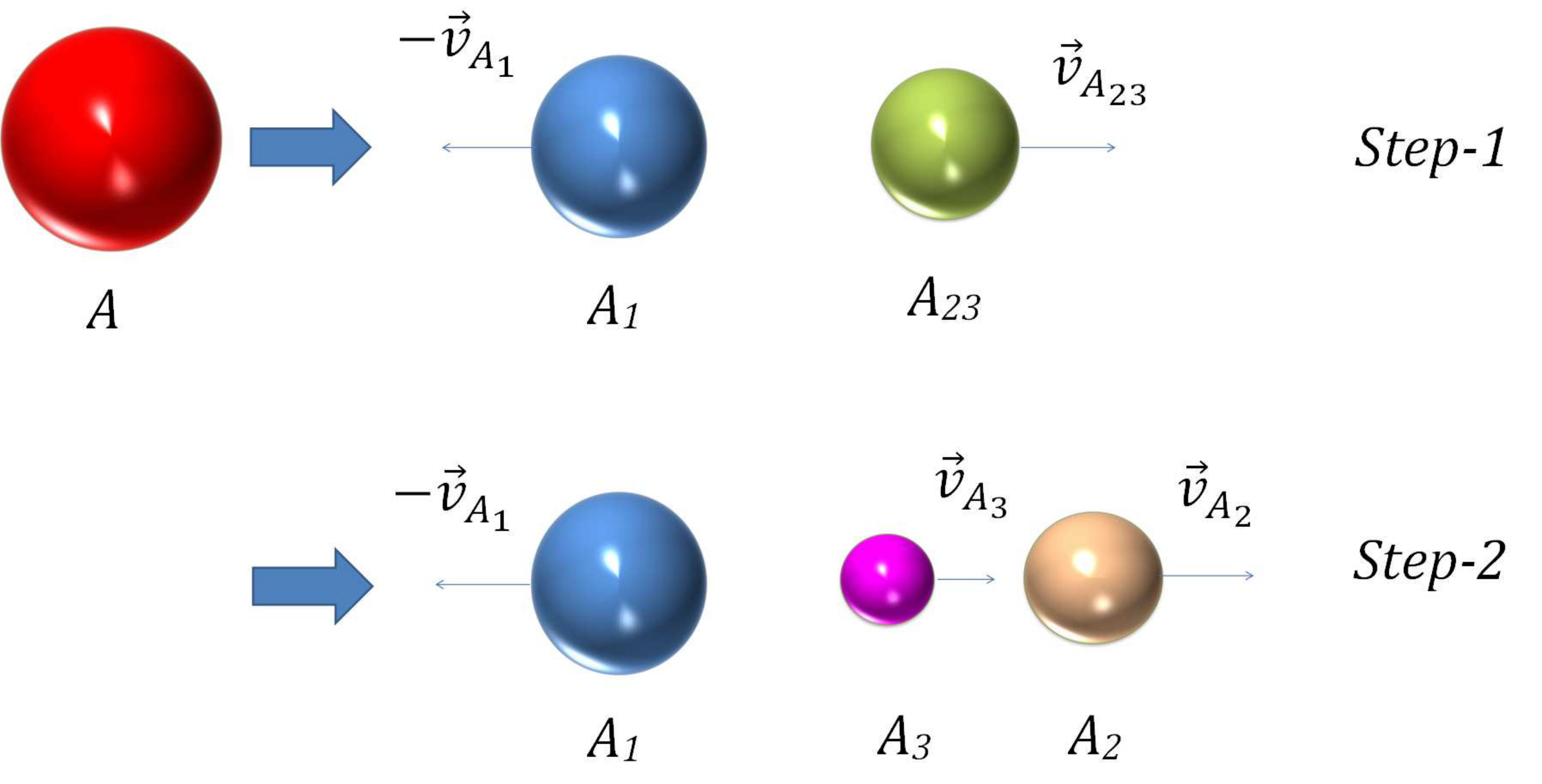}
\caption{ Illustration of the model of a sequential decay for the collinear ternary
              fission proceeding in two steps. In the first step   the
              combination $A_{23}$ + fragment $A_{1}$ are formed, followed (within a very short
              time sequence, in $10^{-21}$ seconds)
              by the fission of $A_{23}$ into $A_{2}$ and $A_{3}$.}
\label{fig:Fig19}
\end{center}
\end{figure}
In fact  in a similar recent study by Vijayaraghavan {\it et al}.~\cite{vijay19} the
trajectories of fragments in a ternary decay have been calculated explicitly for
equatorial and collinear trajectories. For lighter fragments an emission at relative angles of 90$^{\circ}$
is obtained. For third fragments of intermediate mass for an initial configuration
with the fragment on the axis results in a collinear decay.

As a further consideration we look into the kinetic energies of the three fragments with a model
of a  sequential break up of the fissioning nucleus (illustrated in Fig.~\ref{fig:Fig19}).

The collinear decay mode is quite important in the analysis of features of the
fission fragments, in particular their kinetic energies.
Several authors have addressed this point~\cite{vijay14,Manim0}. In Ref.~\cite{Manim0} an
 overall study of the probabilities
of ternary decays is given, comparing oblate and prolate fission. Ternary decays become dominant
 for cases where the central mass becomes larger than   $A_3$ = 30--40.
 Considering the potential energy
as function of the size of the central (smaller) fragment, $A_3$, it has been shown that for larger
central fragments, the  ternary mass splits
dominate the decays of heavy nuclei by  orders of magnitude.  Further the dependence of the
potential energy for ternary mass splits on the geometry of the ternary system
has been analysed by Vijayaraghavan {\it et al}~\cite{vijay14} for different orientation angles
 between the three fragments
 as shown in Fig.~\ref{Fig:angle}.

In order to put the collinear geometry into the perspective of the potential energy of the system,
 we can consider the potential between three clusters in dependence on
an angle between the  axial symmetry axis  of a two-fragment subsystem and
the third fragment, defining different shapes
 of the decaying system (Fig.~\ref{Fig:angle}.) In this work of  Vijyaraghavan~\cite{vijay14}
the potential energies of three fragments in different configurations have  been calculated.
 In  Fig.\ref{Fig:angle1} we show the potential energies of the fragments
for the three-cluster system, where the collinear and oblate arrangements are defined by the angles
shown  in Fig.~\ref{Fig:angle}, by varying the orientation angle for the three clusters.
We note the lowest potential energy for the arrangement with Ca as the central fragment.

 The potential energies are  clearly
 lowest for the collinear geometry. Around the lower region of the potential energy
quantum mechanical
fluctuations around the minimum must be expected. This is an effect of
 the uncertainty principle discussed in Ref.~\cite{Chuvilsk19}.
 This has also
 been considered in Ref.~\cite{Holm17}, where the question of the angular distribution of the
ternary fragments is addressed.
The considerations of
 the macroscopic features using clusters in the ternary nuclear system can, however,
 only give a general view, which describes
mainly the properties of the asymptotic phase space of three fragments
reached by the ternary decay.

For the discussion of the experimental results the kinetic energies of the
 fragments are of importance, they have been calculated in Refs.~\cite{Holm17,vijay12}.
For this study the  collinear cluster tripartition
 is defined to proceed in two steps,
in the first step an intermediate fragment defined as $A_{23}$
is formed with some excitation energy, as shown in  Fig.~\ref{fig:Fig19},
in the second step the
fragments  $A_3$ and  $A_2$ are formed in a binary decay. The fission barrier of the first step is
 small in comparison   with the  barrier of the
 second step, which is considerably higher then the first one.
 Note the fission barriers defined
 here are obtained with the depth of the potential well in the landscape of the potential
  energy surface, {\it i.e.} the value of the barrier relative the
  bottom of the minimum. The absolute value of the barrier is the
  Coulomb barrier, which determines the total kinetic energy of the
  fission products.
  These kinetic energies have been  shown in Fig.~\ref{fig:Fig4},
as function of the mass of the central fragment, $A_3$ and
 for different excitation energies
of the intermediate fragment  $A_{23}$.
A considerable value for the excitation
energy of $A_{23}$ appears for
a large fission probability in the second step. Actually a value of at least 30 MeV must be
 assumed, in this case the kinetic energy of the central fragment  $A_3$ is close to zero,
similar to the other approaches discussed in Ref.~\cite{Holm17,Tashk2016}
and in Ref.~\cite{Karpov16}.
  The kinetic energy $E_{kin_A{_2}}$ of $A_2$ is maximum and reaches values of 150 MeV. This is
actually an artefact of the model as pointed out in Ref.\cite{Pyat17}, because no
 intrinsic excitation of the three fragments are taken into account.
In this reference the energy correlations between the two outer fragments ($A_1$ and $A_3$,
 obtained with the
 experimental setup  COMETA) show that the kinetic energies show several
 groups with much lower kinetic energies.
Most important for
 the experimental circumstances is the fact that the central
fragment usually is lost due to it's low
kinetic energy  by being
absorbed in the target and/or in  the target-backing.

The same result for the
 kinetic energies
has been obtained in the work of Holmwall et al.~\cite{Holm17} in
explicit calculations of the  trajectories. Their approach to calculate the
kinetic energies is based on the ``Almost sequential decay'',
as used in the approach of Ref.~\cite{Tashk2016}.
In this work also the dependence on various
parameters of the decay are shown, namely the dependence on the possible neutron emission
(neutron-multiplicities), the effect of lateral momenta and the variations
in initial geometrical positions of the central clusters.
For deviations from the central position, it is concluded
that they destroy the collinearity in the final channel.
Actually a detailed analysis of the kinetic energies of the fragments and the
dynamical constraints for the decay
as done in Ref.~\cite{Holm17} within a model with existing clusters
(as in many  other approaches) gives results, which prevent the ternary decay,
 in contradiction to the experimental observations.  The authors of Ref.~\cite{Holm17}
 did not include the nuclear interaction in the calculation of the
 nucleus-nucleus potential. As a result the kinetic energy of the outgoing
 fragments had been overestimated.
  As a conclusion we can state that these approaches based on
preformed clusters without deformation of the fragments, fail to describe
the experimental results.

 \begin{figure}[t]  
\vspace{+3 mm}
\begin{center}
\includegraphics[width=0.40\textwidth]{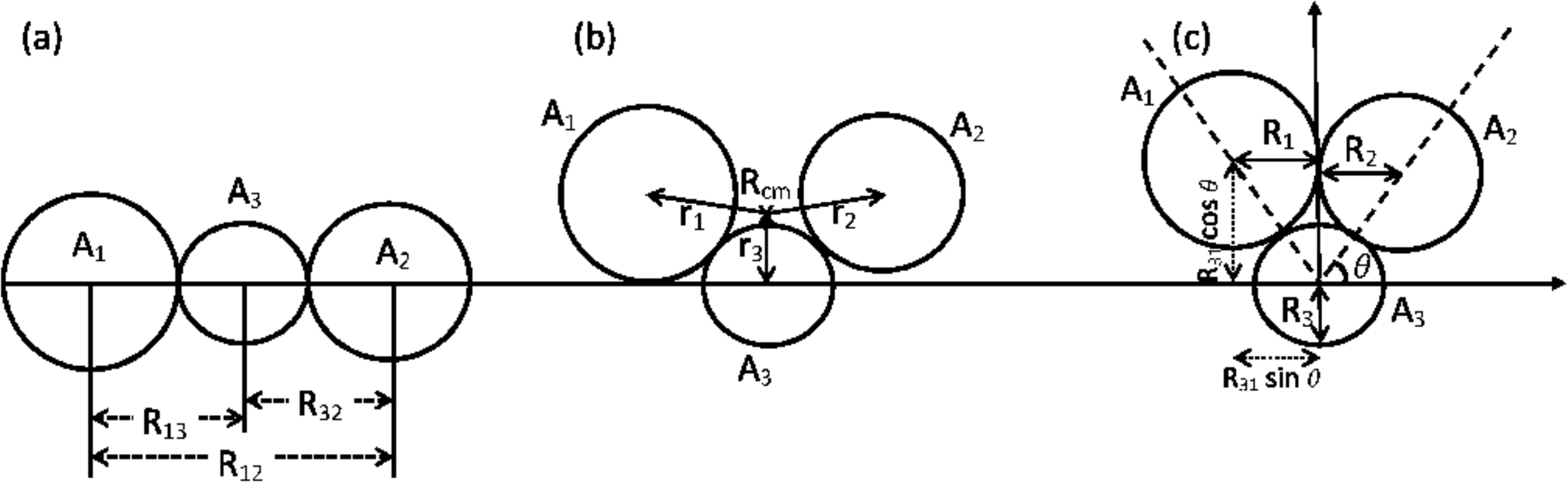}
\hspace{+0 mm}
\vspace{-0mm}
\caption{The coordinates of three clusters defined with an orientation
  angle of fragment $A_{3}$ relative
 to  the  axis between the two other fragments, from Ref.~\cite{vijay14}.
The corresponding potential energies are shown in Fig.\ref{Fig:angle1}.}
\label{Fig:angle}
\end{center}
\end{figure}
\begin{figure}[t]   
\begin{center}
\vspace{+5 mm}
\includegraphics[width=0.45\textwidth]{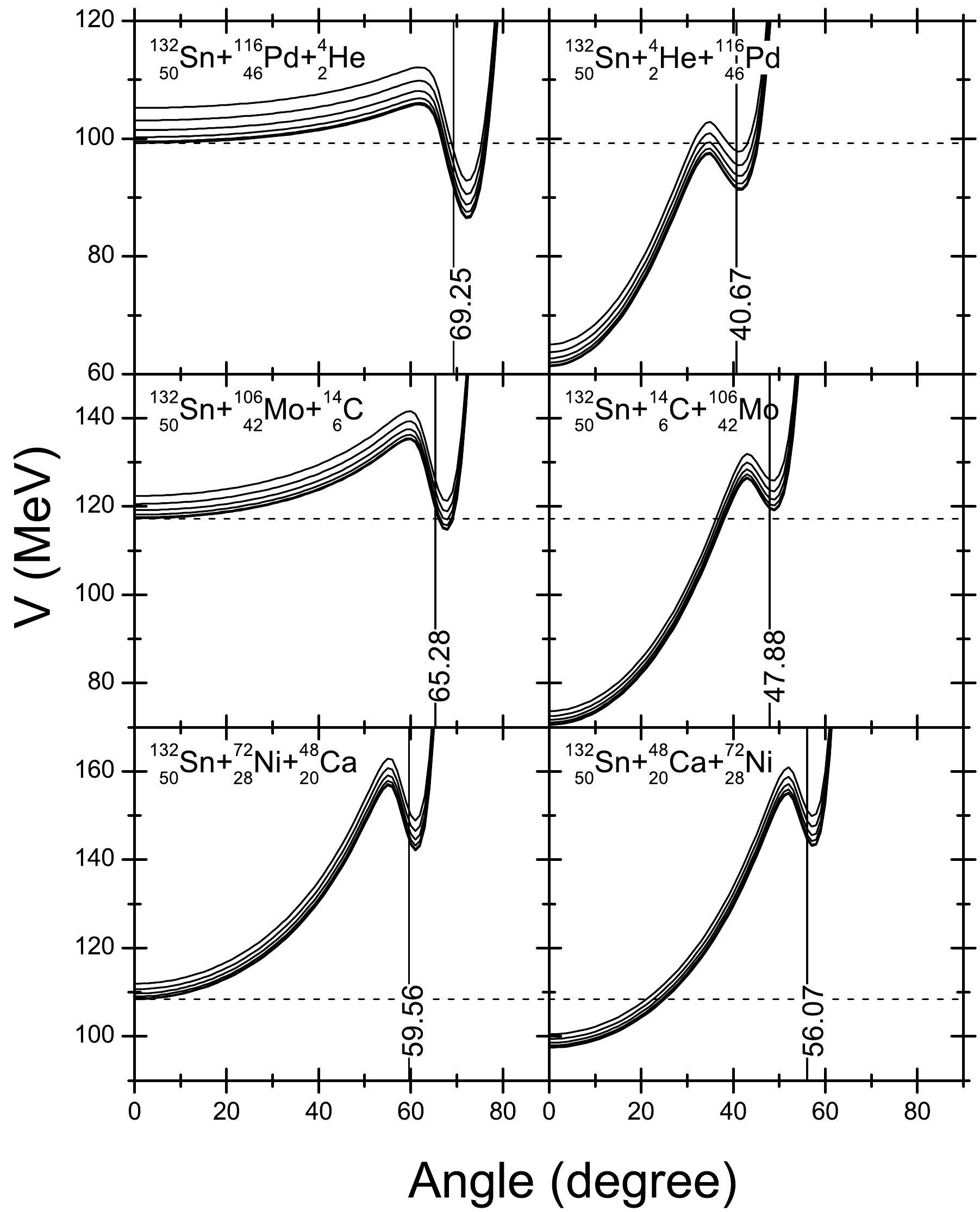}
\hspace{+0 mm}
\caption{The potential energies  of three clusters defined for different orientation
 angles for the three clusters for the combination $A_{23}$ + $A_{1}$ as defined in Fig.\ref{Fig:angle}.
Three cases for the central particle are considered, with $^{4}$He, $^{14}$C and $^{48}$Ca}.
\label{Fig:angle1}
\end{center}
\end{figure}

More relevant are  the
concepts with a continuous deformation path like on hyper-deformation, discussed before
 and in the approach of Karpov\cite{Karpov16},
where the total system with three nuclear centers is used all along the
fission path. In the latter approach
also two barriers exist, which are strongly influenced by the shells, in particular
the second barrier (which is higher) is very strongly reduced by the
 shell effects in the fragments.

Initially ternary fission has been defined by the observation of a third light
particle emitted
perpendicular to the fission axis
in coincidence with the binary fission fragments. The most important feature,
which appeared with the study of \textit{true} ternary fission is the fact that the
decay into three comparable mass fragments becomes collinear.
In  the  experiment by Schall~\textit{et al.}~\cite{Schall}, the question of the
decay from oblate deformed shapes is addressed, typically a  decay with three
similar vectors was expected.
The experiment  has been designed  to detect ternary
fission events with the emission of three heavier fragments at relative angles of
120$^o$. The experiment
 covered a large solid angle
 using several ionisation chambers (detectors) each covering a large angular range of
 (90$^o$) degrees,  designed to observe  over a large solid angle
the {\textit triangular shape of the decay-vectors}.
 This experiment gave  a negative result,
with  an upper limit of the  probability for this decay of $1.0\cdot 10^{-8}$/(binary fission).
 The failure in observation of true ternary fission  in this experiment
is explained by the very small probability of the population of the configuration
which is presented in Fig. \ref{Fig:angle}(c). The potential energy barrier of the last configuration
 is very high and its population needs higher excitation energy for the fission to occur.
 This argument is illustrated in Fig. \ref{Fig:angle1} where the potential energies  of three clusters
 are presented for different orientation angles between lines connecting the centres of mass of
the three clusters.

\subsection{Deformed shell effects, symmetric and very asymmetric binary  fission}
In fission the interplay of the macroscopic (liquid drop) and microscopic (nucleon orbits)
effects are most important~\cite{Strut89,Strut89x,Wilkin76,blons}. A triple humped barrier
is observed in several cases,
giving rise to important phenomena, like fission isomers, multi-modal binary
fission and creating the way to
 ternary fission as illustrated in Fig.\ref{fig:Fig3}.
Binary fission has an asymmetric distribution of the masses with the  isotope  $^{132}$Sn appearing
in almost all fission decays~\cite{wag91}.

In the very asymmetric binary fission
 lighter fragments with masses $A$  = 60--80
 can be observed. In these cases  ``cold'' fragments may appear and
and deformed shell effects are important.
In several works~\cite{Rochman} experimental results on very asymmetric binary fission have
been obtained,
these are single arm  measurements. An example of such experimental results
 is shown in   Fig.~\ref{fig:Fig22}.
In these cases the yields of masses around $^{70}$Ni are enhanced ~\cite{Rochman},
in  particular for the
higher kinetic energies of the fragments.
These yields in the region of the
lighter  fragments of $A = 70$, can be due to special shell effects,
in particular the deformed shells around  $^{70}$Ni ($Z = 28$, $N = 52$), or due to ternary fission..
\begin{figure}[t]  
\begin{center}
\vspace{+2 mm}
\includegraphics[width=0.45\textwidth]{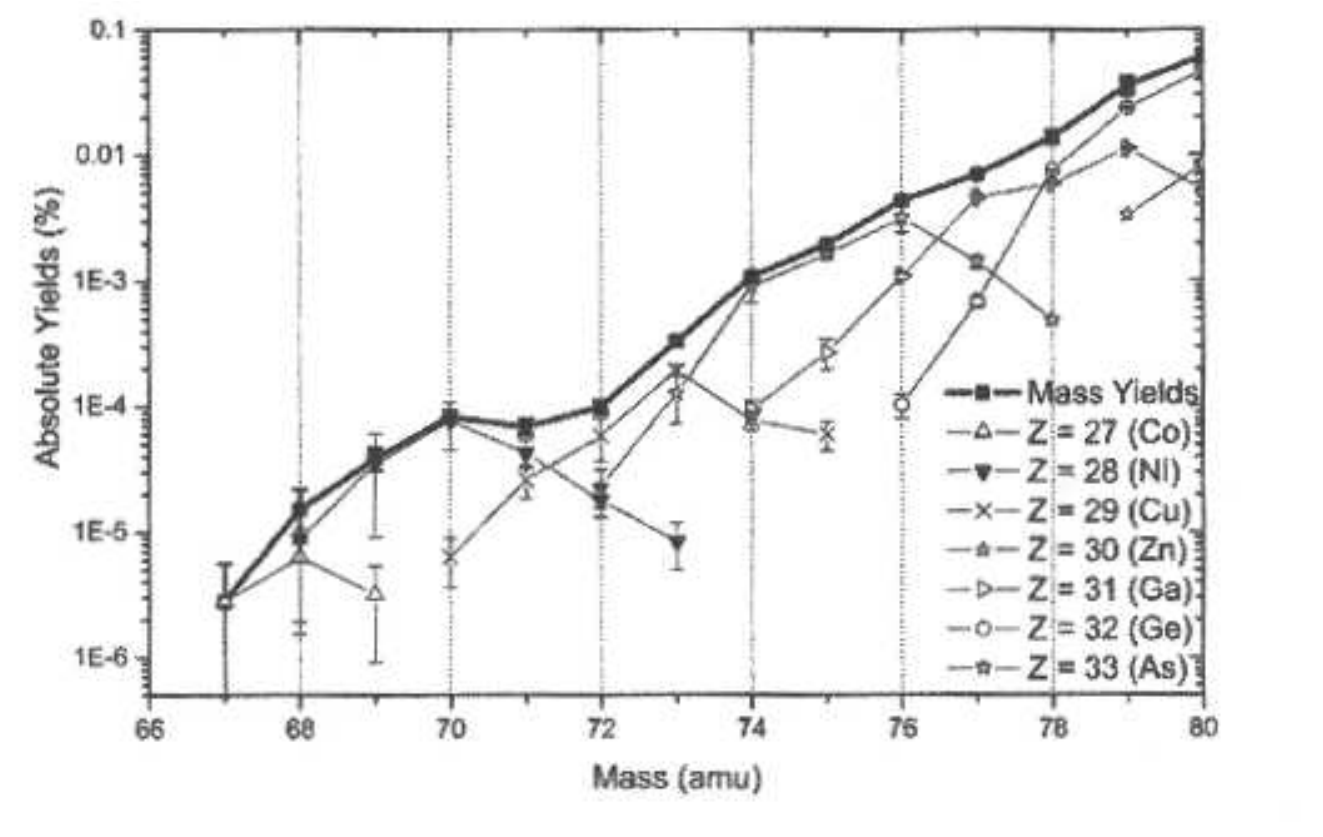}
\vspace{-1.5 mm}
\caption{Very  mass asymmetric  fission yields of the
$^{245}$Cm(n,f) reaction. (adopted from Ref.~\cite{Rochman}).
The fission yield around mass $A=70$ ($Z = 28$) is enhanced.}
\label{fig:Fig22}
\end{center}
\end{figure}

\section{Recent experimental results on ternary
 cluster decays in  $^{236}$U(n,fff) and  $^{252}$Cf(sf)}

Inspecting the recent experimental observations and the numerous theoretical
 predictions ~\cite{Manim1,poe05,Zagreb10}
we can state that in the heavy systems and for ternary fragments with larger charge,
ternary  collinear decay from a prolate configuration is
preferred. We refer to the ternary decays in the present work
 as to  ``true ternary fission'', see also e.g. Zagrebaev \textit{et al.}
in  Refs.~\cite{Zagreb10a,Zagreb10}.
A more recent survey of clustering effects in binary and ternary fission can be found in
the articles by G. Adamian, N. Antonenko and W. Scheid~\cite{Adamian}. In the work by
 D. Poenaru and W. Greiner~\cite{poe10} it is shown, that in heavy nuclei collinear ternary decays
are observed with increasing probabilities for increasing charge of the total system. For
even heavier systems quarterny fission must be considered.

Concerning the experimental evidence for true ternary fission several experiments have been performed
at the FLNR in Dubna, to list most of them:\\
a) with two complete detector telescopes (called FOBOS \cite{Pyat10,Fobos}) with 180 degrees
relative angle for binary coincidences in spontaneous fission
of $^{252}$Cf(sf) and with the same experimental device\\
b) neutron induced fission  $^{236}$U(n,fff), performed  at the reactor in Dubna.\\
c) Experiments of binary fission for $^{252}$Cf(sf) in coincidence with neutrons
 with a system called ``MiniFobos'', containing smaller
detectors of similar structure as with FOBOS, Ref.~\cite{Pyat12}.\\
d) Binary coincidences for $^{252}$Cf(sf), using PIN-diodes (for energy and timing signals)
   called COMETA.

The most complete data set has been obtained with the system described
in Refs.~\cite{Pyat10,Pyat12,Fobos}.
Quite importantly, supporting the claims made in these works, the multi-modal ternary fission decays
predicted from the calculations of the PES's, have been extracted from the complete
 data in a later stage, see Refs.~\cite{voe14,voe15}.

\subsection{The FOBOS Experiments}
For the observation of the ternary decays with the missing mass method
 the decay vectors of two fragments,
which are identified with all parameters (mass, energy and angle)
in a measurement of the two fragments in coincidence
is needed.
In the rather common single arm experiments one fragment is identified with high
precision (mass, energy and angle)
with the partner being known under the assumption of
a binary decay. However,
in the recent experiments, coincidences between two fully identified fragments have
 been measured
 with two FOBOS-detector-telescopes~\cite{Pyat10,Fobos} placed at 180$^o$ (see the
illustration in Fig.\ref{fig:Fig23Fobos1}).
The FOBOS detectors allowed  the measurement of
 time of flight, energy, energy-loss and angle with high precision with a
\textit{very large solid angle}.

\begin{figure*}[t]  
\begin{center}
\vspace{+5 mm}
\includegraphics[width=0.72\textwidth]{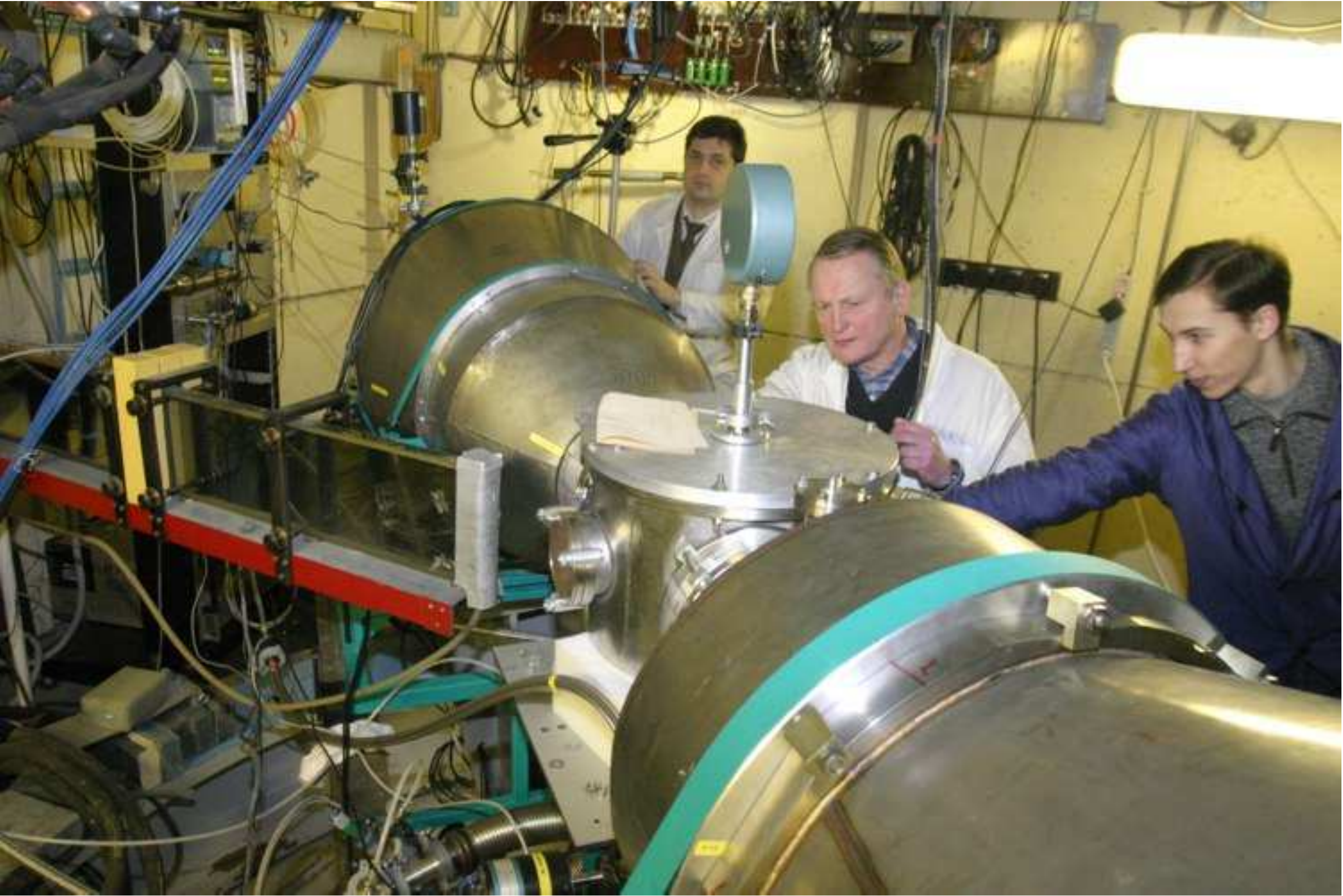}
\vspace{+4 mm}
\caption{The two FOBOS detectors placed at a relative angle of 180$^{o}$ degrees
for the measurement of coincidences between two fragments.
Two of the leading experimentalists, D. Kamanin and Y. Pyatkov (in the center) are depicted.}
\label{fig:Fig23Fobos1}
\end{center}
\end{figure*}

The details of the detector structure are shown in a scheme
in Fig.\ref{fig:Fig24}. In Ref.~\cite{Pyat10} the experimental evidence for
 the missing mass effect had been  explained.
Due to the energy loss and momentum dispersion in the support foils for the low
pressure parallel plate counters in the front
of the Bragg-Ionisation chambers a particular scheme has been considered. In this scheme
the third fragment after dispersion in the foils
 is blocked by the support structure for the foils
of the ionisation chambers.
 Actually in the recent
calculations, cited in this work, it appears that the central (third) fragment,
will have very low kinetic energies,
thus it most probably gets stuck in the material of
 the support of the target or already in the target itself.
 Thus the missing mass effect has also
 been observed with detector systems consisting of solid state
detectors (PIN-Diodes, COMETA-set-up),
with a measurement of masses by the
time of flight (TOF) and energy parameters.

\begin{figure}  
\hspace{+2 mm}
\includegraphics[width=0.4\textwidth]{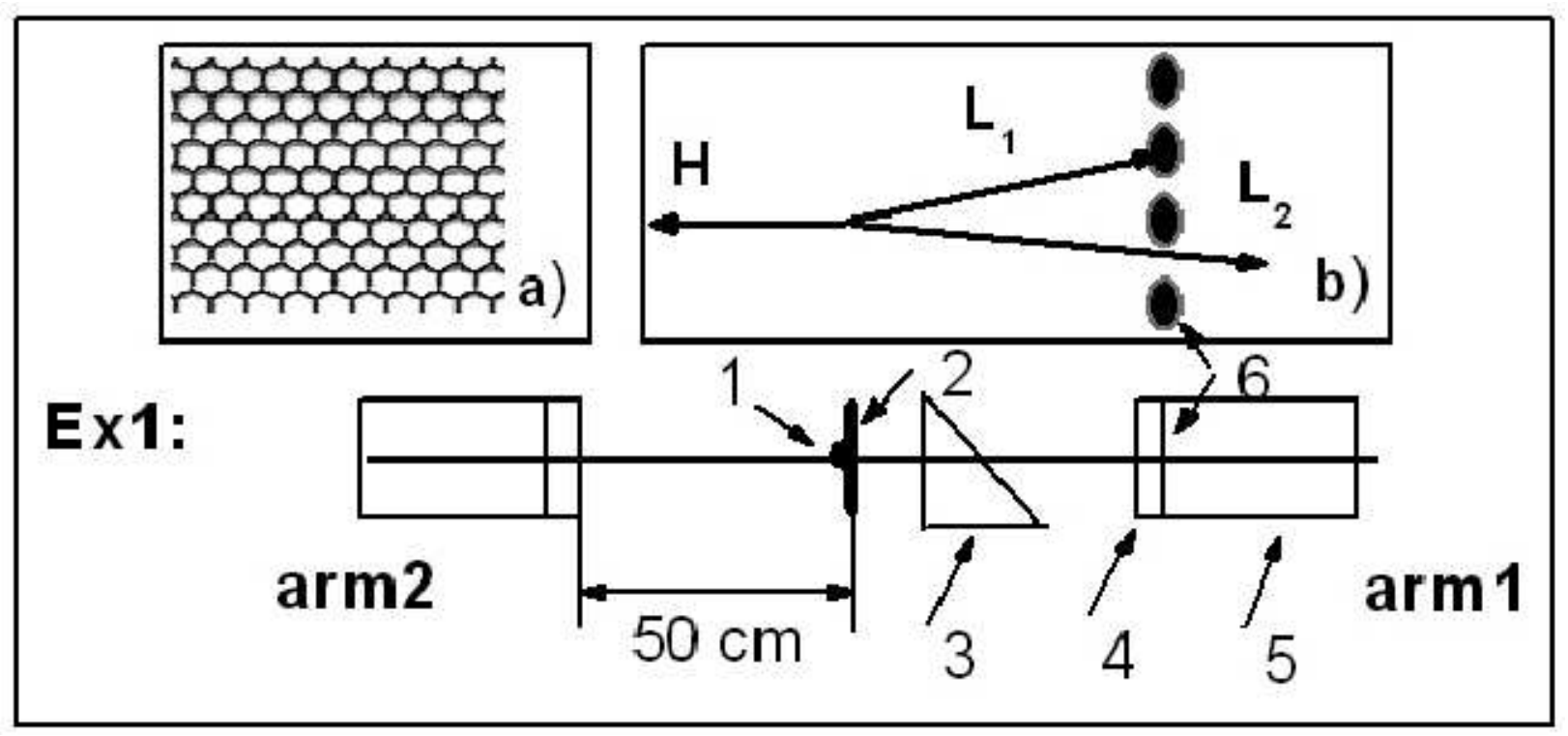}
\vspace{+2 mm}
\caption{Scheme of (Ex1) for coincidence measurements of two
fragments of the collinear decay of $^{252}$Cf. Here: 1 -- Cf
source, 2 -- source backing, 3 -- micro-channel plate (MCP) based
timing "start" detector, 4 -- position sensitive avalanche counter
(PSAC) as "stop" detector, 5 -- ionization chamber (BIC) with the
supporting mesh, 6 -- mesh of the entrance window. The side view of
the mesh is shown in the insert (a), a mesh section is presented in
the insert (b). After passage of the two fragments through the source
backing, two light fragments (\textit{L}$_{1}$ and \textit{L}$_{2}$,
originally fragment \textit{L}), are obtained with a small angle
divergence due to multiple scattering. One of the fragments
(\textit{L}$_{1}$) can be lost hitting the metal structure of the
mesh, while the fragment \textit{L}$_{2}$ reaches the detectors of
the arm1, insert (b). The source backing (2) exists only on one side and
causes an angular dispersion in the direction towards the right side
(arm1). }
\label{fig:Fig24}
\end{figure}

In these experiments the binary decay appears as two strong regions of the
registered  binary fragments in coincidence, as shown in Fig.~\ref{fig:Fig26}.
The determination of all  parameters in coincidence with
two vectors (angles, mass and energy) gives the complete kinematics for the ternary decay,
in these experiments the  determination of the missing mass indicated by ``7'' is well determined.
This  approach has been established, and
the phenomenon of collinear cluster tripartition,  the CCT-decay, has been described,
 Refs.~\cite{Pyat10,Pyat12}.
The CCT-decay  has been observed independently in three different experiments with
 a missing mass peak:
 for the spontaneous
fission of  $^{252}$Cf(sf,fff), and in another experiment (at the neutron
beam of the nuclear reactor at the JINR)
 with the same experimental
set-up for neutron induced fission in $^{235}$U(n$_{\rm th}$,fff),
see Refs.~\cite{Pyat10,Pyat12}. In the latter case with U-nuclei
the missing mass turns out to be
smaller (as expected), namely it is observed as  isotopes of  Si ($Z$=14),
 corresponding to smaller charges and masses due to the smaller total mass
 of the fissioning nucleus, with  $A=236$.
Further observations,  which confirmed these results have been obtained
with very different experimental arrangements with the system called COMETA.
In the fission mode in  the two cases $^{252}$Cf(sf,fff) and  in $^{235}$U(n$_{\rm th}$,fff),
 the decay is dominated  by
fragments which are
strongly bound isotopes (clusters, nuclei with closed shells)
 of  Sn, Ni, and Ca. The latter, Ca (or Si),  as the
smallest third particles, if positioned  along  the line connecting the fragments Sn and Ni,
 they correspond to a minimum value of the potential energy.
The PES calculated for these cases were compared and
 illustrated in Fig.\ref{fig:Fig12_252Cf0n} and in  Fig.~\ref{fig:Fig13_PESU236}.

In the experiments of  Refs.~\cite{Pyat10,Pyat12}, an important experimental
 feature appears, which leads to unsymmetric
experimental observations in the two detector arms. This effect occurs due to the target
 backing (and the material in front of the detector telescopes) pointing only to one
of the detector arms (arm1).
Two of the three collinear fragments (2 and 3), quasi-bound in the fragment $A_{23}$  move
 towards one of the detectors
(called arm1). Fragments  $A_{2}$ and $A_{3}$   are dispersed in angle
while passing through the material of the source/target backing and the entrance foils of the
detector telescopes.
This originally introduced interpretation of the observed effect, can be
almost completely relaxed in view
of the results (obtained later) on the kinetic energies. These calculations
have shown that the central (smaller)
fragments attain very small kinetic energy, and they most probably already get lost
by absorption in the target
and in the target support pointing only into one direction (towards amr1).
\begin{figure}[t] 
\begin{center}
\vspace{+5 mm}
\includegraphics[width=0.42\textwidth]{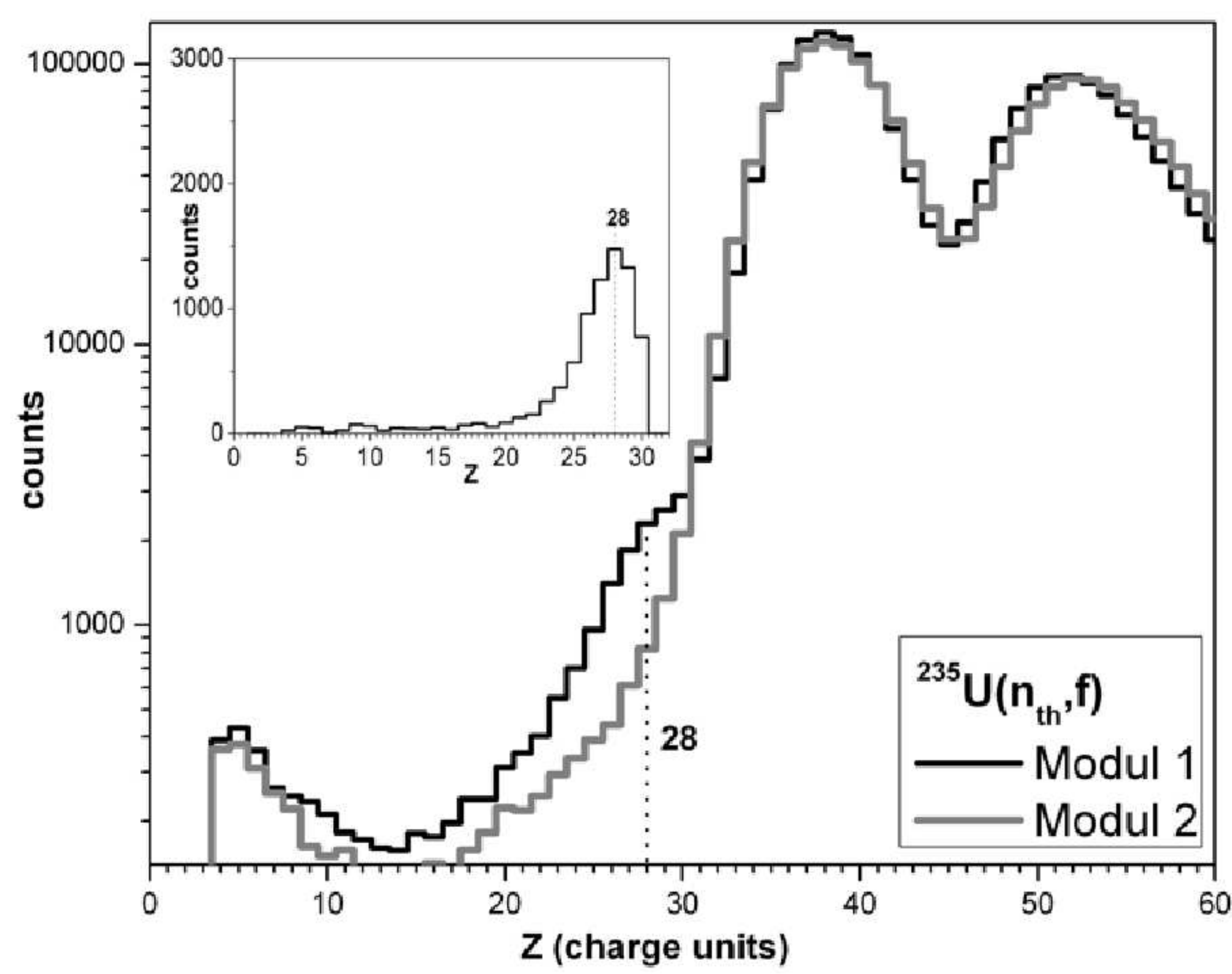}
\caption{
The energy loss signal derived from the
the ionisation chambers of the FOBOS-telescopes.
The signals  have been extracted from
the drift time of the fission fragments tracks in the ionization chambers.
 Two sides relative to the target are
referred to as ``Modul 1'' and ``Modul 2'', only one shows the missing fragment yield,
according to Fig.\ref{fig:Fig26}.}
\label{fig:Fig25}
\end{center}
\end{figure}

 With the knowledge on the kinetic energies derived from the more recent calculations
(see the kinetic energies in Fig.~\ref{fig:Fig4})  we can expect
 that the central fragment $A_{3}$
due  to its very low kinetic energy gets lost in all cited experiments. For the alternative case
 with a small angular dispersion due to
 the target backing material, it gets dispersed in angle and
 it is stopped in arm1 on a support-structure (label 6) of the thin foils in front of the
Bragg-detectors, see Ref.~\cite{Pyat10}. This fact gives the  characteristic side peak
in the mass-mass-correlations with a small maximum (label 7) corresponding
 to a missing mass ($A = 40 -- 48$), shown in Fig.\ref{fig:Fig26}.
This is related to the observation the complementary fragments (complementary to Ca)  the
Ni-isotopes in the missing  mass effect.

In addition,
the signals for the energy-loss have been obtained from the measurement of
the drift time of the fission track in the ionization chamber.
During the measurement of the time-of-flight of  the fission fragments  and their energies,
two more parameters being sensitive to the nuclear charge are added.
The drift time of a track formed after stopping of a fragment in the gas volume of the
Bragg ionization chambers is known to be linked to the fragments nuclear charge.
Special calibration procedures have been worked out for the
nuclear-charge determination of the fission fragments~\cite{Tyukav09}.
The charge resolution is approximately 3.8 units (FWHM) for
the fission fragments in the light-mass peak, the mean values
for each charge are correctly determined.
The spectrum shown in Fig.~\ref{fig:Fig25} shows a clear peak
 with the nuclear charge of the Ni-isotopes  next to the signals of the
two charges due to the two fragments of  binary  fission. At this point
we emphasize, that the distributions (the bumps) in mass of  all three fragments
correspond to a wider range in mass (as opposed to the assumptions in the
model calculations in the examples discussed before in this work.
\begin{figure} 
\vspace{+6 mm}
\begin{center}
\includegraphics[width=0.420\textwidth]{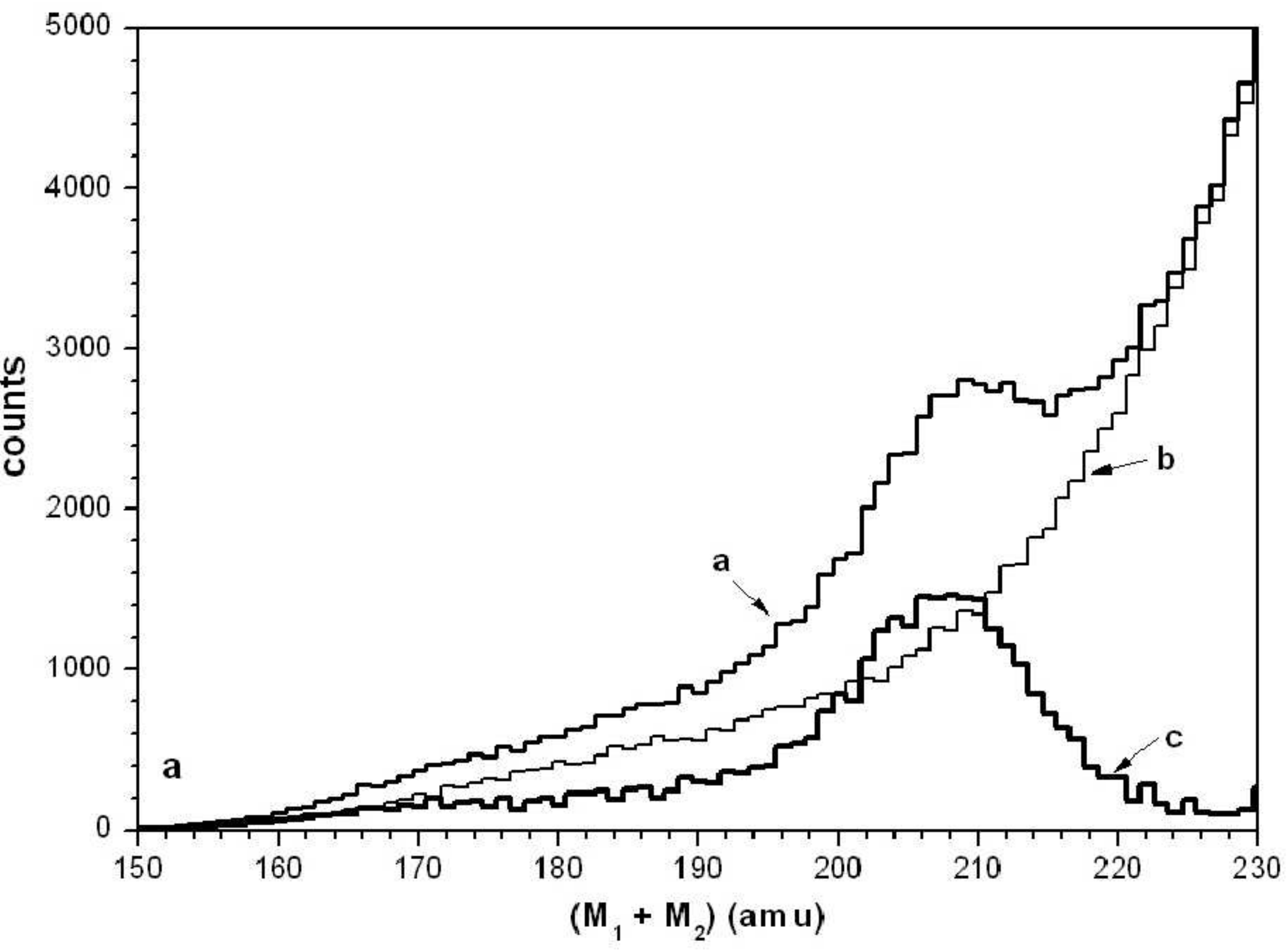}
\vspace{+4 mm}
\includegraphics[width=0.42\textwidth]{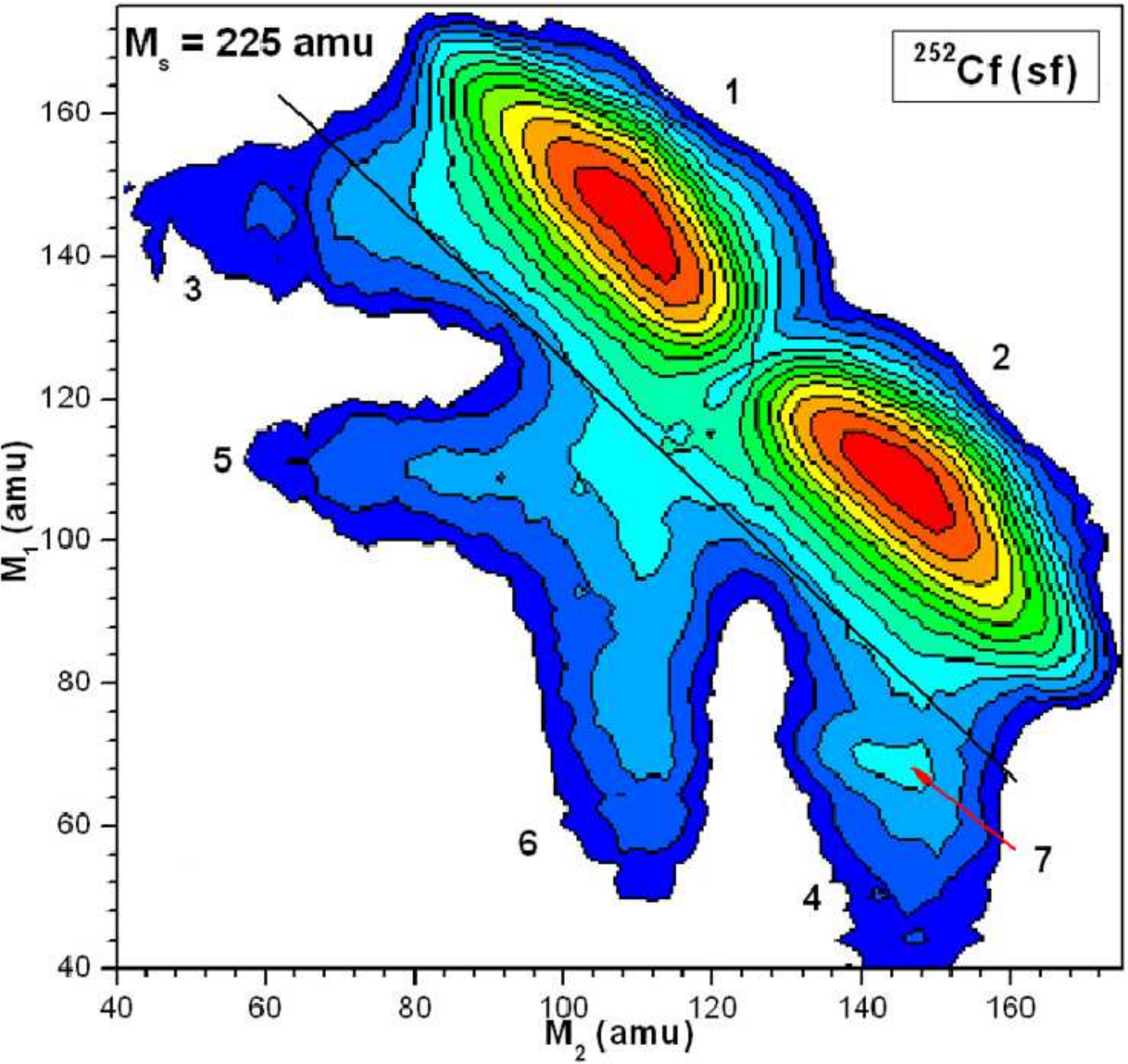}
\vspace{-2 mm}
\caption{ Upper part: the projection of  the mass spectra for
          the detectors in arm1(a) and arm2(b)
         for spontaneous fission in $^{252}$Cf.
        The missing mass spectrum (c) is obtained by the difference.  Lower part:
          shows the original raw data, the coincidences in the two FOBOS-detectors,
          the contour map  of the mass-mass distribution
         (in logarithmic scale the steps between the
         lines are a factor 2.5)
        of the fragments of $^{252}$Cf(sf),
         detected in coincidence in the two opposite arms of the FOBOS
         spectrometer. The specific bump (missing mass) in the yields in arm1 is
          indicated by an arrow, 7.
        The sum line with $M_{s}$ = 225 amu is shown as an illustration.
        }
\label{fig:Fig26}
\end{center}
\end{figure}

For the neutron induced fission of  $^{235}$U(n,fff) as shown in Fig.~\ref{fig:Fig27}
 the same features are observed as before, for
Fig.~\ref{fig:Fig26}, however, with a smaller (charge $Z$=14)
missing mass of  28 - 36 due to the smaller total mass, as described
in Refs.~\cite{Pyat10,Pyat12}. The side peak in this case corresponds to
 some isotopes of Silicon. Therefore the missing mass peak
appears closer to the sum of the masses of the
 binary fragments (sum of the binary masses is now 236).
Due to the support (metallic) grid in front of the thin foils in front of the
Bragg-ionisation chambers, scattered fragments with high intensity
appear in both cases, they form the background shown in Figs.~\ref{fig:Fig26} and Fig.~\ref{fig:Fig27}.
This scattered background can be taken from the other detector arm, where
 this missing mass effect is absent, it defines the background with high counting rates.
The final signal is shown  in the figures by  subtraction
of the counting rates in the two arms.
 Due to the high counting rate in these experiments,  the
 background subtraction can be done with very good precision.
We repeat the fact that the two detector arms give different yields, due to the orientation
of the target-backing to one of the detector arms, which causes the important difference
in the spectra observed, with respect to the detector which ``sees the pure target side''.
The yield in the bump is now  5.1x10$^{-3}$/(binary fission) and it corresponds to the sum
all isotopes in the $A_3$ fragments occuring in the decay.

\begin{figure} 
\begin{center}
\vspace{+6 mm}
\includegraphics[width=0.40\textwidth]{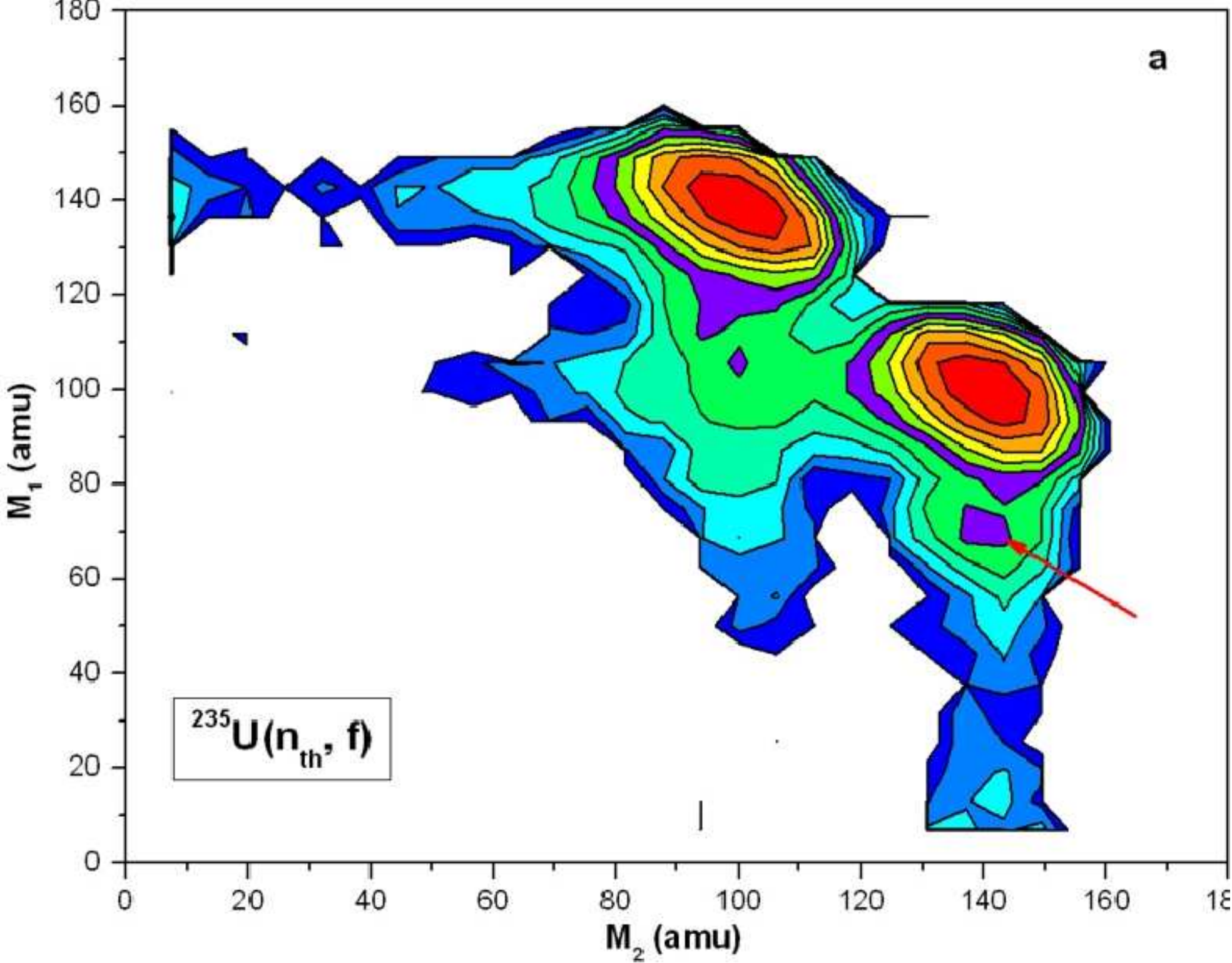}
\vspace{+2 mm}
\caption{ Mass-mass  correlation of the two registered masses measured with the FOBOS detector
      system for the  n-induced fission $^{235}$U(n,fff) at the reactor in DUBNA.
         The extra missing mass peak indicated by an arrow (compare previous figure)
         of the ternary decay (due to a smaller missing mass now
          isotopes of Silicon (Si))
          has now a smaller distance to the binary fission fragments
          compared to the case of Fig.\ref{fig:Fig26},
         due to the smaller total mass of 236.
         Projections of the sum of the masses
         are shown in Fig.\ref{fig:Fig29}}
\label{fig:Fig27}
\end{center}
\end{figure}

\begin{figure} 
\begin{center}
\vspace{+4 mm}
\includegraphics[width=0.36\textwidth]{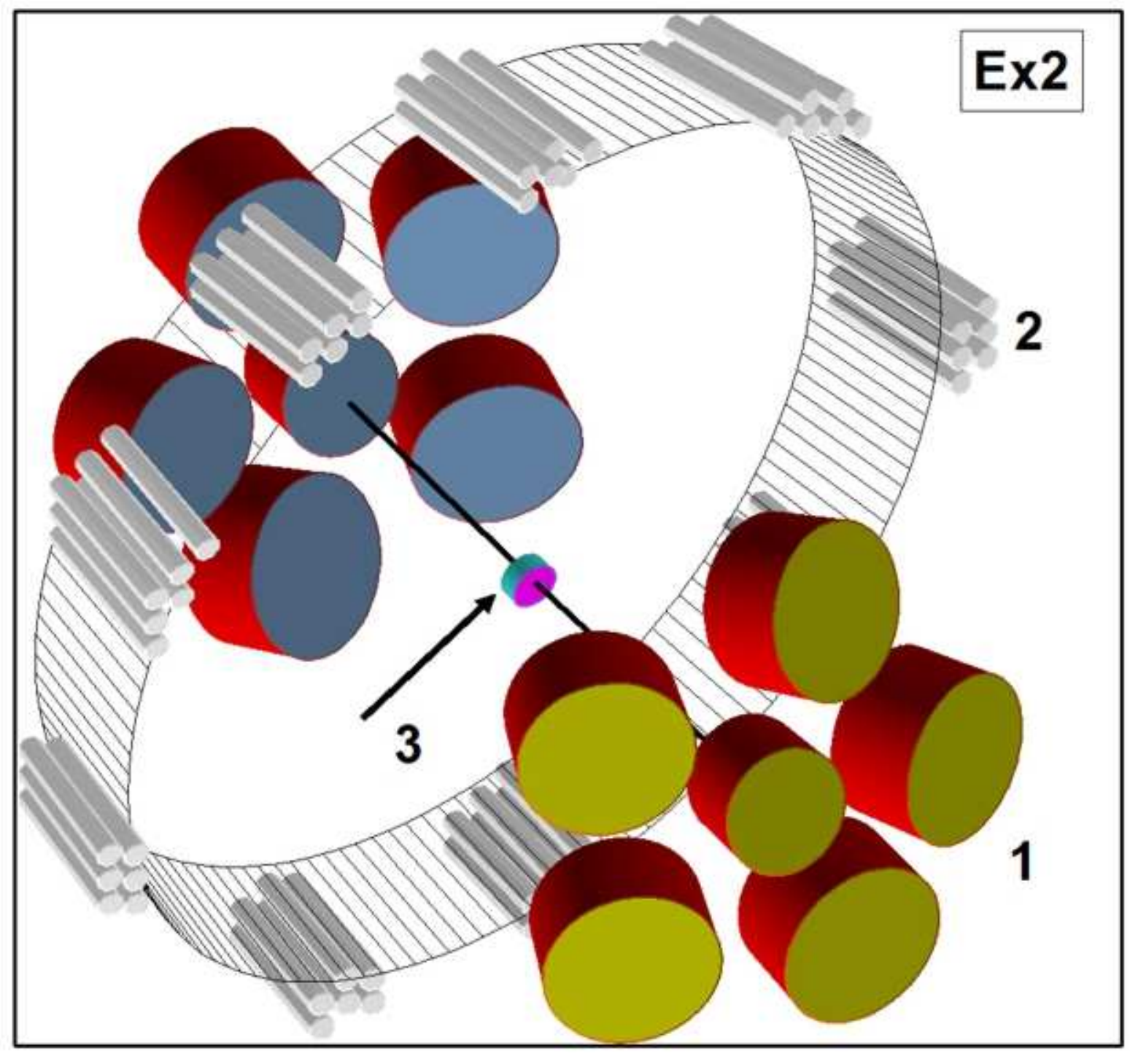}
\vspace{+2.4 mm}
\caption{ Experimental Mini-FOBOS set-up for the measurement of correlations
of the two fission fragments
in coincidence with neutrons. Labels: 1-Minifobos-detectors (Bragg-ionisation chambers),
 2-neutron detectors, 3 - source(target).}
\label{fig:Fig28}
\end{center}
\end{figure}

\subsection{Fission studies with neutron coincidences}
Further studies of fission decays with coincidences of heavy fragments at
relative angles of 180$^o$ and with neutrons
detected in coincidence have been performed as described in Ref.~\cite{Pyat12}.
The experimental arrangement is shown in Fig.~\ref{fig:Fig28}.
The neutrons are registered in neutron detectors with
an  assembly, which surrounds the center, where  the
source for  $^{252}$Cf(ff,fff) is placed.  For this purpose a modified smaller
 FOBOS system has been used,
shown in Fig.~\ref{fig:Fig28},
which allows the observation of binary coincidences in coincidence  with neutrons.

Fig.~\ref{fig:Fig29}  shows
the mass-mass correlation for  $^{252}$Cf fissions in coincidence with
neutrons. We notice the absence of the ``tail'' of
scattered binary fragments seen in the other cases (Figs.\ref{fig:Fig26} and \ref{fig:Fig27}).
We compare in the lower part the projections on the mass scales denoted as ($M_1$ and $M_2$),
 the first showing the bumps
observed in the Ni-region in the  three experiments. For $^{252}$Cf(sf), Ex2, denotes the result
with neutron coincidences, where no tail of scattered binary fragments appear, for the case of
 $^{252}$Cf(sf) they are visible. The spectra show the sum of
masses without the mass of the missing fragment.

Quite remarkable is the observation of the yields  in the experimental correlation
plot with a definit multiplicity ($n$=2), shown in Fig.~\ref{fig:Fig30n=2} related to
decays with missing lighter fragments as central
 nuclei with the neutron rich isotopes of Ne, O, and C.
In fact, with these lighter fragments we anticipate an increased neutron emission
 from the two neck-ruptures,
 with two heavier fragments at the outside borders of the chains.
  The projection onto the axis with $M_{1}$ is also shown on the left
side. We see an enhanced yield
for masses with $A = M_3 = 30$ and $A = M_3$ = 70-80, which indicate the emission of lighter
central masses with $Z_3$=8,10 as well as for  $Z_3$=20-28.

In the work in Ref.~\cite{Pyat12}, the neutron detectors have been arranged perpendicular to
the fission axis of the binary coincidences,
defined by the two arrays of MiniFOBOS detectors.
This orientation has been chosen to have an increased efficiency for
the spontaneous neutrons emitted during fission, potentially from the
two necks with the lighter fragments, this explains the observations in Fig.\ref{fig:Fig30n=2}.

\begin{figure} 
\begin{center}
\vspace{+4 mm}
\includegraphics[width=0.45\textwidth]{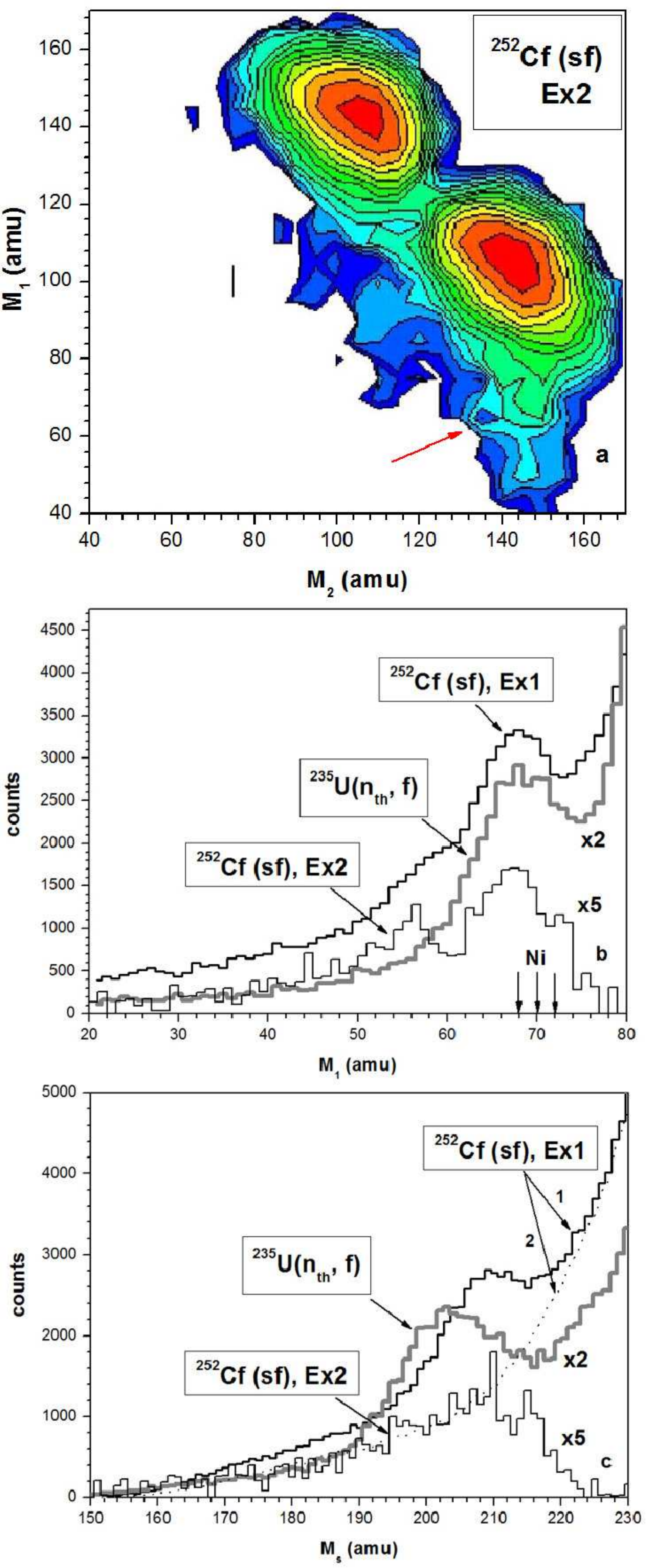}
\caption{
      a) Contour map of the  mass-mass correlation of the MINI-Fobos experiment with coincident neutrons.
      Note the absence of the scattering tails observed in the other
        cases shown in Fig.\ref{fig:Fig26} and
      Fig.\ref{fig:Fig27};
       Projections (onto the M$_1$ (b) and of the sum,  M$_s$,  of registered masses (c))
      of the yields for comparison with different experiments as indicated;
     For Ex2 yields of masses from the experiment with neutron coincidences, (MiniFobos)
      is substracted.
        }
\label{fig:Fig29}
\end{center}
\end{figure}

a) Contour map of the mass-mass
distribution (logarithmic scale, with lines approximately a step
factor of 1.5) from a coincidence in the two opposite arms of
Ex2. The bump in the spectrometer arm (arm1) facing the
backing of the Cf source is marked by the arrow. b) Projections
onto the M1-axis for comparison with the experiment Ex1,
and with the results of the 235U(nth, f) reaction [1]. Positions
of the magic isotopes of Ni are marked by the arrows (see
text of sect. 4.2 for details). c) Projections onto the direction
Ms = M2 +M1. The result for Ex1 is presented by two curves
marked by the arrows 1 and 2 (dotted) for the arm1 and arm2,
respectively. For Ex2 the yield of arm2 is subtracted
\begin{figure} 
\begin{center}
\vspace{+4 mm}
\includegraphics[width=0.45\textwidth]{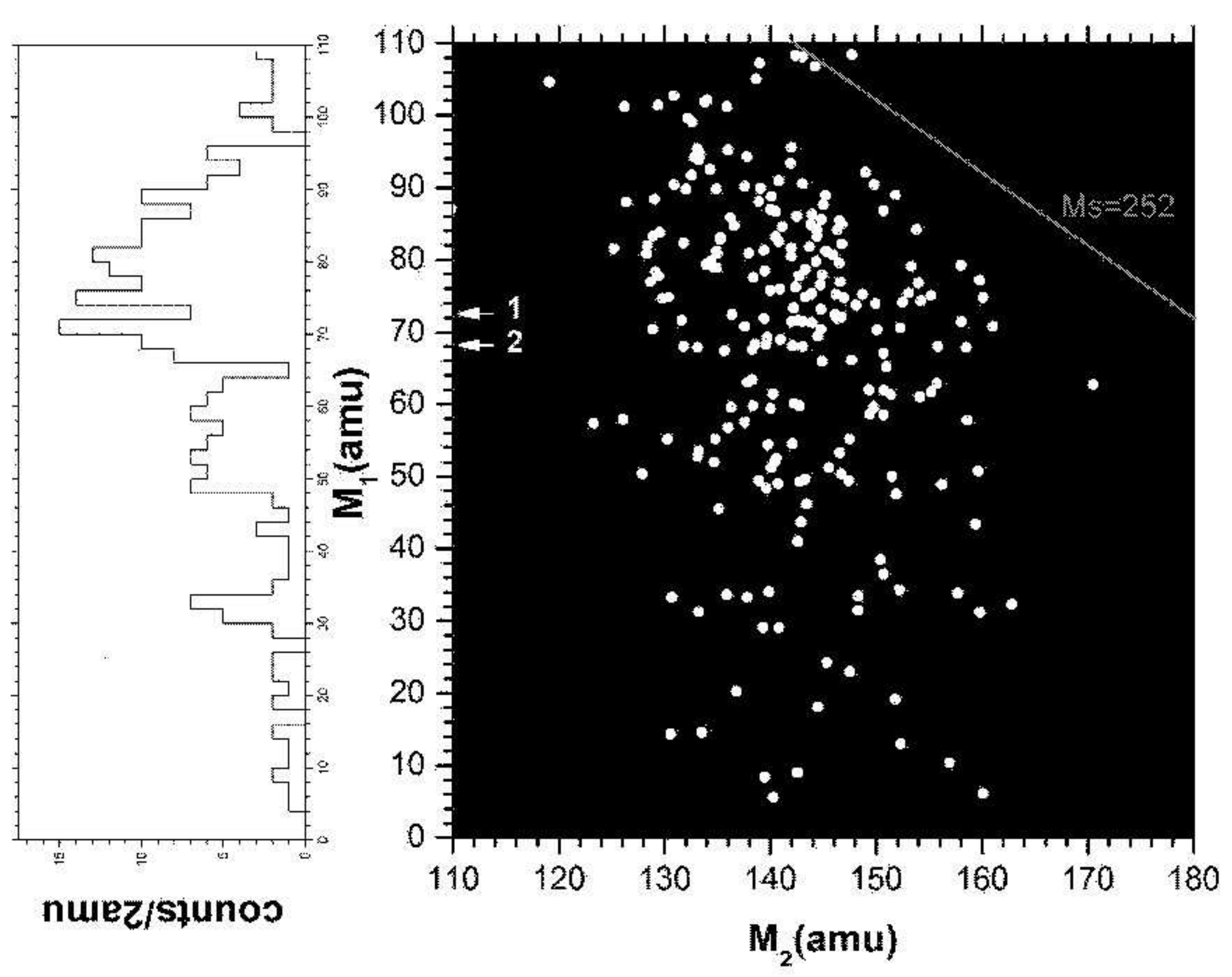}
\caption{The  yields of masses from the experiment with a chosen multiplicity in
         neutron coincidences (n=2), (MiniFobos),
        shown in a correlation of (M1-M2) and projections on the axis of
            M$_1$ for fragments with missing light fragments M$_3$ (with charges Z). }
\label{fig:Fig30n=2}
\end{center}
\end{figure}

\begin{figure} 
\begin{center}
\vspace{+2 mm}
\includegraphics[width=0.45\textwidth]{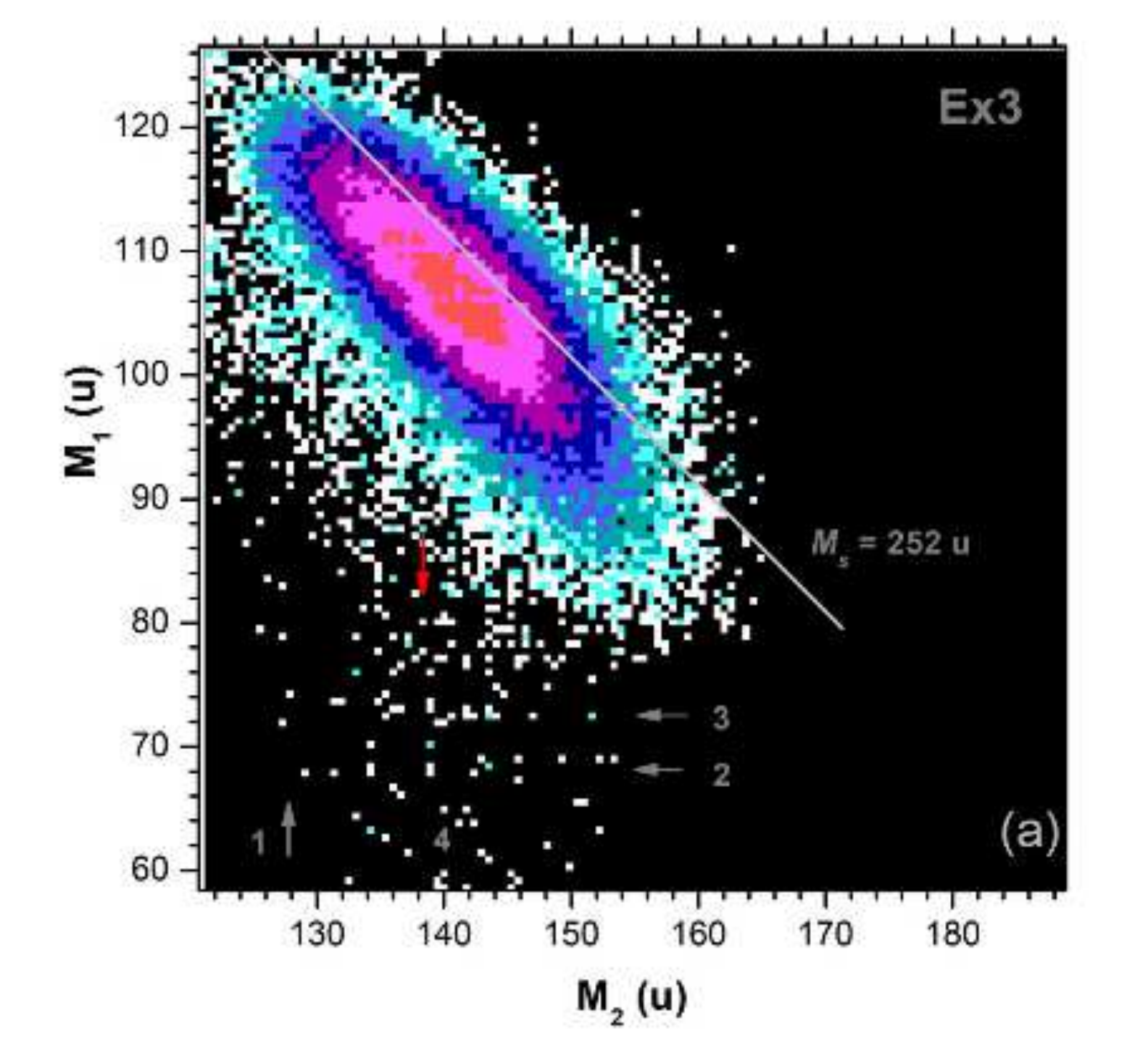}
\caption{ Binary coincidences of fragments registered with the COMETA set-up.
The region of the heavy fragments around masses $A$ = 120 -- 160 in the arm1
is shown for the
dominant binary mass. In addition we observe corresponding
to the ``Ni-bump'', now Ni isotopes well separated in mass
(from Ref.~\cite{Pyat12}).
The Ni-isotopes with mass 68 and 72 (indicated by labels 2 and 3
in the 2D-plot of M$_1$ - M$_2$)
 appear in the projection on the  M$_1$ axis, as
 shown in Fig.~\ref{fig:Fig31b}.
The sum-line   M$_s$ of the fragments is shown.}
\label{fig:Fig31a}
\end{center}
\end{figure}

 With the knowledge from the calculations of the kinetic energies
described before (shown in figure ~\ref{fig:Fig4}),
 we have concluded that Ni-isotopes  can be observed as fission fragments
with the missing mass of Ca-isotopes in the ternary decay of $^{252}$Cf.
 As before in Ex1 and Ex2 the ``Ni''-yield can be observed in the COMETA-stup,
  because the central fragment is  absorbed in the source and the
source-backing. The experimental result is shown in Fig.~\ref{fig:Fig31a} and
with the projection on the mass scale for the observed Ni-isotopes in Fig.~\ref{fig:Fig31b}.
Their overall (summed over the isotopes) yield is  equal to 2.5 x 10$^{-4}$ per
 binary fission. A further discussion of the
COMETA-experiment with their results is given in Ref.~\cite{Pyat17}, where energy spectra
in correlations between the two heavier fragments (FF$_1$ and FF$_3$) are discussed.
 In this work three different energy groups appear,
which are interpreted by three different decay scenarios, with FF$_2$ and FF$_1$ in different
geometrical arrangements. A big variety of geometrical arrangements of
the intermediate system  (FF$_1$+FF$_2$) (in the form of pictograms)
 are proposed, which
can produce the different correlated energy groups observed.
 Most importantly the  number of different
groups in the energy correlations can be explained,
the intermediate system consisting of the fragments FF$_1$ and FF$_2$
gives different energies, in one case the rotation of the intermediate system of  (FF$_1$+FF$_2$)
is assumed, which gives particular low values of the energies of  FF$_1$ and FF$_3$, as observed.
 Further a  critical assessment of other theoretical works is given.

\begin{figure} 
\begin{center}
\vspace{+2 mm}
\includegraphics[width=0.40\textwidth]{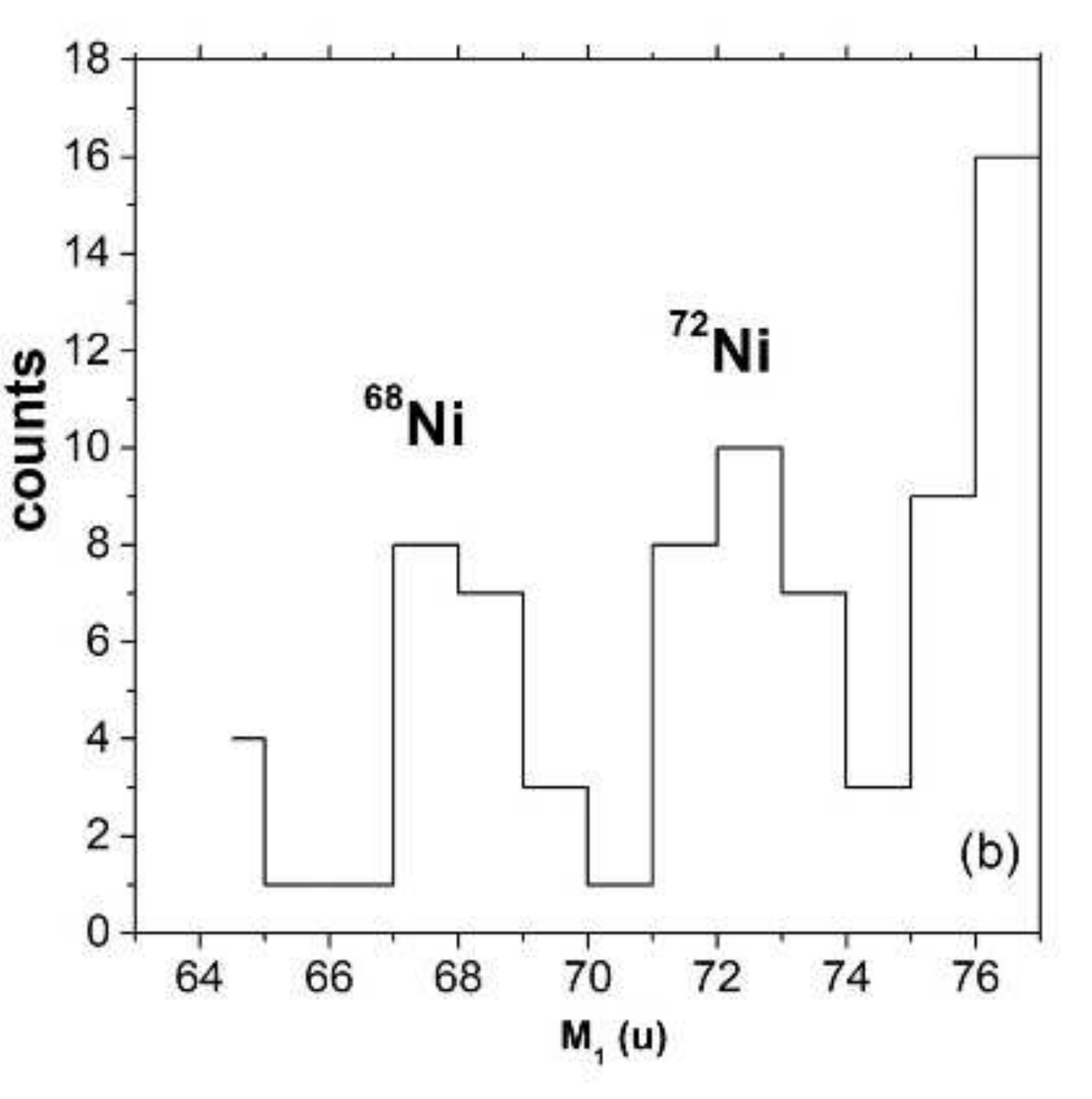}
\vspace{-2 mm}
\caption{Projection of the masses observed for  the binary coincidences
 of fragments with masses in the Ni-region (see Fig.~\ref{fig:Fig31a}),
 registered in the arm1 (from Ref.~\cite{Pyat12}). }
\label{fig:Fig31b}
\end{center}
\end{figure}

\section{Ternary decays, potential energy surfaces, multi-modal decays}
\label{sect4}
\subsection{Ternary decays, potential energy surfaces}
 Ternary fission into fragments with comparable masses is a process,
which occurs in heavy nuclei under
conditions  of large values of the fissibility parameter: $X$, for the ratios $Z^2/A >$ 31.
The decay into  three heavier fragments (true ternary fission) is
found to be  collinear, as observed in the recent experiments
and discussed in the previous sections. This important dynamical aspect of true ternary
 fission has in fact  been often predicted  in the last decades~\cite{DG,Manim1,poe05}.
The ternary decay has $Q$-values which are larger than for  binary channels,
for the favoured  ternary channels this amounts to additional 20 -- 30 MeV
see Fig.~\ref{fig:Fig7}.
This fact is observed by inspecting the potentials  and the potential energy
 surfaces
for the decay which can be calculated as function of the masses of the fragments formed in the decay.

\begin{figure} 
\vspace{+5 mm}
\includegraphics[width=0.48\textwidth]{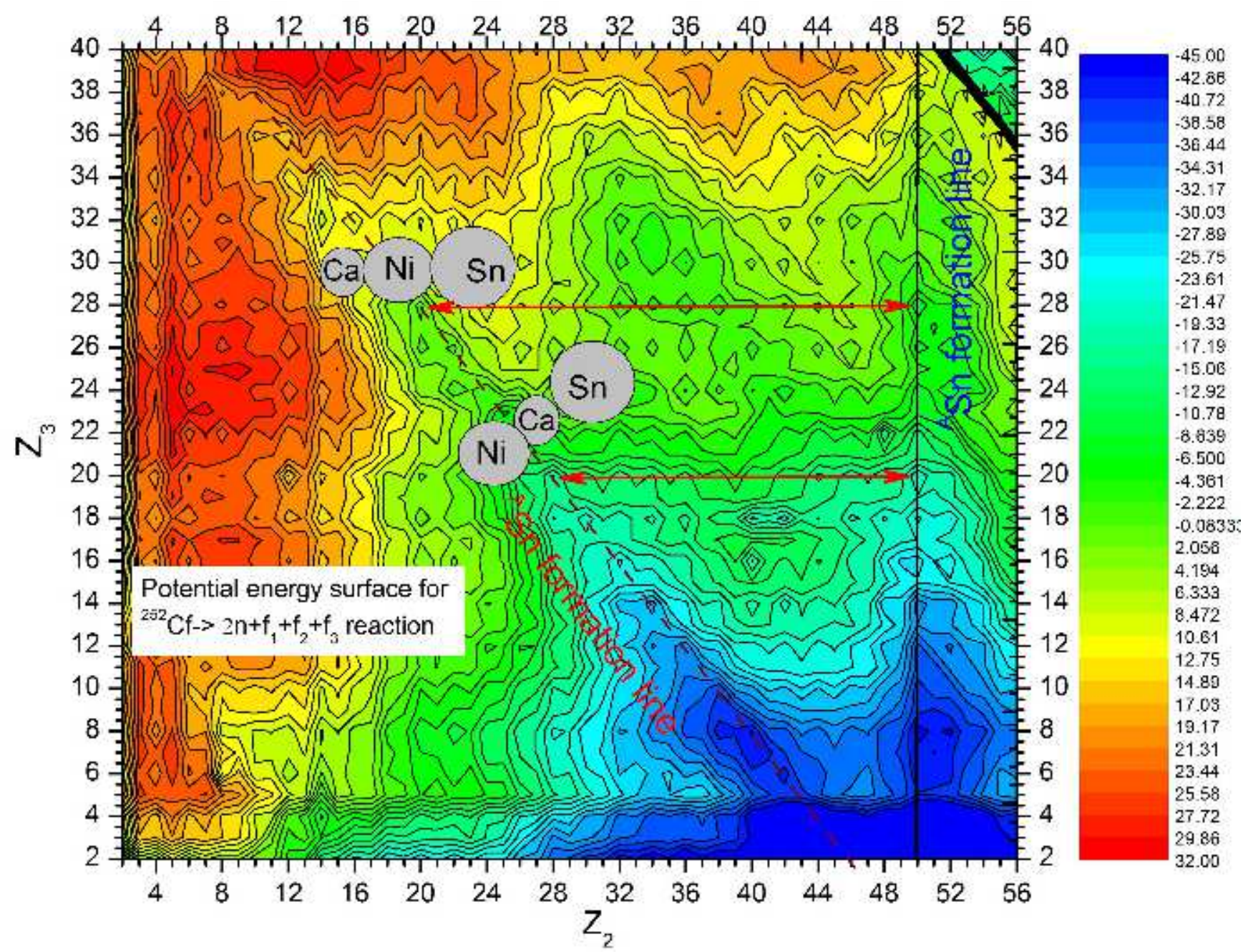}
\vspace{1 mm}
\caption{
The potential energy surface  for the ternary decay of $^{252}$Cf(2n,sf),
with the emission of two neutrons,
fragments ${A3 - A2}$
with dominating ${A2}$ nuclei with $Z_2$ = 50. The combination of $Z_2$= 50
and $Z_3$ = 20, or /and $Z_1$ = 28
are clearly seen to be favored.}
\label{fig:Fig33}
\end{figure}

Binary coincidences are obtained between fragments $A_{2}$, in arm Nr.2
and $A_{1}$ in arm Nr.1
with the missing mass method. The masses and their vectors of two fragments are measured,
thus  the mass $A_3$ is uniquely determined, it corresponds  to Ca isotopes
in the case of  $^{252}$Cf.
This new exotic  decay  can be understood as the breakup of very prolate
deformed elongated hyper-deformed
shapes, as discussed for hyper-deformation in  $^{236}$U in  Ref.~\cite{U236Hyperd},
and shown in Fig.\ref{fig:Fig3}.
The decay is considered with two sequential neck ruptures~\cite{vijay12,voe12}, as
illustrated in Fig.\ref{fig:Fig19}. Actually
the central fragment ${\it A}_3$ has extremely low kinetic energy (see Fig.\ref{fig:Fig4})
 and is mostly lost.
The kinetic energies of the fragments have been calculated for the  sequential kinematics as
shown in Fig.\ref{fig:Fig4}.
We find that the central fragment attains very low kinetic energies, if we assume that the
 nucleus $A_{23}$ has some intermediate excitation energy of (10--30 MeV).

The main effect in the missing mass FOBOS-experiments with  the binary coincidences
is the difference in the counting rates (mass spectra) in the two arms of the coincidence arrangement.
Two fragments of the ternary decay travel in one, arm1, (see Ref.~\cite{Pyat10}) through the
dispersive media, the  source backing and the foils of the start detector. Thus the dispersive effect
(angular dispersion of  1-2$^o$) of the two fragments from the ternary decay
 is only present in arm1(!) with the
target/source backing  pointing to the detector of arm1.  Thus the missing mass effect appears
in the {\it counting rate difference} and the difference in the mass spectra of arm1 and arm2
(the difference {\it N}(arm1)-{\it N}(arm2)). In Fig.\ref{fig:Fig26}
 two mass spectra of arm1 and arm2
and their difference are shown for the case of  $^{252}$Cf,
 as well as the raw data. The yield derived from the peak in the difference is 4.7 x 10$^{-3}$

For the consideration of the absolute and relative probabilities, the phase space (see below)
and the barriers have to be discussed.
With the PES for $^{252}$Cf we observe several places favored for ternary decays.
In order to judge the importance of the various decay channels, the internal decay barriers
have been calculated. In the sequential decay mechanism of the ternary fission, the
splitting system goes through
two corresponding barriers at each step of fission, these barriers are the same for
a symmetric decay.
These barriers are shown in Fig.~\ref{barriers}, where the potential energy of the
 ternary system $^{70}$Ni+$^{50}$Ca+$^{132}$Sn is shown in a two dimensional plot
for the collinear configuration.
The motion along $R_{13}$ ($R_{23}$)  at the fixed value of $R_{23}$ ($R_{13}$)
is an example for the sequential fission. The case, when $R_{13}$ and $R_{23}$
are increased simultaneously, the simultaneous decay of the ternary system into three parts
 at the same time will occur. In Fig. \ref{fig:Fig7}, the motion along
 the diagonal line corresponds to the last kind of decay.
It is seen from Fig. \ref{barriers} that the barrier for the simultaneous
 fission is much higher than the barriers which appear at the sequential fission.
 This figure allows to judge the relative importance
of the decay channels. The symmetric decay with
$^{84}$Ge +  $^{84}$Se + $^{84}$Ge  occurs if  the symmetric configuration is populated
 in the corresponding valley of the
potential surface.
This figure further shows an important fact: the initially observed
CCT-decay with the closed shell nuclei (Sn,Ca,Ni) has two  different internal
 barriers, which are the lowest barriers.
Variations of the neutron number
adds further favored channels, so the total yield becomes quite large.
\begin{figure} 
\vspace{+4 mm}
\hspace{+2 mm}
\includegraphics[width=0.48\textwidth]{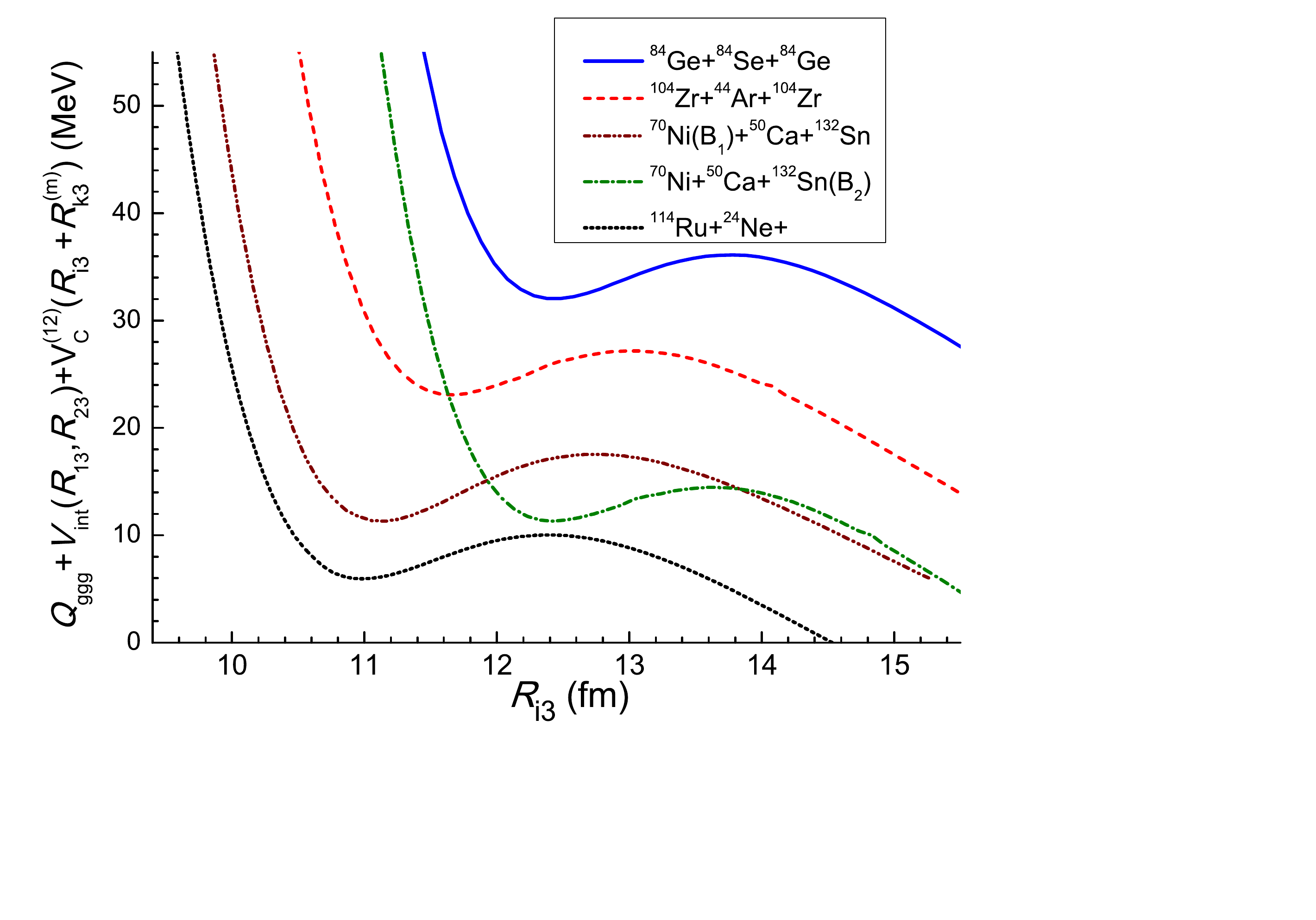}
\vspace{-6 mm}
\includegraphics[width=0.48\textwidth]{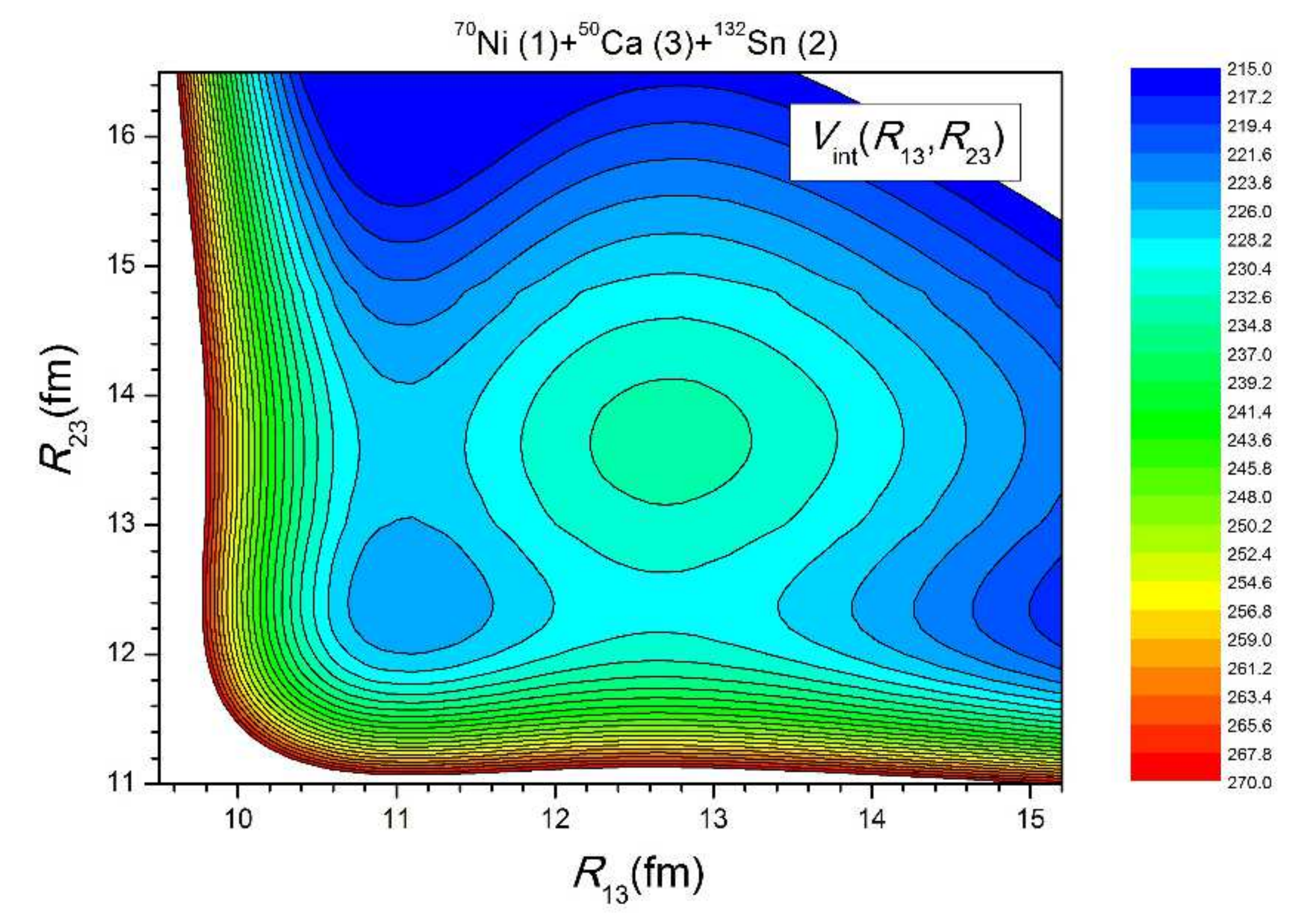}
\vspace{+6 mm}
\caption{ Upper part: Comparison of the barriers of symmetric ternary collinear cluster
 decays of $^{252}$Cf and other ternary decay modes.
The most symmetric decays have the highest barriers. Lower part:
 the barriers in a 3D plot for
the non symmetric decay
into (Ni+ Ca + Sn), in a correlation of two characteristic distances R$_{13}$ and R$_{23}$.}
\label{barriers}
\end{figure}

 The other symmetric channels
with $^{44}$Ar and $^{24}$Ne have lower barriers, we expect that the
probability increases
 for smaller values  on the charge and the mass of
the central fragment. This hierarchy corresponds to the experimental observations.
The barriers of the most important ternary collinear mass splits are shown in
the upper panel of Fig.~\ref{barriers}.
The probabilities for the observation of different modes
depend on the formation probability of fragments and on
 the barriers for the collinear decay channels.
 For mass-symmetric decays one barrier will be  sufficient to characterize the channel,
however, for the CCT-decay observed in Ref.~\cite{Pyat10,Pyat12},
with  $^{132}$Sn + $^{50}$Ca + $^{70}$Ni,
fragments as the dominant channel, two barriers are relevant, see the bottom
panel of Fig.~\ref{barriers}.
The experimental observations clearly reflect with their  probabilities
the influence of the barriers:, the most symmetric mass partition  has the lowest yield,
the CCT decay is of intermediate probability, the CCT decay with a light fragment, like $^{24}$Ne,
in the middle,
 would have the highest yield~\cite{voe15,Pyat14}.

\subsection{Potential energy surfaces and the multi-modal ternary decays}
The fission processes is  a statistical decay of the total (compound) nucleus, CN,
as already discussed in 1939 by N. Bohr and J.A. Wheeler~\cite{Bohr39}.
Actually the ternary decay occurs with two neck ruptures in a short time-sequence,
 as discussed in Ref.~\cite{Tashk2011}.
The probabilities of the
ruptures  are governed by the internal  prescission barriers, by the phase space and   by
  the PES  in
 each configuration. The PES are obtained
as described in Refs.~\cite{Tashk2011,Nasirov14}, and are shown in Figs.\ref{fig:Fig12_252Cf0n}
  and \ref{fig:Fig13_PESU236}.
The total  phase space of these statistical decays is determined by:\\
 i) the energy balance and thus the details of the potential energy surface, PES, namely,
 its valleys and hills, ii) the internal barriers for the two necks, and
 iii) the $Q_{ggg}$-values, the latter determining  the kinetic energies and the
number of possible fragment (isotope) combinations,\\
 iv) the excitation energy range in the individual fragments, v) their momentum range,\\
 vi) the number of excited states (or the density of states) in each of the fragments, the
 combinations consisting of 2(or 3) isotopes,  and by \\
 vii) the spin ({\it $J$}) multiplicity in
these excited states  with  spins up to (6-8)$^+$ (phase space factor $(2J+1)$).

We have seen that the PES for the case of neutron induced fission,  $^{235}$U(n,fff),
 shown in  Fig.\ref{fig:Fig13_PESU236}, shows different details
in structure compared to the $^{252}$Cf(sf) case due
to the absence of shells in the favored fragments. Details of this decay have  been discussed by
 Tashkhodjaev {\it et al.} in Ref.\cite{Tashk2011}, and we
show in Fig.\ref{fig:Fig35a} the yield of fragments
obtained in this work. A much wider distribution in masses of fission fragments
 is observed as compared to the case of the spontaneous fission in  $^{252}$Cf(sf).

\begin{figure} 
\hspace{+4 mm}
\vspace{+10 mm}
\includegraphics[width=0.45\textwidth,angle=0]{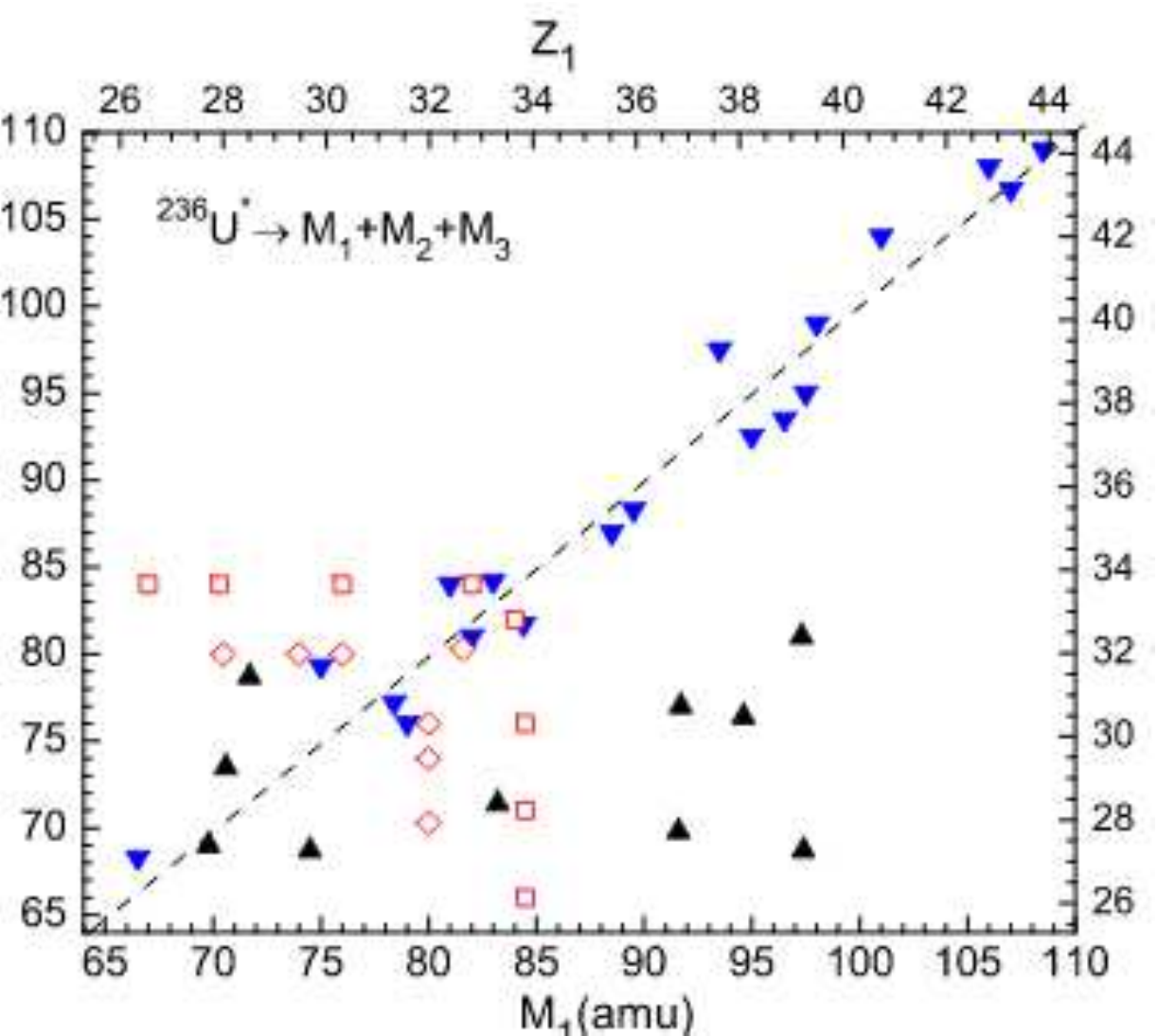}
\vspace{-7 mm}
\caption{
Results of the calculation by Tashkhodjaev et al.,~\cite{Tashk2011},
for the fragments yields of sequential ternary fission in  $^{235}$U(n,fff).
 The wider (as compared to  $^{252}$Cf(sf)) overall spread in masses of fragments is obtained,
 as predicted from the PES
in  Fig.\ref{fig:Fig13_PESU236}. Up triangles, results of the experimental data selected.
Red diamonds and red squares
the result of symmetric (sequential, with approximately equal momenta)
 decays  $^{235}$U$^*$ $->$ ($^{154}$Nd$^{*}$ + $^{82}$Ge$^*$) $->$ $^{80}$Ge + $^{72}$Ni
and  ($^{150}$Ce$^{*}$$->$ $^{82}$Zn + $^{72}$Ni)
+  $^{82}$Ge$^*$.
Fragments with comparable masses have been selected.}
\label{fig:Fig35a}
\end{figure}

 In this work attentions is paid to the fact of the
sequential nature of the ternary fission decay, which produces in the first step
two fragments with excitation energies,  $^{154}$Nd$^{*}$ at 25.3 MeV and  $^{82}$Ge$^*$,
which  leads to the  evaporation of neutrons.
In the second step  the fission of $^{154}$Nd$^*$ into  $^{72}$Ni +  $^{80}$Ge$^*$ will occur.
These predictions, with a yield of 1.5x10$^{-4}$/(binary fission) are in very good
agreement with the experimental observations.

\begin{figure}[t] 
\hspace{+0 mm}
\vspace{+10 mm}
\includegraphics[width=0.45\textwidth,angle=0]{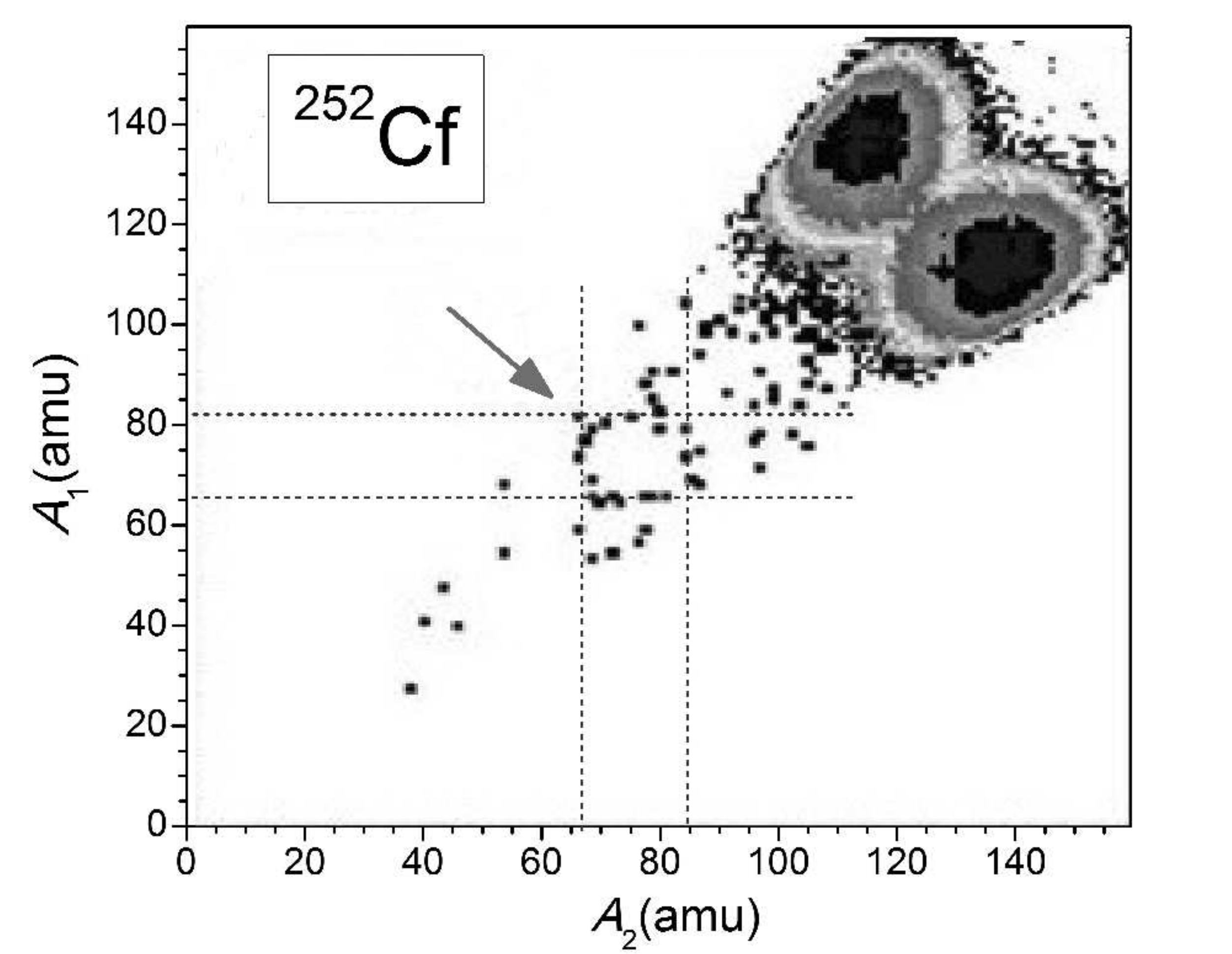}
\vspace{-9 mm}
\caption{ Binary coincidences of fragments in  the spontaneous ternary
          decay (FFF) of  $^{252}$Cf(sf) selected for the symmetric
           decay, from the inclusive data,  measured  in Ref.~\cite{Pyat10}.
          The outer fragments ({\it A}$_1$, {\it A}$_2$, with masses {\it M}$_1$, {\it M}$_2$)
           are selected with the condition for the velocities
          and the momenta $P_i$: ($V_1\approx V_2$, and $P_1\approx P_2$).
          Remnants of the binary fragments in coincidences  are seen.
          The region of the FFF-decays
          is marked. Missing fragments ({\it A}$_3$) here are again
           isotopes of nuclei with ({\it Z}$_3$ = 30 - 36).
          Scattered points above this region are true coincidences  of other
          symmetric ternary decays, from Ref.\cite{voe14}.}
\label{fig:Fig36}
\end{figure}

\subsection{Multi-modal ternary decays}
\label{multi}

In Fig.~\ref{fig:Fig30n=2}, we show the  yield in an experiment
Ref.~\cite{Pyat12} with coincident neutrons (multiplicity n=2).
This experiment selects a region of lighter missing masses of ternary fragments.
These light neutron-rich  fragments are typically created at the center of the system
 with the emission of neutrons. The area is seen in the lower part
 (blue area) of the PES's (in Figs.8 and 9). The projection on the  A1 - axis
shows that these are isotopes of Neon, Oxygen and Carbon.
The examples indicate, that the multiple modes of the ternary fission decays,
can successfully be
predicted with the PES's, however,
it can be very difficult to extract the corresponding yield from the raw data.

For the ternary decays  of  $^{252}$Cf with a collinear arrangement of the three fragments the
 PES's, the contour-plot in Fig.~\ref{fig:Fig27} show distinct minima for various charge
combinations (multi-modal ternary decays)
 with $\Sigma_Z$ = 98:\\
i) for CCT  \emph {$Z_3$} =  20, and {\it Z}$_1$ = 28, this is the main CCT-mode
 observed in Refs.~\cite{Pyat10} and\\
ii) less pronounced are  charge combinations \emph {$Z_3$} = 28, and  \emph {$Z_1$} = 20,
as observed in Ref.~\cite{Pyat10,Pyat12}.
The complementary  fragments with $Z_2$ are isotopes of Sn ({\it Z} = 50).
The PES shows a pronounced valley with charge {\it $Z_2$} = 50 and
neutron number $N_2$=126 , due to the closed shell for
the number of protons and neutrons. Because of the dominance of
the Coulomb interaction, the proton shells are
the most important. The individual  probabilities vary over several orders
 of magnitude, see also Ref.~\cite{voe15}.\\
iii) The ternary fissions with {\it $Z_3$} = 18, are predicted.\\
iv) We observe a pronounced region of minima for the symmetric charge combinations with
 three comparable fragments, for the FFF-decays (this decay is marked as FFF):\\
 for  ($Z_1$ = 32, 34, 32), and ($Z_3$ = 34, 32, 32), the fragment $Z_2$ has an
 equivalent role as the other two  $Z$-values $Z_2$ = 32, 32, 34 (since
 $Z_1+Z_2+Z_3=98$)  they can be interchanged
 -- we have an almost symmetric ternary decay.
Because of this fact,  permutations of
 the labels including $Z_2$ in the figure of the PES (see also Fig.\ref{fig:Fig7})
  will produce similar results, and a symmetric shape of the coincident events in the experiment.
A quite equivalent decay pattern is observed in the ternary fission in the  $^{236}$U(n,fff) reaction,
where the dominant decays are connected to the  proton-shells.
 We show in Fig.\ref{fig:Fig13_PESU236}
the PES for this case, the observed decays (Ref.\cite{voe14}) are analogous to the
 spontaneous fission in
 $^{252}$Cf(sf). Several favored fragment combinations are observed, including the
             symmetric decay discussed in Refs.~\cite{voe14,voe15}.

\begin{figure} 
\vspace{5mm}
\includegraphics[width=0.430\textwidth]{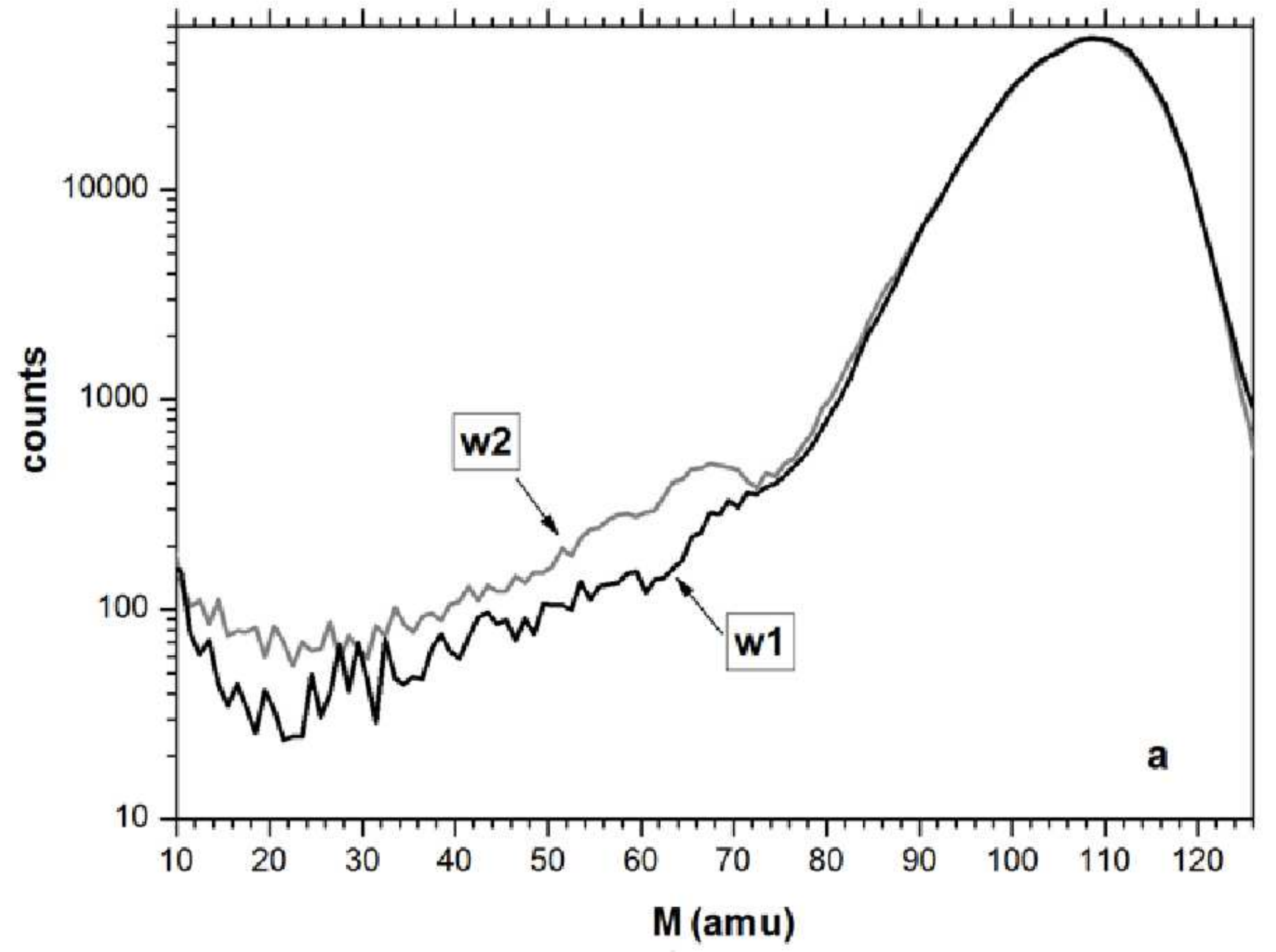}
\caption{Extraction of the yield from the data of Ref.~\cite{Pyat10,Pyat14}: the ``Mo-Ar''-mode.
        The two different mass spectra, obtained from the two arms (arm1 and arm2).
          The data are projected to the
          A$_1$ axis for arm1 and arm2
          measured  in Ref.~\cite{Pyat10} with binary coincidences.
                    }
\label{fig:Fig37}
\end{figure}

For $^{252}$Cf(sf)  we can select this symmetric decay mode  from
 the overall data of Ref.~\cite{Pyat10,Pyat12}
by choosing conditions on the momenta (velocities) of $Z_2$ and $Z_1$,
with ($P_1 = P_2$), see Fig.~\ref{fig:Fig36}. This choice produces
 an   rectangular two-dimensional field of the experimental binary coincident events
in the favored region (similar result will be obtained by the reflection of the labels),
see also Fig.~\ref{fig:Fig7}.

\begin{figure}[t] 
\vspace{01mm}
\includegraphics[width=0.430\textwidth]{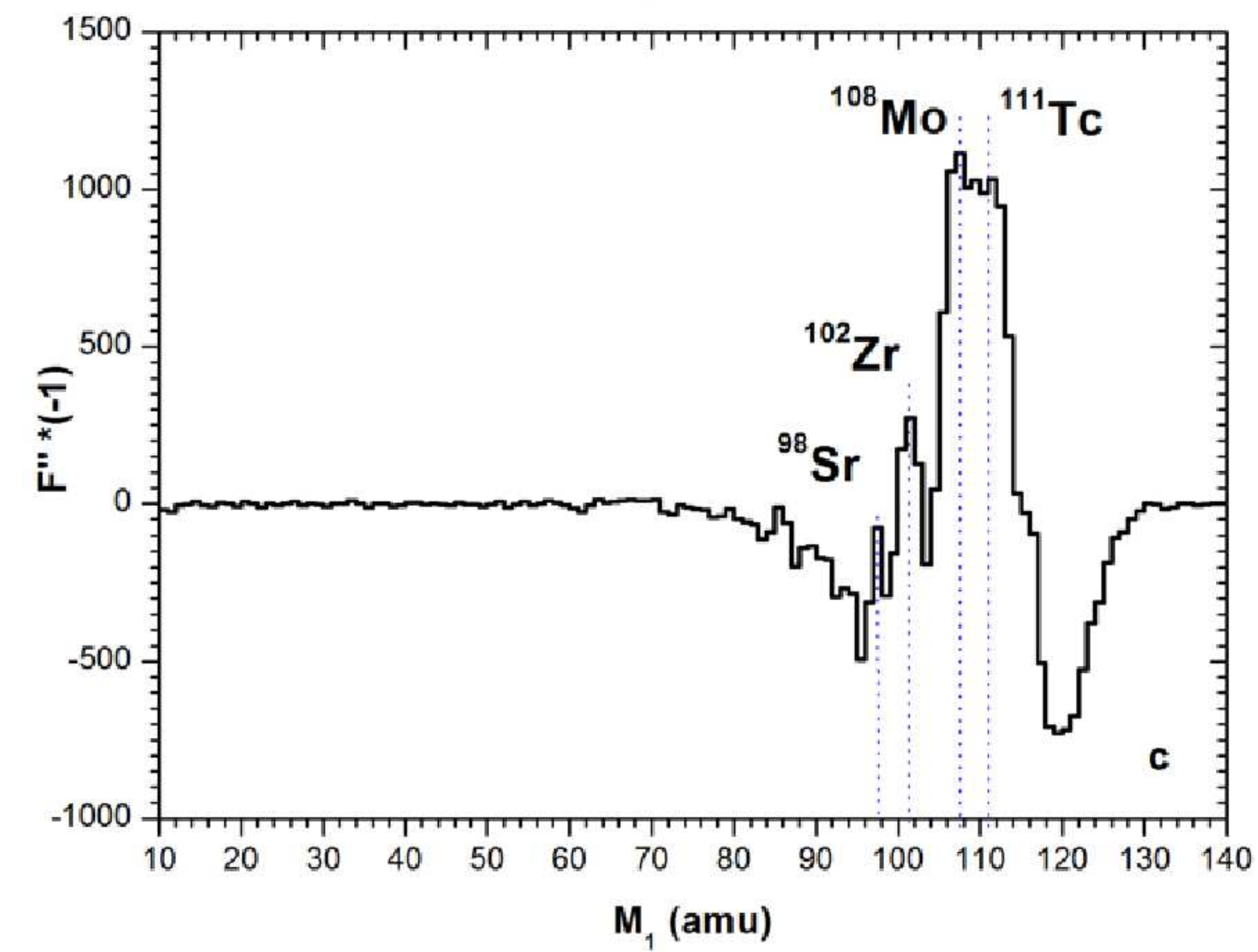}
\caption{The  derivative of the  difference of
          the two spectra shown in Fig.\ref{fig:Fig37}, events in the region of $^{106}$Mo appear.
          The complementary missing fragments, ($A_3$),
          here are isotopes of Ar with ($Z_3$ = 18).
          }
\label{fig:Fig38}
\end{figure}

An unexpected favored ternary decay mode is connected to the charge $Z = 18$ seen in the PES
of both cases  $^{252}$Cf(sf) and  $^{236}$U(n,fff)
as a blue dip. A special procedure had to be chosen to obtain the yield of this ``Mo-mode''.
The original spectra show only a small difference in the two detectors.
Making a first derivative of this difference we observe a strong peak for the
mass region of $A_3$ = 102 -- 110 (shown in Fig.\ref{fig:Fig38}).

 \section*{Summary}
In this survey we have discussed several theoretical approaches to the ternary fission  decay
of heavy nuclei. From principal considerations we conclude that the decay will be sequential.
This conclusion comes from arguments with the phase space, this is generally much smaller
(three decay vectors) for spontaneous three-body final channels than in
a sequence of two two-body decays. In the latter two barriers appear in the
respective two-body channels. For  this case the kinetic energies have been
calculated for the three fragments in a collinear geometry. In this
case the central (smaller) fragment
attains energies close to zero (0), giving a binary coincidence event with a missing mass.
 Thus in the experiments discussed, the central fragments
are generally lost by absorption in the target and/or in the target backing.
Some detailed predictions of the
 sequential ternary  decay are consistently described in the theoretical analysis
in  two  of the studied cases.

From the two experimental cases of the observed ternary fission in heavy nuclei with U and Cf, which have
 large $Q$-values for a split into
three nuclear fragments with comparable masses studied in the present survey,
 we can conclude
that the appearance of deformed shells in the total system and in the  fragments
(for certain proton and neutron numbers)
 turns out to be decisive for the occurrence of true ternary fission at
low excitation energy. In all cases
the fragments with $Z$ = 50 (Tin isotopes) dominate the outcome of the decay. The ternary decay
in  $^{252}$Cf(sf) is a unique case, because of deformed shells and because in all
 three fragments  closed shells for proton
 numbers $Z$ = 20, 28 and 50 and for neutron numbers (20, 28, 50, 126) appear,
these induce the decisive lowering of the second barrier.
In the case of  $^{252}$Cf(sf) the shells in the lighter fragments appear with, $Z$ = 8,20,28.
For the discussion of the decay modes the potential energy surfaces (PES's) are shown.

In the survey of the experimental data, we notice that for completely independent experimental setups
with binary coincidences for two fragments (for $^{252}$Cf(sf)) and for
the completely different physical case,  $^{235}$U(n,fff) with a smaller total mass and charge,
 we have the same physical phenomenon, the observation of the missing mass in the
 coincidence of two
binary fragments, with a comparable  mass for the third fragment.
The latter ($^{236}$U(n,fff)) case shows the same  feature with  different and smaller (as expected)
 ternary fragment masses.
 The decays are observed  with dominant fragments with $Z$ = 50,
 which are consistent with the independent and different theoretical
predictions, mainly based on  the PES's. Based on
 the PES's different ternary decays, multi-modal decays, can be predicted. These have been extracted
 by using corresponding conditions (gates in the experimental data)
in the analysis of the experimental data.
In the analysis of the various decay properties we find that the models based
 on preformed clusters
show deficiencies, which are only partially removed by introducing deformation in
the fragments. The approaches with
three preformed clusters allow the calculation of the PES`s, which determine the phase space
for the decay after scission, once the fragments have been formed.
In an detailed analysis of ternary fragmentations,
based on an approach with three clusters by Holmval et al.\cite{Holm17}
it is actually shown
that in the case of $^{252}$Cf(sf) ternary fragmentation
 is not possible if nuclear interaction between fragment and deformed shell effects are not considered.

The approach of Karpov~\cite{Karpov16} based on the three-center shell model,
 with strongly deformed shapes
contains the maximum of necessary macroscopic and microscopic
features and gives the best description for the dynamics with the formation of  two necks,
and finally of the three fragments. This approach can be  connected to
 the concept of \textit{hyper-deformation} with strongly deformed (elongated)
shapes, which show the path towards ternary decays.
The present work shows the uniqueness
 of the case of  $^{252}$Cf(sf), namely that the
ternary fragmentation occurs
\textit{due the shell effects in the decaying nucleus and in all three fragments}.
Already in the dominant
 case of fragments  with $Z$=50, the decay barriers are
high in the liquid drop potential energies, fission can only be observed due
 the shell effects in the fragments with $Z$ = 20,  28 and $Z$ = 50. For the case
of $^{235}$U(n,fff) the excitation energy brought in by the neutron capture is decisive
to induce ternary fragmentation.

From our analysis we conclude that the CCT-decay observed in Ref.~\cite{Pyat10}, the collinear
cluster tripartition can be
considered as a manifestation of \textit{hyper-deformation} in heavy  deformed nuclei.

 \section*{Acknowledgments}
We thank A. Karpov, D. Kamanin and Y. Pyatkov for many useful discussions and preparation of figures.

 \section*{Dedication}
The authors dedicate this survey to the  memory of two colleges, who passed away in recent years,
Walther Greiner (passed away in 2016) of Frankfurt (Germany) and Valerij Zagrebaev (passed away in 2015)
 of FLNR  in Dubna, (Russian Federation)
(e.g. Refs. \cite{Zagreb10a,Zagreb10}). They had contributed  many ideas and
 important steps for the development
of various aspects in the physics of true ternary fission.


\end{document}